\documentclass[12pt]{amsart}
\usepackage[margin=1in]{geometry}

\usepackage{fullpage}
\usepackage{amssymb,amsfonts}
\usepackage[all,arc]{xy}
\usepackage{enumerate}
\usepackage{enumitem}
\usepackage{mathrsfs}
\usepackage{tabulary}
\usepackage{mathtools}
\usepackage{tikz}
\usepackage{tikz-cd}
\usepackage{physics}
\usepackage{graphicx}
\usepackage[font={small}]{caption}
\usepackage{float}
\usepackage{subcaption}
\usepackage[colorlinks=true]{hyperref}
\usepackage{rotating}
\usepackage{stmaryrd}
\usepackage{standalone}
\usepackage{relsize}
\usepackage{accents}
\usepackage{tabularx}
\usepackage[toc]{appendix}

\usetikzlibrary{knots}
\usetikzlibrary{decorations.pathreplacing}
\usetikzlibrary{hobby}
\usetikzlibrary{positioning,fit}
\usetikzlibrary{math}
\usetikzlibrary{calc}
\usetikzlibrary{shapes.multipart}
\usetikzlibrary{decorations.markings}
\usetikzlibrary{braids}
\usetikzlibrary{patterns}
\usepackage{longtable}
\usepackage{lscape}

\DeclareMathOperator{\bF}{\mathbb{F}}

\DeclareMathOperator{\Z}{\mathbb{Z}}

\DeclareMathOperator{\R}{\mathbb{R}}
\DeclareMathOperator{\C}{\mathbb{C}}
\DeclareMathOperator{\A}{\mathbb{A}}
\DeclareMathOperator{\id}{id}
\DeclareMathOperator{\im}{im}
\DeclareMathOperator{\Hom}{Hom}
\DeclareMathOperator{\basis}{\mathcal{B}}
\newcommand{\til}[1]{\widetilde{#1}}
\newcommand{\ol}[1]{\overline{#1}}

\newcommand{\lr}[1]{\vert {#1} \vert}
\renewcommand{\b}[1]{\overline{#1}}
\newcommand{\dist}{\hat{d}}
\newcommand{\F}{\mathcal{F}} 

\renewcommand{\o}{\otimes}
\newcommand{\eps}{\varepsilon}
\newcommand{\m}
{
\begin{tikzpicture}[baseline=-3pt]
\draw (0,0) circle [radius=.25];
\node (0,0) {$\mathsmaller{-}$};
\end{tikzpicture}
}

\newcommand{\p}
{
\begin{tikzpicture}[baseline=-3pt]
\draw (0,0) circle [radius=.25];
\node (0,0) {$\mathsmaller{+}$};
\end{tikzpicture}
}
\newcommand{\circled}[1]
{
\begin{tikzpicture}[baseline=-3pt]
\draw (0,0) circle [radius=.25];
\node (0,0) {$\mathsmaller{#1}$};
\end{tikzpicture}
}
\newcommand{\circledd}[1]
{
\begin{tikzpicture}[baseline=-3pt]
\draw (0,0) circle [radius=.25];
\filldraw (-0.25,0) circle [radius=.04];
\node (0,0) {$\mathsmaller{#1}$};
\end{tikzpicture}
}
\newcommand{\circledm}[1]
{
\begin{tikzpicture}[baseline=-3pt]
\draw (0,.17) to [closed, curve through ={(.2,.25) (.5,0) (.2,-.25) (0,-.17) (-.2,-.25) (-.5,0) }] (-.2,.25);
\node (0,0) {$\mathsmaller{#1}$};
\end{tikzpicture}
}
\newcommand{\circledmd}[1]
{
\begin{tikzpicture}[baseline=-3pt]
\draw (0,.17) to [closed, curve through ={(.2,.25) (.5,0) (.2,-.25) (0,-.17) (-.2,-.25) (-.5,0) }] (-.2,.25);
\filldraw (-0.5,0) circle [radius=.04];
\node (0,0) {$\mathsmaller{#1}$};
\end{tikzpicture}
}

\newcommand{\RIIleftover}
{
\begin{tikzpicture}[baseline = -3, scale = 0.3]
    \draw (.75,1) to [curve through = {(-3/8,0)}] (.75,-1);
    \draw[black,preaction={draw,line width=3pt, white}] (-0.75,1) to [curve through = {(3/8,0)}] (-.75,-1);
    \draw[dotted] (0,0) circle (1.25); 
\end{tikzpicture}
}

\newcommand{\RIIrightover}
{
\begin{tikzpicture}[baseline = -3, scale = 0.3]
    \draw(-0.75,1) to [curve through = {(3/8,0)}] (-.75,-1);
    \draw (.75,1)[black,preaction={draw,line width=3pt, white}]  to [curve through = {(-3/8,0)}] (.75,-1);
    \draw[dotted] (0,0) circle (1.25); 
\end{tikzpicture}
}
\newcommand{\RIIcomplexa}[2]{
\begin{tikzpicture}[baseline = -3, scale =0.15, every node/.style={scale=0.6}];
    \draw (1.3,-2) to [curve through= {(0.3,-1.25) (.8,0) (0.3,1.25) }](1.3,2);
    \draw (-1.8,-1.6) to [closed, curve through = {(-0.3,-1.2) (-.8,0) (-0.3,1.2) (-1.8,1.6)}] (-2.5,0);
    \node at (-1.6,0) {#2};
    \node at (1.65,0) {#1};
\end{tikzpicture}
}
\newcommand{\RIIcomplexb}[2]{
\begin{tikzpicture}[baseline = -3, scale =0.15, every node/.style={scale=0.6}];
    \draw (0,-1.2) to [closed, curve through = {(-0.7, 0) (0,1.2)}] (0.7,0);
    \draw (1.2,-2) to [curve through = { (.2,-1.5) (-1.75,-1.6) (-2.5,0)  (-1.75,1.6) (.2,1.5)}] (1.2,2);
    \node at (0,0) {#1};
    \node at (1.6,1.1) {#2};
\end{tikzpicture}
}
\newcommand{\RIIcomplexc}[1]{
\begin{tikzpicture}[baseline = -3, scale =0.15, every node/.style={scale=0.6}];
    \draw (1.4,2) to [curve through ={(.2,1.5) (-1.75,1.6) (-2.5,0) (-1.75,-1.6) (-0.25,-1.4) (-0.7, 0.1) (0,1.1) (.7,0.1) (0.25,-1.4)}] (1.4,-2);
    \node at (1.5,1.1) {#1};
\end{tikzpicture}
}
\newcommand{\RIIcomplexd}[1]{
\begin{tikzpicture}[baseline = -3, scale =0.15, every node/.style={scale=0.6}];
    \draw (1.4,-2) to [curve through = {(.2,-1.5) (-1.75,-1.6) (-2.5,0) (-1.75,1.6) (-0.25,1.4) (-0.7, -0.1) (0,-1.1) (.7,-0.1) (0.25,1.4)}] (1.4,2);
    \node at (1.5,1.1) {#1};
\end{tikzpicture}
}
\newcommand{\RIIcomplexdisjoint}[2]{
\begin{tikzpicture}[baseline = -3, scale =0.2, every node/.style={scale=0.6}];
    \draw (-2.6,0) circle (1.1);
    \draw (1,-2) to [curve through = {(-.6,0)}] (1,2);
    \node at (-2.6,0) {#2};
    \node at (0.8,0) {#1};
\end{tikzpicture}
}
\newcommand{\RIIcomplexmerged}[0]{
\begin{tikzpicture}[baseline = -3, scale =0.2, every node/.style={scale=0.6}];
    \draw (-0.65,0) circle (1.2);
    \draw[black,preaction={draw,line width=3pt, white}] (1,-2) to [curve through = {(-.6,0)}] (1,2);;
\end{tikzpicture}
}

\newtheorem{theorem}{Theorem}[section]
\newtheorem{corollary}[theorem]{Corollary}
\newtheorem{proposition}[theorem]{Proposition}
\newtheorem{lemma}[theorem]{Lemma}

\theoremstyle{definition}
\newtheorem{definition}[theorem]{Definition}
\newtheorem{example}[theorem]{Example}

\newtheorem{claim}[theorem]{Claim}

\theoremstyle{remark}
\newtheorem{remark}[theorem]{Remark}

\title{Khovanov homology and quantum error-correcting codes}
\address{Department of Mathematics, Columbia University, New York, NY 10027}

\author[R. Akhmechet]{Rostislav Akhmechet}
\email{\href{mailto:akhmechet@math.columbia.edu}{akhmechet@math.columbia.edu}}

\author[M. Harned]{Milena Harned}
\email{\href{mailto:mdh2192@columbia.edu}{mdh2192@columbia.edu}}

\author[P. Konda]{Pranav Venkata Konda}
\email{\href{mailto:pvk2108@columbia.edu}
{pvk2108@columbia.edu}}

\author[F. Liu]{Felix Shanglin Liu}
\email{\href{mailto:fl2671@columbia.edu}{fl2671@columbia.edu}}

\author[N. Mudumbi]{Nikhil Mudumbi}
\email{\href{mailto:nm3497@columbia.edu}{nm3497@columbia.edu}}

\author[E. Shao]{Eric Yuang Shao}
\email{\href{mailto:e.shao@columbia.edu}{e.shao@columbia.edu}}

\author[Z. Xiao]{Zheheng Xiao}
\email{\href{mailto:zx2377@columbia.edu}{zx2377@columbia.edu}}

\begin{document}

\begin{abstract}
Error-correcting codes for quantum computing are crucial to address the
fundamental problem of communication in the presence of noise and imperfections.
Audoux used Khovanov homology to define families of quantum error-correcting codes with desirable
properties. We explore Khovanov homology and some of its many extensions, namely reduced, annular, and $\mathfrak{sl}_3$ homology, to generate new families of quantum codes and to establish several properties about codes that arise in this way, such as behavior of distance under Reidemeister moves or connected sums.
\end{abstract}


\maketitle

\tableofcontents

\section{Introduction}

Classical error-correcting codes are an important object concerned with mitigating inevitable corruptions
during the storage and transmission of data. They are also important in the generalization to
quantum computing, where quantum decoherence makes handling such corruptions critical. CSS codes, a
specific class of error-correcting codes, have been studied since the end of the 20th century, first
described by Calderbank, Shor, and Steane (\cite{Calderbank_Shor} and \cite{Steane}), where codes are generated by associated pairs
of $\bF_2$-matrices $\mathbf{H}_X$ and $\mathbf{H}_Z$. From this process, we obtain infinite
families of error-correcting codes with potentially desirable properties. CSS codes are associated the parameters known as \emph{length} (denoted $n$), \emph{dimension} denoted $k$), and \emph{distance} (denoted $d$). 

Homology is a fruitful area for searching for such matrices (and eventually obtaining codes). In particular, it is well known that
there exists a bijection between length 3 chain complexes and CSS codes \cite{Audoux,Audoux-Couvreur}. One homology theory that
has begun to produce CSS codes from chain complexes (such as in \cite{Audoux} and \cite{Audoux-Couvreur}) is Khovanov homology,
first developed by Khovanov in \cite{Khovanov} in his categorification of the Jones polynomial. Given a
link diagram, one can produce its Khovanov chain complex by computing its cube of resolutions.
From this chain complex, one can compute the Khovanov homology which, with the chain complex, contains enough data to create CSS codes. There is also a \emph{reduced} version of Khovanov homology, which is determined by a choice of basepoint on the link diagram; this reduced theory was used to produce the families in \cite{Audoux}. We note that while Khovanov homology is independent of the choice of diagram representing a particular link, the Khovanov chain complex, and hence the CSS parameters, are diagram dependent.

In this paper, we investigate the structural modifications one can make to link diagrams
to generate desirable infinite families of CSS codes. Specifically, we investigate how these modifications impact the code parameters (particularly the distance parameter, which is the
homological weight in topological terms). We focus on Reidemeister moves, connected sums,
and tensor products of links and chain complexes. We also explore the CSS codes derived from
selected examples of link families, both in Khovanov homology and in generalizations of Khovanov homology theory, in particular annular Khovanov
homology  (a triply-graded homology theory) and $\mathfrak{sl}_3$ link homology (\cite{Khovanov_sl3}). 

Many computations of parameters were initially done through a computer, using a program written in \texttt{C++} whose source code can be found in \cite{program}.

The organization and main results of this paper are as follows. In Section \ref{sec:background}, we give the necessary background on homological algebra and Khovanov homology. In Section \ref{sec:properties of distance from Khovanov homology}, we establish several properties of distance in Khovanov homology. Our first main result is the following.
\begin{theorem}[detailed version in Theorem \ref{thm:reduced equals unreduced}]
    The distance of a Khovanov chain complex is the same in its unreduced and reduced forms (under the basis introduced in \cite{Audoux}).
\end{theorem}
We prove this by explicitly constructing a weight-preserving isomorphism. In Section \ref{sec:behavior under RII and RIII moves}, we give several examples to answer \cite[Question 2.6]{Audoux}, whether Reidemeister II moves double the distance and whether Reidemeister III moves preserve the distance, in the negative. Our examples are summarized in Figure~\ref{fig:RII/IIIcexs}. 

\begin{theorem}
     Reidemeister II moves do not double the distance and Reidemeister III moves do not preserve the distance, in general.
\end{theorem}
In the process, in Lemma \ref{lemma:dist=2condition} we introduce a necessary condition to have the distance equal 2, which helps verify some of the computations. We are able to provide a positive answer in the following special case, which we view as a generalization of the unlink code from \cite{Audoux}. The following is stated as Corollary \ref{cor:RII for disjoint diagrams}. 
\begin{theorem}
    Reidemeister II moves do double distance when applied to two disjoint diagrams. 
    \begin{equation*}
        \begin{aligned}
            \begin{tikzpicture}[baseline = -2];
            \draw (-3/8,-5/8) rectangle ++(-3/4,5/4) node[pos=.5] {$D_1$};
            \draw (3/8,-5/8) rectangle ++(3/4,5/4) node[pos=.5] {$D_2$};
            \draw (3/8,-1/2) to [curve through = {(1/8,0)}] (3/8, 1/2);
            \draw (-3/8,-1/2) to [curve through = {(-1/8,0)}] (-3/8, 1/2);
        
            \draw [->] (1.5,0) -- (2.5,0) node[midway, above] {RII};
            \begin{scope}[shift = {(4,0)}]
                \draw (3/8,-1/2) to [curve through = {(-1/8,0)}] (3/8, 1/2);
                \draw[black,preaction={draw,line width=3pt, white}] (-3/8,-1/2) to [curve through = {(1/8,0)}] (-3/8, 1/2);
                \draw (-3/8,-5/8) rectangle ++(-3/4,5/4) node[pos=.5] {$D_1$};
                \draw (3/8,-5/8) rectangle ++(3/4,5/4) node[pos=.5] {$D_2$};
            \end{scope}
        \end{tikzpicture}
        \end{aligned}
    \end{equation*}
\end{theorem}

In Section \ref{sec:annular}, we consider the \emph{annular} Khovanov homology \cite{{Asaeda_2004}} with an additional grading stemming from placing a puncture somewhere disjoint from the link diagram. We show that for $(1,1)-$tangles in which a puncture is placed next to the arc forming the closure, there exists an isomorphism between the annular chain complex and reduced chain complex if we were to remove the puncture and place a basepoint on this arc instead (see Proposition~\ref{tanglesprop}). We also investigate some examples where the distance grows, both rigorously and experimentally. 

In Section \ref{sec:tensor}, we consider the CSS codes obtained from a connected sum of two link diagrams, corresponding to a tensor products of their respective Khovanov chain complexes. Given the naive upper bound on the distance of a tensor product of chain complexes (Corollary \ref{cor: ConnectUpperBound}), we try to find link diagrams for which this upper bound can be matched with the same lower bound. To this end, we introduce an algebraic method for computing distances in tensor-product complexes. In Section \ref{sec:Hopf link connected sum} we inductively apply this result on the Hopf link to obtain the following.
\begin{theorem}
   The $\ell^{\text{th}}$ \textit{iterated Hopf link code},
      \begin{center}
     \includestandalone[scale=0.5]{HopfLinkSequence}
\end{center}
    has parameters $
\llbracket n_\ell; k_\ell; d_\ell\rrbracket$ where
\[ n_\ell \sim \frac{\sqrt{3}\cdot 6^{2\ell}}{2\sqrt{2\pi\ell}}, \ \ k_\ell = \binom{2\ell}{\ell}, \ \ d_\ell = 2^{\ell}.
\]
Here and later, $\sim$ denotes asymptotic convergence as $\ell$ tends to infinity.
\end{theorem}
 We then use this method in Section \ref{sec:torus knot connected sum} to compute appropriate lower bounds when one of the two complexes is a $(2,n)$ torus link for $n > 2$. We note that our computations are consistent with a conjecture \cite[Conjecture 18]{ZengPryadko}, which to our knowledge is still open, stating that the naive upper bound is always attained for a tensor product of complexes; see also Remark \ref{remark: tensor}. 

In Section \ref{sec:sl3 homology}, we use the $\mathfrak{sl}_3$ link homology from \cite{Khovanov_sl3} to construct CSS codes over $\bF_3$.\footnote{These are therefore $q$-ary CSS codes, associated with ``qudits'' rather than qubits.} Just as Khovanov homology categorifies the Jones polynomial, $\mathfrak{sl}_3$ link homology categorifies the $\mathfrak{sl}_3$ link polynomial. Unlike the cube of resolutions for Khovanov homology, where each resolution consists of disjoint circles, the $\mathfrak{sl}_3$ case introduces more complicated trivalent graphs, known as \textit{webs}. These webs pose a greater computational challenge, as there is no universal choice of basis for webs. To proceed, we need to choose a family of particularly ``nice" bases that make the computations more manageable. Nonetheless, we successfully compute the $\ell$th unknot code from \cite{Audoux}
\begin{equation*}
  D_{\ell,\ell} = \begin{aligned}
  \includestandalone[scale = 0.8]{unknot_ll}
  \end{aligned}
  \end{equation*}
using specific a choice of basis $\mathcal{B}$. The central result of this section is the following theorem. 
\begin{theorem}[Theorem \ref{main theorem sl3}] For any $\ell\in \mathbb{N}$ and a specific choice of basis $\mathcal{B}$, the $\ell$th unknot code with basis $\mathcal{B}$ has parameters $[[ n_\ell; k_\ell; d_\ell^{\mathcal{B}}]]$, where 
    \[
    n_{\ell} \sim \frac{15 \cdot 25^\ell}{2 \sqrt{6 \pi \ell}}, \ \ k_{\ell} = 3, \ \ d_{\ell} = 3^{\ell}.
    \]
\end{theorem}
We hope to study further examples of codes arising from $\mathfrak{sl}_3$ homology (and more generally, $\mathfrak{sl}_N$ link homology) in future work.

\vskip.5em
\noindent
{\bf{Acknowledgments}.} R.A. thanks Joshua Sussan, Benjamin Audoux, and Leonid Pryadko for helpful discussions. We thank Alvaro Martinez for advising us, Professor George Dragomir for organizing the REU, and Columbia Department of Mathematics as well as Columbia Undergraduate Scholars Program for the financial support. F.L. and E.S. thank the Columbia mathematics department for financial support. M.H. Z.X and P.K. thank the I. I. Rabi Scholars program for their financial support. N.M. thanks the Columbia Science Research Fellows program for financial support.

\section{Background}
\label{sec:background}
\subsection{Homological algebra and CSS codes} \label{sec:hom alg}

We begin by setting up conventions regarding homological algebra. All vector spaces we consider are finite-dimensional. We usually work over the field of two elements $\bF_2$ except in Section \ref{sec:sl3 homology}, where we work over $\bF_3$. For a vector space $V$, its dual $V^*$ is given by 
\[
V^* = \Hom(V,\bF).
\]
If $W$ is another vector space and $f: V\to W$ is a linear map, then there is a corresponding dual map $f^* : W^* \to V^*$ given by $f^*(\varphi) = \varphi \circ f$. Given a basis $\basis$ for $V$ and a vector $b\in \basis$, define $b^* \in V^*$ by $b^*(c) = \delta_{b,c}$ for all $c\in \basis$ and extending linearly. The \emph{dual basis} $\basis^*$ for $V^*$ is given by $\basis^* = \{ b^* \mid b\in \basis\}$. 

\begin{definition}
\label{def:weight}
Let $V$ be a vector space with a fixed basis $\basis$. For $x\in V$, define the \emph{$\basis$-weight} of $x$, denoted $\vert x \vert_{\basis}$ to be the number of elements  of $\basis$ that appear when writing $x$ as a linear combination of vectors in $\basis$.  When the basis is understood, we will simply say the \emph{weight} of $x$ and denote it by $\vert x \vert$. 
\end{definition}

\begin{definition}
    An \emph{$\eps$-chain complex}, for $\eps=\pm 1$, is a sequence $C=(C^i)_{i\in \Z}$ of vector spaces together with a sequence of linear maps $\partial=(\partial^i)_{i\in \Z}$ with each  $\partial^i: C^i \to C^{i+\eps}$ such that $d^{i+\eps} \circ d^i = 0$. Each of the maps $\partial^i$ is called a \emph{differential}. We will always assume that at most finitely many $C^i$ are nonzero. The vector space $C^i$ is called the $i$-th chain group and is said to be in \emph{homological grading $i$}. The \emph{$i$-th homology group} $H^i(C)$ is defined to be 
    \[
H^i(C) = \dfrac{\ker(\partial^i)}{\im(\partial^{i-\eps})}.
    \]
    The \emph{dual complex} $(C^*,\partial^*)$ is the $-\eps$-chain complex obtained by taking the linear dual of each vector space and differential,
    \[
    (C^*)^i = (C^i)^*, \ \ (\partial^*)^i = (\partial^i)^*.
    \]
    Typically we will simply write \emph{chain complex} without specifying $\eps$. 
    
    We also set $(\b{C},\b{\partial})$ to be the $\eps$-chain complex obtained from $C^*$ by negating all homological gradings, 
    \[
    \b{C}^i = (C^*)^{-i}, \ \ \b{\partial}^i = (\partial^*)^{-i}.
    \]
    
    If $(C,\partial_C)$ and $(D,\partial_D)$ are two $\eps$-chain complexes, then a \emph{chain map} $f: C\to D$ consists of linear maps $(f^i : C^i \to D^i)_{i\in \Z}$  such that $ \partial_D^i f^i = f^{i+\eps} \partial_C^i$ for all $i$. A chain map $f$ induces a map $f_*$ on homology groups by $f_*^i : H^i(C) \to H^i(D)$, $[x]\mapsto [f^i(x)]$. We say $f$ is a \emph{quasi-isomorphism} if each $f_*^i$ is an isomorphism. 
\end{definition}

Define the \emph{homological distance} $\dist^i(C)$ and the \emph{CSS} or \emph{code} distance $d^i(C)$ to be
\begin{align}
    \begin{aligned}
    \label{eq:distance}
        \dist^i(C) &= \min\{ \vert x \vert_{\basis} \mid x \in \ker(\partial^i)\setminus \im(\partial^{i-\eps}) \},\\
        d^i(C) &= \min \{\dist^i(C), \dist^i(C^*) \}.
    \end{aligned}
\end{align}
where $\dist^i(C^*)$ is computed with respect to $\basis^*$. In other words, $\dist^i(C)$ is the minimal weight of a representative of a nonzero homology class in $H^i(C)$, and $d^i(C)$ is the minimal weight of a representative of a nonzero homology class in both $H^i(C)$ and $H^i(C^*)$.

A length three chain complex is a chain complex whose chain groups are all zero outside of three consecutive homological gradings:
\[
\cdots \to 0 \to C^{i-\eps} \xrightarrow{\partial^{i-\eps}} C^i \xrightarrow{\partial^i} C^{i+\eps} \to 0 \cdots.
\]
CSS codes \cite{Calderbank_Shor, Steane} are in bijection with length three chain complexes. To define a CSS code, it thus suffices to specify a chain complex together with a homological grading. The parameters $\llbracket n;k;d \rrbracket$ where $n$ is the \emph{length}, $k$ is the \emph{dimension}, and $d$ is the \emph{minimal distance}, are given by 
\[
n = \dim(C^i),\ \ k = \dim(H^i(C)),\ \ d = d^i(C). 
\]

We also recall the following operation on chain complexes. 
\begin{definition}
    Let $(C, \partial_C)$ and $(D, \partial_D)$ be two $\eps$-chain complexes. The $\eps$-chain complex $(C\o D, \partial)$ is given by 
    \[
(C\o D)^k = \bigoplus_{i+j=k} C^i \o D^j,\ \ \partial^k(x\o y)= \partial_C^i(x) \o y + (-1)^i x\o \partial_C^j(y) \text{ for } x\o y \in   C^i \o D^j. 
    \]
\end{definition}
The Künneth Theorem implies that the natural map 
\begin{equation}
    \label{eq:Kunneth theorem}
\bigoplus_{i+j=k} H^i(C)\o H^j(D) \to H^{k}(C\o D), \ \ [x]\o [y] \mapsto [x\o y]
\end{equation}
is an isomorphism for all $k$. We have the following naive upper bound on the homological distance.
\begin{proposition}
    \label{prop: TensorUpperBound}
    Given chain complexes $C$ and $D$, we have that $$\dist^m(C\otimes D) \leq \min_{i \in \mathbb{Z}} \{ \dist^i(C)\dist^{m-i}(D)\}$$ for all $m\in \Z$.
    Note: we are considering the minimum distance of an empty set to be $\infty$. Thus, the minimum in the above expression will naturally ignore degrees that do not have homology.
\end{proposition}
\begin{proof}
    Clearly, the term $H^i(C) \otimes H^{m-i}(D)$ will produce a term of weight $\dist^i(C)\dist^{m-i}(D)$, since we can simply take the tensor product of two minimally weighted terms in each homology. If we set terms from all other contributions to the direct-sum to zero, then we obtain a nontrivial term in $H^m(C\otimes D)$ of weight $\dist^i(C)\dist^{m-i}(D)$. Since all of these weights are achievable, their minimum provides an upper bound on distance.
\end{proof}

\begin{remark}
\label{remark: tensor}
   In \cite[Conjecture 18]{ZengPryadko} the authors conjecture that equality in Proposition \ref{prop: TensorUpperBound} is always attained. To our knowledge, this question is still open. Our results in Section \ref{sec:tensor} support the conjecture. 
\end{remark}
\subsection{Khovanov homology}
\label{sec:Khovanov homology}
Links are disjoint embeddings of circles into $\R^3$. Moreover, every link has a link diagram, which is a generic projection of the link from $\R^3$ into $\R^2$ such that the preimage of every point in the diagram is either unique or consists of exactly two points on the original link. In the event of the latter, we have a crossing in the diagram, and can assign one strand to be the overstrand and the other to be the understrand. It is a classic theorem by Reidemeister that two link diagrams describe the same link if and only if they are related by a sequence of Reidemeister moves, depicted in Figure \ref{fig:R moves}.
\begin{figure}
    \centering
    \begin{tikzpicture}
        \begin{scope}[shift = {(-4.3,0)}]
            \begin{scope}[shift = {(-2,0)}]
                \node[dotted,circle,draw=white, fill=white, inner sep=0pt,minimum size=1cm] (a) at (.1,0){};
                \begin{knot}[end tolerance = 5pt, consider self intersections]
                    \strand (-.3,.4) to [curve through = {(-.2,.2) (.4,-.2)  (.4,.2) (-.2,-.2)}](-.3,-.4);
                \end{knot}
            \end{scope}
            \begin{scope}[shift = {(.2,0)}]
                \node[dotted,circle,draw=white, fill=white, inner sep=0pt,minimum size=.9cm] (b) at (-.1,0){};
                \draw (-.3,.4) to [curve through = {(0,0)}] (-.3,-.4);
            \end{scope}
            \begin{scope}[shift = {(2,0)}]
                \node[dotted,circle,draw=white, fill=white, inner sep=0pt,minimum size=1cm] (c) at (.1,0){};t
                \begin{knot}[end tolerance = 5pt, consider self intersections]
                    \strand (-.3,-.4) to [curve through = {(-.2,-.2) (.4,.2)  (.4,-.2) (-.2,.2)}](-.3,.4);
                \end{knot}
            \end{scope}
            \node[fit = (a)] (A){};
            \node[fit = (b)] (B){};
            \node[fit = (c)] (C){};
        
            \draw[->] (B) -- node [midway, above] {RI$^-$}(A);
            \draw[->] (B) -- node [midway, above] {RI$^+$}(C);
        \end{scope}

        \begin{scope}[shift = {(0,0)}]
            \begin{scope}[shift = {(.2,0)}]
                \node[dotted,circle,draw=white, fill=white, inner sep=0pt,minimum size=.9cm] (a) at (-.1,0){};
                \draw (-.25,.4) to [curve through = {(-.1,0)}] (-.25,-.4);
                \draw (.25,.4) to [curve through = {(.1,0)}] (.25,-.4);
            \end{scope}
            \begin{scope}[shift = {(2,0)}]
                \node[dotted,circle,draw=white, fill=white, inner sep=0pt,minimum size=1cm] (b) at (.1,0){};t
                \begin{knot}[end tolerance = 0.05pt, consider self intersections]
                    \strand (-.25,.4) to [curve through = {(.1,0)}] (-.25,-.4);
                    \strand (.25,.4) to [curve through = {(-.1,0)}] (.25,-.4);
                \end{knot}
            \end{scope}
            \node[fit = (a)] (A){};
            \node[fit = (b)] (B){};
            \draw[->] (A) -- node [midway, above] {RII}(B);
        \end{scope}

        \begin{scope}[shift = {(5,0)}]
            \begin{scope}[shift = {(-.3,0)}]
                \node[dotted,circle,draw=white, fill=white, inner sep=0pt,minimum size=1.2cm] (a) at (0,0){};
                \begin{knot}[end tolerance = 10pt]
                    \strand (.4,.4) to (-.4,-.4);
                    \strand (-.4,.4) to (.4,-.4);
                    \strand[in angle = 180, out angle = 0] (-.5,0) to [curve through = {(0,.3)}] (.5,0);
                    \flipcrossings{1,2,3};
                \end{knot}
            \end{scope}
            \begin{scope}[shift = {(2,0)}]
                \node[dotted,circle,draw=white, fill=white, inner sep=0pt,minimum size=1.2cm] (b) at (0,0){};t
                \begin{knot}[end tolerance = 10pt]
                    \strand (.4,.4) to (-.4,-.4);
                    \strand (-.4,.4) to (.4,-.4);
                    \strand[in angle = 180, out angle = 0] (-.5,0) to [curve through = {(0,-.3)}] (.5,0);
                    \flipcrossings{1,2,3}
                \end{knot}
            \end{scope}
            \node[fit = (a)] (A){};
            \node[fit = (b)] (B){};
            \draw[->] (A) -- node [midway, above] {RIII}(B);
        \end{scope}
    \end{tikzpicture}
    
    \caption{
    The three Reidemeister moves, where we distinguish two variants of RI.
    }
 \label{fig:R moves}
\end{figure}

\begin{figure}
    \centering
    \begin{subfigure}{0.45\textwidth}
    \centering
        \begin{tikzpicture}[scale=1.2]
\draw[thick,<-] (0,.5) -- (.5,0);
\draw[thick, preaction={draw, line width=6pt, white}] (0,0) -- (.5,.5);
\draw[->] (0,0) -- (.5,.5);
\node at (.25,-.5) {positive};

\begin{scope}[shift={(3,0)}]
\draw[thick,->] (0,0) -- (.5,.5); 
 \draw[preaction={draw, line width=6pt, white}] (0,.5) -- (.5,0);
 \draw[thick,<-] (0,.5) -- (.5,0);
\node at (.25,-.5) {negative};
\end{scope}
\end{tikzpicture}
        \caption{The sign of a crossing.}
        \label{eq:crossing signs}
    \end{subfigure}
    \begin{subfigure}{0.45\textwidth}
        \centering
     \begin{tikzpicture}[scale=.5]
\draw[ ] (0,2) -- (2,0);

\draw[preaction={draw, line width=12pt, white}] (0,0) -- (2,2); 

\draw[->] (-.2,1) -- (-1,1);
\draw[->] (2.2,1) -- (3,1);

\node[draw=none, anchor = south] at (-.6,1) {$0$};
\node[draw=none, anchor = south] at (2.6,1) {$1$};

\draw[ ] (-1.2,0) .. controls (-2.2,1) .. (-1.2,2);
\draw[ ] (-3.2,0) .. controls (-2.2,1) .. (-3.2,2); 

\draw[ ] (3.2,0) .. controls (4.2,1) .. (5.2,0);
\draw[ ] (3.2,2) .. controls (4.2,1) .. (5.2,2); 

\end{tikzpicture}
        \caption{The $0$- and $1$-resolution of a crossing.}
        \label{eq:two smoothings}
    \end{subfigure}
\caption{Conventions in Khovanov homology.}
\end{figure}

Moreover, we can also have oriented links by giving an orientation to each circle (component) that is embedded in $\R^3$. This naturally induces an oriented link diagram, and there are analogous oriented Reidemeister moves. Each crossing in an oriented diagram is either positive or negative according to Figure \ref{eq:crossing signs}.

Let $D$ be a diagram for an oriented link $L \subset \R^3$. We will define the Khovanov chain complex $C(D)$ over $\bF_2$. First, we define the \emph{cube of resolutions}. Label the crossings of $D$ by $1,\ldots, n$. There are two ways to resolve a crossing, called the \emph{0-resolution} and \emph{1-resolution}, as in {Figure \ref{eq:two smoothings}. For a sequence $u = (u_1,\ldots, u_n)\in \{0,1\}^n$, resolve the $i$-th crossing according to $u_i$. The result is a collection $D_u$ of disjoint circles in $\R^2$. Elements of $\{0,1\}^n$ are considered as vertices of an $n$-dimensional cube, and we label the vertex $u$ by the planar diagram $D_u$.

Let $u=(u_1,\ldots, u_n)$ and $v=(v_1,\ldots, v_n)$ be vertices which are the same except in the $i$-th entry, where $u_i = 0$ and $v_i=1$. We consider there to be an \emph{edge} from $u$ to $v$ in the $n$-dimensional cube and say that $v$ is an \emph{immediate successor} of $u$. The two resolutions $D_u$ and $D_v$ are identical except near the $i$-th crossing. There is a surface cobordism $S_{u,v}$ from $D_u$ to $D_v$ which is a saddle near the $i$-th crossing and the identity away from the crossing. We view the edge from $u$ to $v$ in the cube as labeled by $S_{u,v}$. 

Consider the algebra $A := \bF_2[X]/(X^2)$  over $\bF_2$ with the two $\bF_2$-linear maps \emph{multiplication}, denoted by $m:A\o A \to A$, and \emph{comultiplication} 
$\Delta :A \to A\o A$ given by
\[
\Delta(1) = 1\o X + X \o 1, \ \ \Delta(X) = X\o X.
\]
We now define the following assignment $\F$.\footnote{While we do not need this terminology, $\F$ is the (1+1)-dimensional topological quantum field theory (TQFT) associated with the Frobenius algebra $A$.} If $Z\subset \R^2$ is a collection of $k$ disjoint circles, then $\F(Z) = A^{\o k}$. If $v$ is an immediate successor of $u$ in the cube, then $D_v$ is obtained from $D_u$ by either merging two circles into one or splitting one circle into two. We define $\F(S_{u,v}) : \F(D_u) \to \F(D_v)$ to be multiplication in the first case and comultiplication on the second; these maps are applied to the tensor factors corresponding to the circles involved in the merge/split, and are the identity on the other factors. 

The chain complex $C(D)$ is defined by 
\begin{equation}
\label{eq:Khovanov cx}
C^i(D)= \bigoplus_{\vert u \vert = i+ n_-} \F(D_{u})
\end{equation}
where $\lr{u} = \sum_i u_i$, and $n_-$ and $n_+$ are, respectively, the number of negative and positive crossings in $D$.
The differential $\partial$ on the summand $\F(D_u)$ is defined to be the sum of all edge maps $\F(S_{u,v})$ as $v$ ranges over all immediate successors of $u$.
Properties of $m$ and $\Delta$ together with the fact that we work over $\bF_2$ implies that $\partial^2=0$. The homology of $C(D)$ is written as $Kh(D)$.

\begin{theorem}\emph{(\cite{Khovanov}}\label{thm:[[D]] is invariant}
If diagrams $D$ and $D'$ are related by a Reidemeister move, then there is a quasi-isomorphism from $C(D)$ to $C(D')$.
\end{theorem}

We will also make use of \emph{reduced} Khovanov homology, defined in \cite{Khovanov_patterns}. A \emph{pointed} link diagram $D^\bullet$ means a link diagram $D$ together with a basepoint, denoted by a dot $\bullet$, on $D$ that is disjoint from its crossings. Note that in any resolution of a pointed link diagram, exactly one circle contains the basepoint. The reduced Khovanov complex $C(D^\bullet)$ is the subcomplex\footnote{It is straightforward to verify that this is indeed a subcomplex.} of $C(D^\bullet)$ spanned by all elements where the marked circle in every resolution is labeled by $X$. For a resolution $D_u$, we let $\til{\F}(D_u)\subset \F(D_u)$ denote the vector space assigned to it. We sometimes drop the $\bullet$ if it is clear that $D$ is pointed. An analogue of Theorem \ref{thm:[[D]] is invariant}  holds for pointed diagrams when the Reidemeister move happens away from the basepoints.

\begin{proposition}[{\cite[Proposition 2]{Khovanov_patterns}}]
    \label{prop:connected sum}
    For pointed link diagrams $D_1^\bullet, D_2^\bullet$, there is an isomorphism $C(D_1^\bullet \# D_2^\bullet) \cong C(D_1^\bullet) \o C(D_2^\bullet)$, where $\#$ denotes the connected sum operation depicted in Figure \ref{fig:connectsumprocess}.
\end{proposition}

\begin{figure}
        \begin{center}
        \begin{tikzpicture}[baseline = -2];
          \draw[color=white] (-3/8,-1/2) to [curve through = {(-1/8,0)}] (-3/8, 1/2);
    \draw (-3/8,-5/8) rectangle ++(-3/4,5/4) node[pos=.5] {$D_1$};
    \draw (3/8,-5/8) rectangle ++(3/4,5/4) node[pos=.5] {$D_2$};
    \draw(-3/8,-1/2) to [curve through = {(-1/8,0)}] (-3/8, 1/2);
    \draw (3/8,-1/2) to [curve through = {(1/8,0)}] (3/8, 1/2);
    \filldraw (1/8,0) circle (1pt);
    \filldraw (-1/8,0) circle (1pt);

    \draw [->] (1.5,0) -- (3,0) node[midway, above] {\#};
    \begin{scope}[shift = {(4.5,0)}]
        \draw (-3/8,-5/8) rectangle ++(-3/4,5/4) node[pos=.5] {$D_1$};
        \draw (3/8,-5/8) rectangle ++(3/4,5/4) node[pos=.5] {$D_2$};
        \draw (3/8,-1/2) to [curve through = {(0,-1/8)}] (-3/8,-1/2);
        \draw (3/8,1/2) to [curve through = {(0,1/8)}] (-3/8,1/2);
        \filldraw (0,1/8) circle (1pt);
    \end{scope}
\end{tikzpicture}
\end{center}
        \caption{A connected sum being taken between two pointed diagrams.}
        \label{fig:connectsumprocess}
    \end{figure}

The reader may notice that (quantum) gradings, which are key to Khovanov homology categorifying the Jones polynomial, have not been discussed. This is because we will use a non-homogeneous basis, following \cite{Audoux}. To that end, set $\m :=1, \p := 1+ X \in A$. With respect to the basis $\{\m, \p\}$, multiplication and comultiplication are given by \eqref{eq:multiplication new basis} and \eqref{eq:comultiplication new basis}, respectively. 
\begin{equation}\label{eq:multiplication new basis}
\circled{\pm} \otimes \circled{\mp} \mapsto \p,\ \  \circled{\pm} \o \circled{\pm} \mapsto \m
\end{equation}
\begin{equation}\label{eq:comultiplication new basis}
\m\mapsto \m\o \p + \p \o \m,\  \  \p \mapsto \m\o \m + \p \o \p.
\end{equation}
More succinctly, multiplication and comultiplication are also given by \eqref{eq:multiplication compact form} and \eqref{eq:comultiplication compact form} respectively:
\begin{equation}\label{eq:multiplication compact form}
    \circled{\varepsilon}\o \circled{\eta} \mapsto \circledm{-\varepsilon\eta},
\end{equation}
\begin{equation}\label{eq:comultiplication compact form}
    \circled{\varepsilon} \mapsto \m \o\circled{-\varepsilon} + \p \o \circled{\varepsilon},
\end{equation}
where $\varepsilon,\eta\in \{-,+\}$.

For a link diagram $D$, by a \emph{basis element} of $C(D)$ we mean a simple tensor $b_1\o \cdots \o b_k \in \F(D_u)$ for some resolution $D_u$ where each $b_i \in \{\m, \p\}$. If we are considering reduced homology of a based diagram, we assume the marked circle always corresponds to the first tensor factor, and require that a basis element be labeled by $X$ on the marked circle and by elements in $\{\m, \p\}$ on all other circles. We will write $\dist^i(D), d^i(D), \dist^i(D^\bullet)$, and $d^i(D^\bullet)$ in place of $\dist^i(C(D)), d^i(C(D)), \dist^i(C(D^\bullet))$, and $d^i(C(D^\bullet))$, respectively, and it is understood that weights are computed with respect to the above basis. A key property of this basis is that no basis element is sent to zero by $m$ or $\Delta$, whereas $m(X\o X) = 0$.

Let $\b{D}$ denote the \emph{mirror image} of a link diagram $D$, obtained by switching the over- and under-crossing data at each crossing. 
\begin{proposition}[{\cite[Proposition 32]{Khovanov}}]
\label{prop:mirror complex}
    There is a weight-preserving isomorphism $ C(\b{D}) \cong \b{C}(D)$.
    \end{proposition}
The above isomorphism in \cite{Khovanov} preserves the quantum grading and is given on each tensor factor in each resolution by the map $A\to A^*: 1\mapsto X^*, X\mapsto 1^*$. As in \cite[Proposition 2.5]{Audoux},  $\circled{\pm} \mapsto \circled{\mp}^*$ also assembles into an isomorphism of chain complexes $ C(\b{D}) \xrightarrow{\sim} \b{C}(D)$ and $C(\b{D}^\bullet) \xrightarrow{\sim} \b{C}(D^\bullet)$. This yields 
\begin{equation}
\label{eq:dual distance in Khovanov homology}
d^i(D) = \min\{\dist^i(D), \dist^i(\b{D})\},\ \ d^i(D^\bullet) =  \min\{\dist^i(D^\bullet), \dist^i(\b{D}^\bullet)\}
\end{equation}
allowing one to stay within the world of Khovanov homology while computing the CSS distance.

\section{Properties of distance from Khovanov homology}
\label{sec:properties of distance from Khovanov homology}

\renewcommand{\arraystretch}{1.5}
\begin{table}
    \centering
     \begin{tabulary}{\textwidth}{|c|J|}
        \hline
  Notation  &  Meaning \\ \hline
  $C$, $C^*$ & A chain complex $C$ and its dual complex $C^*$ \\ \hline 
      $\dist^i(C)$ & Homological distance, $\min\{ \vert x \vert_{\basis} \mid x \in \ker(\partial^i)\setminus \im(\partial^{i-\eps}) \}$, with respect to a basis $\basis$ of $C$ \\ \hline 
        $d^i(C)$ & CSS distance, $\min \{\dist^i(C), \dist^i(C^*) \}$ \\ \hline 
        $n^i(C)$ & Length, $\dim(C^i)$ \\ \hline
        $k^i(C)$ & Dimension, $\dim (H^i(C) )$ \\ \hline
        $D_u$ & Resolution of an $n$-crossing link diagram $D$ according to $u\in \{0,1\}^n$ \\ \hline 
        $A$ & $\bF_2[X]/(X^2)$, a two-dimensional $\bF_2$-vector space with distinguished basis $\{\m= 1, \p = 1+X\}$ \\ \hline
        $m : A\o A \to A$ & Multiplication, corresponding to two circles merging \\ \hline
        $\Delta : A\to A\o A$ & Comultiplication, corresponding to one circle splitting into two \\ \hline
        $\F(D_u)$ & $A^{\otimes n}$, where $n$ is the number of circles in $D_u$ \\ \hline 
        $\til{\F}(D_u)$ & Subspace of $\F(D_u)$ where the circle marked by $\bullet$ is labeled $X$ \\ \hline
        $C(D)$, $Kh(D)$ & The Khovanov chain complex of a link diagram $D$ and its homology \\ \hline 
              $C(D^\bullet)$, $Kh(D^\bullet)$ & The reduced Khovanov chain complex of a pointed link diagram $D^\bullet$ and its homology \\ \hline
              $\dist^i(D)$ & Shorthand for $\dist^i(C(D))$, and similarly for $d^i(D), \dist^i(D^\bullet)$, and $d^i(D^\bullet$) \\ \hline
    \end{tabulary}
    \caption{A summary of the notation in CSS codes and Khovanov homology.}
    \label{table:summary of notation Khovanov homology}
\end{table}

In this section we study properties of distance of CSS codes arising from Khovanov homology. A summary of notation is provided in Table \ref{table:summary of notation Khovanov homology}. 

\subsection{Distance in reduced  homology}
We show that the code distance with Audoux's basis is unchanged in reduced Khovanov homology. To do this, we will construct a weight-preserving isomorphism of chain complexes $R: C(D^\bullet) \oplus C(D^\bullet) \to C(D)$.\footnote{We note that \cite[Corollary 3.2.C.]{Shumakovitch} establishes $Kh(D^\bullet)\oplus Kh(D^\bullet) \cong Kh(D)$, but we need an explicit chain map realizing this isomorphism.} This will be done by first constructing two linear maps (not chain maps) $r_0, r_1: C(D^\bullet) \to C(D)$, which we then upgrade to a chain map $r:C(D^\bullet) \to C(D)$. Finally this chain map $r$ will be used to construct the full isomorphism $R$.

To begin, let $D$ be a link diagram with chosen basepoint $\bullet$ and let the linear map $r_0: C(D^\bullet) \to C(D)$ be defined as follows. If $v$ is a vertex in the cube of resolutions and $D_v$ is the associated resolution picture, then $r_0$ modifies a single element $x = \circledd{X} \o \circled{\varepsilon_1} \o \cdots \o \circled{\varepsilon_n} \in \til{\F}(D_v)$ by replacing the circle labeled $X$ with the sign product $\pi = \varepsilon_1 \cdots \varepsilon_n$ so that $r_0(x) = \circledd{\pi} \o \circled{\varepsilon_1} \o \cdots \o \circled{\varepsilon_n}$. For example, $r_0(\circledd{X} \o \m \o \p) = \circledd{-} \o \m \o \p$. This then extends naturally to a linear map.

Similarly, we define $r_1: C(D^\bullet) \to C(D)$ defined by replacing the circle labeled $X$ with the negative sign product $\pi' = - \varepsilon_1 \cdots \varepsilon_n $ so that $r_1(x) = \circledd{\pi'} \o \circled{\varepsilon_1} \o \cdots \o \circled{\varepsilon_n}$. For example, $r_1(\circledd{X} \o \m \o \p) = \circledd{+} \o \m \o \p$. 

We now pause to make an observation. Let $\sigma: C^i(D) \to C^i(D)$ denote the operation of merging a $\p$ with the marked circle under multiplication such that that $\circledd{\pm} \o \circled{\varepsilon_1} \o \cdots \mapsto\circledd{\mp} \o \circled{\varepsilon_1} \o \cdots $. This has the effect of swapping the sign of the marked circle. Moreover, let $\iota: C(D^\bullet) \to C(D)$ denote the inclusion map defined by $(\circledd{X} \o \circled{\varepsilon_1} \o \cdots) \mapsto (\circledd{-} \o \circled{\varepsilon_1} \o \cdots) + (\circledd{+} \o \circled{\varepsilon_1} \o \cdots)$. Then we have the relations $r_1 = \sigma r_0$ and $\iota = r_0 + r_1$.

However, $r_0$ and $r_1$ are not chain maps as they do not commute with the differential, but they may combined to create an actual chain map as follows. First, we denote $v_0$ as the vertex in the cube of resolutions corresponding to 0-resolving all the crossings of our diagram. Then, for any vertex $v$ in the cube, we define $n_v \in \qty{0,1}$ to be the number of multiplication maps needed to go from $v_0$ to $v$ modulo 2. This number is well-defined since any two paths from $v_0$ to $v$ change the number of circles by a constant amount within the same number of steps, so the total number of multiplication maps remains constant on any path. Then, if $x \in \til{\F}(D_v)$ we can define $r:\til{\F}(D_v) \to \F(D_v)$ by
\begin{align}
    r(x) = r_{n_v}(x).
\end{align}
Explicitly, if $r = \circledd{X} \o \circled{\varepsilon_1} \o \cdots \o \circled{\varepsilon_n}$ is an element at a vertex $v$, then $r(x) = \circledd{\pi} \o \circled{\varepsilon_1} \o \cdots \o \circled{\varepsilon_n}$, where $\pi = (-)^{n_v} \varepsilon_1 \cdots \varepsilon_n$. By construction, this extends naturally from a map at a given vertex to the entire chain complex. 

\begin{lemma}
\label{lem:the r map}
    The map $r:C(D^\bullet) \to C(D)$ is indeed a chain map.
\end{lemma}
\begin{proof}
To show that $r$ is a chain map, we must show that it commutes with the differential. There are four cases to consider: (co)-multiplication on unmarked circles and (co)-multiplication with the marked circle. In each of these cases, $r$ commutes with the differential as shown in Figure \ref{fig:rischainmap}. To make the argument more explicit, we describe each of the four cases in detail. 

\begin{figure}[h]
    \centering
    \begin{subfigure}{0.45\textwidth}
        \centering
        \includestandalone[scale = 0.6]{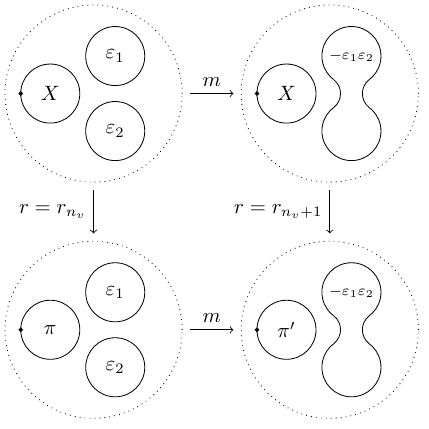}
        \caption{Multiplication of unmarked circles}
    \end{subfigure}
    \hfill
    \begin{subfigure}{0.45\textwidth}
        \centering
        \includestandalone[scale = 0.6]{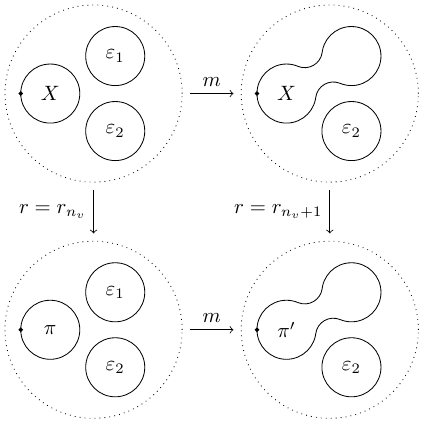}
        \caption{Multiplication with marked circle}
    \end{subfigure}
    \newline
    \begin{subfigure}{0.45\textwidth}
        \centering
        \includestandalone[scale = 0.6]{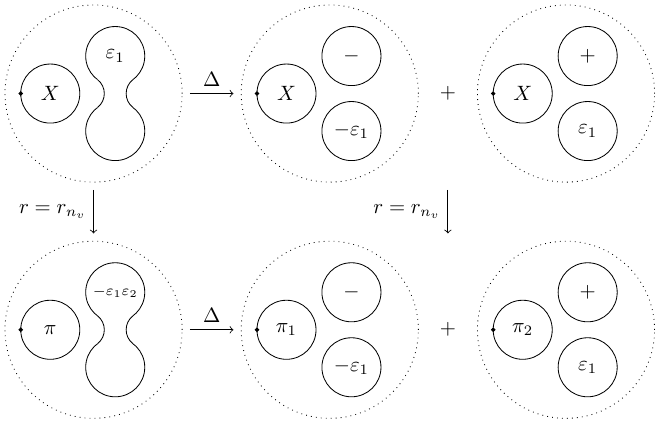}
        \caption{Comultiplication of unmarked circle}
    \end{subfigure}
    \hfill
    \begin{subfigure}{0.45\textwidth}
        \centering
        \includestandalone[scale = 0.6]{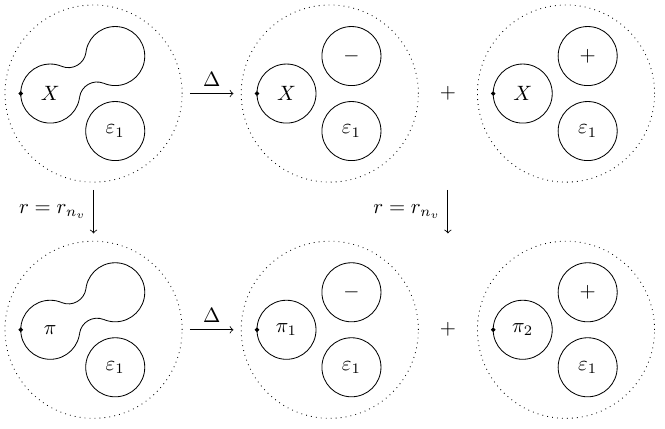}
        \caption{Comultiplication of marked circle}
    \end{subfigure}
    \caption{We show that $r$ commutes with the 4 different types of edge maps of the differential.}
    \label{fig:rischainmap}
\end{figure}

\paragraph{\textbf{1. Multiplication of unmarked circles:}}

Let $x = \circledd{X} \o \circled{\varepsilon_1} \o \cdots \o \circled{\varepsilon_n}$ be a single element in $C(D^\bullet)$ at a vertex $v$ in the cube of resolution. We consider the part of the differential map that multiplies two of the unmarked circles. Without loss of generality, suppose the multiplication merges the circles $\circled{\varepsilon_1}$ and $\circled{\varepsilon_2}$ to yield a circle $\circledm{-\varepsilon_1 \varepsilon_2}$ as described in \eqref{eq:multiplication compact form}. Then, noting that this multiplication takes $x$ to a vertex $v'$ in the cube of resolutions such that $n_{v'} \equiv n_v + 1\pmod 2$, we have
\begin{align*}
    \partial r x &= \partial r_{n_v} x= \partial( \circledd{\pi} \o \circled{\varepsilon_1}\o\circled{\varepsilon_2} \o \cdots \o \circled{\varepsilon_n}) =\circledd{\pi} \o \circledm{-\varepsilon_1\varepsilon_2} \o \cdots \o \circled{\varepsilon_n}, \\
    r \partial x &= r_{n_{v'}}(\circledd{X} \o \circledm{-\varepsilon_1\varepsilon_2} \o \cdots \o \circled{\varepsilon_n} ) = \circledd{\pi'} \o \circledm{-\varepsilon_1\varepsilon_2} \o \cdots \o \circled{\varepsilon_n},
\end{align*}
where $\pi' = (-)^{n_v + 1} (-\varepsilon_1 \varepsilon_2) \cdots \varepsilon_n = (-1)^{n_v} \varepsilon_1 \varepsilon_2 \cdots \varepsilon_n = \pi$. Thus, $\partial r = r\partial$.
\paragraph{\textbf{2. Multiplication with marked circle:}}

Let $x$ be as before as consider the part of the differential map that merges an unmarked circle with the marked circle. Without loss of generality, let the multiplication merge the circles $\circledd{X}$ and $\circled{\varepsilon_1}$ to yield $\circledmd{X}$. Then, noting that this multiplication takes $x$ to a vertex $v'$ with $n_{v'} \equiv n_v +1\pmod 2$, we have
\begin{align*}
    \partial r x &= \partial(\circledd{\pi} \o \circled{\varepsilon_1} \o \circled{\varepsilon_2} \cdots \o \circled{\varepsilon_n}) = \circledmd{-\pi\varepsilon_1} \o \circled{\varepsilon_2} \cdots \o \circled{\varepsilon_n} \\
    r \partial x &= r( \circledmd{X} \o \circled{\varepsilon_2}\o\cdots \o\circled{\varepsilon_n})=\circledmd{\pi'} \o \circled{\varepsilon_2}\o\cdots \o\circled{\varepsilon_n},
\end{align*}
where $\pi' = (-)^{n_v+1} \varepsilon_2 \cdots \varepsilon_n = (-\varepsilon_1) (-)^{n_v}  \varepsilon_1 \cdots \varepsilon_n = -\varepsilon_1 \pi$. Thus, $\partial r = r \partial$.

\paragraph{\textbf{3. Comultiplication of unmarked circle:}}

Now consider the part of the differential that is comultiplication on an unmarked circle, say $\circledm{\varepsilon_1}$ (the shape of the circle just suggests how the differential will act). Then the differential takes $x$ to a vertex $v'$ such that $n_{v'} \equiv n_v \pmod 2$, so
\begin{align*}
    \partial r x &=\partial( \circledd{\pi} \o \circledm{\varepsilon_1}\o\cdots) = (\circledd{\pi} \o \circled{-} \o \circled{-\varepsilon_1} \o\cdots) + (\circledd{\pi} \o \circled{+} \o \circled{\varepsilon_1} \o\cdots),\\
    r \partial x &= r( (\circledd{X} \o \circled{-} \o \circled{-\varepsilon_1} \o\cdots) + (\circledd{X} \o \circled{+} \o \circled{\varepsilon_1} \o\cdots) )\\
    &= (\circledd{\pi_1} \o \circled{-} \o \circled{-\varepsilon_1} \o\cdots) + (\circledd{\pi_2} \o \circled{+} \o \circled{\varepsilon_1} \o\cdots), 
\end{align*}
where $\pi_1 = (-)^{n_v} (-)(-\varepsilon_1) \cdots  = (-)^{n_v} \varepsilon_1\cdots  = \pi$ and $\pi_2 = (-)^{n_v} (+)(\varepsilon_1)\cdots = \pi$. Thus, $\partial r = r\partial$.

\paragraph{\textbf{4. Comultiplication of marked circle:}}

Consider the part of the differential that consists of comultiplication on the marked circle, so $\circledmd{X}$ splits into $\circledd{X} \o \m + \circledd{X} \o\p$. Then the differential takes $x$ to a vertex $v'$ such that $n_{v'} \equiv n_v \pmod 2$, so 
\begin{align*}
    \partial r x &= \partial(\circledmd{\pi} \o \circled{\varepsilon_1} \o\cdots) =  (\circledd{-\pi} \o \m \o\circled{\varepsilon_1} \o\cdots )+(\circledd{\pi} \o \p\o\circled{\varepsilon_1} \o\cdots ),\\
    r\partial x &= r((\circledd{X} \o \m \o\circled{\varepsilon_1} \o\cdots )+(\circledd{X} \o \p\o\circled{\varepsilon_1} \o\cdots )) \\
    &= (\circledd{\pi_1} \o \m \o\circled{\varepsilon_1} \o\cdots )+(\circledd{\pi_2} \o \p\o\circled{\varepsilon_1} \o\cdots ),
\end{align*}
where $\pi_1 = (-)^{n_v} (-) \varepsilon_1 \cdots = -\pi$ and $\pi_2 = (-)^{n_v} (+)\varepsilon_1 \cdots = \pi$. Thus, $\partial r = r\partial$.

Now by linearity of the differential, we see that $r$ indeed commutes with the differential as a whole, so $r$ is a chain map.
\end{proof}

We are now in a position to explicitly define an explicit chain map $R: C(D^\bullet) \oplus C(D^\bullet) \to C(D)$, which we do by setting $R = (r, \sigma r)$. Note that this map is surjective since every element of $C(D)$ lies either in the image of $r$ or $\sigma r$. Also note that the images of $r$ and $\sigma r$ are disjoint and that both these maps are weight preserving, so then $R$ is actually a weight-preserving isomorphism that induces a weight preserving isomorphism on homology $R: Kh^i(D^\bullet) \oplus Kh^i(D^\bullet) \to Kh^i(D)$. We now state and prove our main result.
\begin{theorem}
\label{thm:reduced equals unreduced}
Let $D$ be a link diagram and $D^\bullet$ be a corresponding pointed link diagram. Then $d^i(D) = d^i(D^\bullet)$ for all $i\in \Z$.
\end{theorem}
\begin{proof}
We begin by considering a minimally weighted, nontrivial element of homology in the reduced complex $x \in C^i(D^\bullet)$. Then, under the weight-preserving isomorphism $R:Kh^i(D^\bullet) \oplus Kh^i(D^\bullet) \to Kh^i(D)$, we have that $R(x, 0) \in C^i(D)$ is a nontrivial element of homology in $C(D)$. This means that $\dist^i(D^\bullet) = |x| = |R(x,0)| \geq \dist^i(D)$.

To prove the reverse direction, we now consider a minimally weighted, nontrivial element of homology in the unreduced complex $x \in C^i(D)$. Under the inverse isomorphism $R^{-1}$, we have $R^{-1}(x) = (y,z) \in C^i(D^\bullet) \oplus C^i(D^\bullet)$, which must also represent a nontrivial element in $Kh^i(D^\bullet) \oplus Kh^i (D^\bullet)$. This means that either $y$ or $z$ is nontrivial in reduced homology. Without loss of generality, suppose that $y$ is nontrivial so that $|y| \geq \dist^i(D^\bullet)$. As an aside, we note that this implies $z$ must be trivial (it is in fact 0), since otherwise we have $R(y,0)$ is a nontrivial element of homology in the unreduced complex of weight smaller than $x$, which is clearly impossible. Then, since the inverse isomorphism is also weight preserving, we have $\dist^i(D) = |x| = |R^{-1}(x)| = |y| + |z| \geq \dist^i(D^\bullet)$. This completes the proof that $\dist^i(D) = \dist^i(D^\bullet)$ at any homological degree. Since this same argument applies for the diagram of the mirror image, the result indeed applies for code distances, not just homological distances.
\end{proof}

\subsection{Behavior under connected sums}

Note that over the field $\bF_2$, the code formed by considering the reduced Khovanov homology exactly halves the length $n$ and dimension $k$, while $d$ remains constant. The placement of the point does not matter.

With equality of distance between reduced and unreduced Khovanov homology, we now examine some properties of connected sums.

\begin{theorem}
\label{thm:connectSumToDisjointUnionParameters}
    The parameters of the CSS code obtained from $D_1 \# D_2$ are independent of where the connected sum is taken, and they satisfy
    \begin{alignat*}{3}
        &n^i(C(D_1 \# D_2)) &&= \frac{1}{2} n^i(C(D_1) \otimes C(D_2)) &&= \frac{1}{2} n^i(C(D_1 \sqcup D_2)) \\
        &k^i(C(D_1 \# D_2)) &&= \frac{1}{2} k^i(C(D_1) \otimes C(D_2)) &&= \frac{1}{2} k^i(C(D_1 \sqcup D_2)) \\
        &d^i(C(D_1 \# D_2)) &&= d^i(C(D_1) \otimes C(D_2)) &&= d^i(C(D_1 \sqcup D_2)).
    \end{alignat*}
\end{theorem}
\begin{proof}
    We consider two diagrams $D_1$ and $D_2$ as well as pointed variants $D_1^\bullet$ and $D_2^\bullet$, where the two basepoints are chosen arbitrarily. Then we have the following sequence of isomorphisms
    \begin{align*}
    C(D_1 \# D_2) \oplus C(D_1 \# D_2) &\cong \left( C(D_1^\bullet \# D_2^\bullet) \right)^{\oplus 4} \\
    &\cong \left( C(D_1^\bullet) \otimes C(D_2^\bullet)\right)^{\oplus 4} \\
    &\cong \left( C(D_1^\bullet) \oplus C(D_1^\bullet) \right) \otimes \left( C(D_2^\bullet) \oplus C(D_2^\bullet) \right) \\
    &\cong C(D_1) \otimes C(D_2) \\
    &\cong C(D_1 \sqcup D_2),
    \end{align*}
    where the first and fourth isomorphisms are done using the map $R$, the second isomorphism is from \cite[Proposition 2.7]{Audoux}, and the last isomorphism is from \cite[Corollary 12]{Khovanov}. Then, from examining dimensions, we establish that 
    \[n^i(C(D_1 \# D_2)) = \frac{1}{2} n^i(C(D_1) \otimes C(D_2)) = \frac{1}{2} n^i(C(D_1 \sqcup D_2)).\]
    Since an isomorphism of chain complexes induces an isomorphism on homology, we immediately obtain
    \begin{align*}
    Kh(D_1 \# D_2) \oplus Kh(D_1 \# D_2)  \cong   Kh(D_1) \otimes Kh(D_2) \cong Kh(D_1 \sqcup D_2).
    \end{align*}
     Then, by examining dimensions again, we have 
     
     \[k^i(C(D_1 \# D_2)) = \frac{1}{2} k^i(C(D_1) \otimes C(D_2)) = \frac{1}{2} k^i(C(D_1 \sqcup D_2)).\]

    To prove the claim on distances, we look more carefully at the explicit isomorphisms. Note that by Theorem \ref{thm:reduced equals unreduced}, it suffices to show that
    \begin{equation*}
        d^i(C(D_1^\bullet \# D_2^\bullet)) = d^i(C(D_1^\bullet) \otimes C(D_2^\bullet)) = d^i(C(D_1^\bullet \sqcup D_2^\bullet)).
    \end{equation*}
    The isomorphism $C(D_1^\bullet \# D_2^\bullet) \to C(D_1^\bullet \sqcup D_2^\bullet)$ is given by a saddle cobordism on the part where the connected sum occurs. In particular, it is weight preserving, so we immediately have 
    \begin{align*}
         d^i(C(D_1^\bullet \# D_2^\bullet)) = d^i(C(D_1^\bullet \sqcup  D_2^\bullet)).
    \end{align*}
    Next, the isomorphism $C(D_1^\bullet \sqcup D_2^\bullet) \to C(D_1^\bullet) \otimes C(D_2^\bullet)$ is given by natural projection maps that are also weight-preserving. Hence, we have
    \[
        d^i(C(D_1^\bullet \sqcup  D_2^\bullet)) = d^i(C(D_1^\bullet) \otimes C(D_2^\bullet))
    \]
    as desired. The equation on unreduced distances then immediately follows.
\end{proof}
Note that the last assertion on distance assumes we already know what the distance is for the picture $D_1 \sqcup D_2$ without relating it the individual distances of $D_1$ and $D_2$. However, using Proposition \ref{prop: TensorUpperBound}, we can also obtain a naive upper-bound on distances of a connected sum of link diagrams in terms of the distances for the individual diagrams. We directly have
\begin{corollary}
    \label{cor: ConnectUpperBound}
    Given pointed diagrams $D_1$ and $D_2$, we have that $$\dist^m(D_1 \# D_2) \leq \min_{i \in \mathbb{Z}} \{ \dist^i(D_1) \dist^{m-i}(D_2)\}.$$
\end{corollary}
Of course, given Theorem \ref{thm:reduced equals unreduced}, we are free to work with un-pointed diagrams for the connected sum as well.

\subsection{Behavior under Reidemeister II/III moves}
\label{sec:behavior under RII and RIII moves}
Audoux \cite[Question 2.6]{Audoux} asks whether or not RII moves always double the distance and whether or not RIII moves always preserve the distance. We answer both parts of this question in the negative through the counterexamples shown in Figure \ref{fig:RII/IIIcexs}. However, Corollary \ref{cor:RII for disjoint diagrams} gives a positive answer in the RII case under special circumstances. Note that the homological degree of the distances are not given in the below examples since we only look at unlink diagrams, so it is implied we only look at homological degree 0 (the homology is trivial everywhere else). We now describe these counterexamples in some more detail.

\begin{figure}[h]
    \centering
    \hfill
    \begin{subfigure}[t]{0.3\textwidth}
        \centering
        \begin{tikzpicture}[scale = 0.4]
\begin{scope}[rotate = -90]
\node[circle,draw=white, fill=white, inner sep=0pt,minimum size=1.1cm] (a) at (0,0){};
\begin{knot}[
xshift = 1cm,
clip width = 3,
end tolerance = 0.1 cm,
consider self intersections=true
]
\strand (-1,0.5) to[closed, curve through={(-2,0) (-1,-0.5) (0.4,0.3) (0,1.7) (-1.8,2) (-3,0) (-1.8,-2) (0,-1.7) }] (0.4, -0.3);
\node at (-1,4) {$\hat{d} = 2$};
\end{knot}
\end{scope}

\begin{scope}[shift = {(0,-4.5)},rotate = -90]
\node[circle,draw=white, fill=white, inner sep=0pt,minimum size=1cm] (b) at (0.2,0){};
\begin{knot}[
xshift = 1cm,
clip width = 3,
end tolerance = 0.05 cm,
consider self intersections=true
]
\strand (-1,0.5) to[closed, curve through={(-2,0) (-1,-0.5) (0.4,0.3) (0,1.7) (-1.8,2) (-1,0) (-1.8,-2) (0,-1.7) }] (0.4, -0.3);
\node at (-1,4) {$\hat{d} = 2$};
\end{knot}
\end{scope}

\begin{scope}[shift = {(0,-9)},rotate = -90]
\node[circle,draw=white, fill=white, inner sep=0pt,minimum size=1cm] (c) at (0.2,0){};
\begin{knot}[
xshift = 1cm,
clip width = 3,
end tolerance = 0.1 cm,
consider self intersections=true
]
\strand (-1,-0.5) to[closed, curve through={(-2,0) (-1,0.5) (0.4,-0.3) (0,-1.7) (-1.8,-2) (1,0) (-1.8,2) (0,1.7) }] (0.4, 0.3);
\node at (-1,4) {$\hat{d} = 4$};
\flipcrossings{3};
\end{knot}
\end{scope}

\node[fit = (a)] (A){};
\node[fit = (b)] (B){};
\node[fit = (c)] (C){};

\draw[->] (A) -- node [midway,right] {RII}(B);

\draw[->] (B) -- node [midway,right] {RIII}(C);

\end{tikzpicture}
        \caption{Homological distance is not doubled under RII nor preserved under RIII}
        \label{fig:RIIRIIcex}
    \end{subfigure}
    \hfill
    \begin{subfigure}[t]{0.3\textwidth}
        \centering
        \includestandalone{RIIcounterexample}
        \caption{RII does not double code distance (a nontrivial element in homology for the bottom link is shown)}
        \label{fig:RIIcex}
    \end{subfigure}
    \hfill
    \begin{subfigure}[t]{0.3\textwidth}
        \centering
        \includestandalone{RIIIcounterexample}
        \caption{RIII does not preserve code distance}
        \label{fig:RIIIcex}
    \end{subfigure}
    \hfill
    \newline
    \hfill
    \begin{subfigure}[t]{\textwidth}
        \centering
           \begin{tikzpicture}        
        \begin{scope}
            \begin{scope}[shift = {(-.5,0)}]
                \node[dotted,circle,draw=black, fill=white, inner sep=0pt,minimum size=1.6cm] (a) at (.4,-0.7){};
                \pic[braid/.cd, scale = 0.4, gap = 0.15]{braid={s_2^{-1} s_1^{-1} s_2}};
            \end{scope}
    
            \begin{scope}[shift = {(2,0)}]
                \node[dotted,circle,draw=black, fill=white, inner sep=0pt,minimum size=1.6cm] (b) at (.4,-0.7){};
                \pic[braid/.cd, scale = 0.4, gap =0.15] {braid={s_1 s_2^{-1} s_1^{-1}}};
            \end{scope}
        \end{scope}

        \begin{scope}[shift = {(6.5,0.5)}]
            \begin{scope}[shift = {(-0.5,0)}]
                \node[dotted,circle,draw=black, fill=white, inner sep=0pt,minimum size=1.96cm] (c) at (.43,-0.9){};
                \pic[braid/.cd, scale = 0.4, xshift = 2,gap = 0.15] {braid={s_2^{-1} s_1^{-1} s_2 s_2 }};
                \node at (0.43,-2.3) {$d = 2$};
            \end{scope}

            \begin{scope}[shift = {(2.5,0)}]
                \node[dotted,circle,draw=black, fill=white, inner sep=0pt,minimum size=1.96cm] (d) at (.43,-0.9){};
                \pic[braid/.cd, scale = 0.4, gap = 0.15] {braid={ s_1 s_2^{-1} s_1^{-1} s_2 }};
                \node at (0.43,-2.3) {$d = 4$};
            \end{scope}

            \node[fit = (a)] (A){};
            \node[fit = (b)] (B){};
            \draw[->] (A) -- node [midway,above] {RIII} (B);

            \node[fit = (c)] (C){};
            \node[fit = (d)] (D){};
            \draw[->] (C) -- node [midway,above] {RIII} (D);

        \end{scope}
        
    \end{tikzpicture}
        \caption{RIII does not preserve code distance}
        \label{fig:RIIIcexbraid}
    \end{subfigure}
    \caption{Examples of how distance behaves under RII and RII}
    \label{fig:RII/IIIcexs}
\end{figure}

\begin{example}
    Figure \ref{fig:RIIRIIcex} depicts a minimal counterexample that shows homological distances are not doubled under RII, nor are they preserved under RIII. To verify the computations, note that the distances for the top and bottom diagram can be found using \cite[Corollary 2.11]{Audoux}, which describes how the homological distance changes under the different RI moves. The homological distance for the middle diagram can be found by hand.
\end{example}

However, this example fails to show whether or not Audoux's question still holds for code distances, since when we take mirror images, the corresponding homological distances for the diagrams in Figure \ref{fig:RIIRIIcex} are then 1, 2, and 2 from top to bottom. Before we introduce more counterexamples about the code distance, we introduce a helpful lemma to verify some of the computations on distance. This lemma gives a necessary condition for the homological distance to equal 2.

\begin{lemma}
    Given a link diagram $D$ with $n$ crossings ($n_-$ negative crossings and $n_+$ positive crossings), if $\dist^i(D) = 2$ and $i \le n-n_- -2 $, then there exists a vertex $v$ in the cube of resolutions at homological degree $i$ such that all the outgoing differential maps from $v$ are multiplication maps that merge the same two circles in the resolution corresponding to vertex $v$.
    \label{lemma:dist=2condition}
\end{lemma}
\begin{proof}
    Let $v_1$ and $v_2$ be two vertices (possibly the same) in the same homological degree in the cube of the resolutions and let $D_{v_1}$ and $D_{v_2}$ be the corresponding resolutions. Then any element in the kernel of weight 2 must be of the form $g_1 + g_2$, where $g_j \in \F(D_{v_j})$ for generators $g_j$.
    
    Note that the condition $i \le n - n_- - 2$ is equivalent to the condition that there are at least two outgoing edges for every vertex in the cube of resolutions at homological degree $i$. This implies that $v_1= v_2$ since if $v_1 \neq v_2$, then the set of immediate successors of $v_1$ are distinct from the set of immediate successors of $v_2$. Thus, $\partial g_1$ and $\partial g_2$ cannot cancel and $g_1 + g_2$ cannot be in the kernel. 
    
    Now that $v_1 = v_2 = v$, we derive necessary conditions for when the distance is 2. First, note that none of the outgoing edge maps from $v$ are comultiplication, since comultiplication is injective.

    We now claim that all the outgoing multiplication maps must in fact be merging the same two circles. For sake of contradiction, suppose there exist two outgoing multiplication maps from $v_1$ that merge different circles. Let $g_1 = \circled{\varepsilon_1} \o\circled{\varepsilon_2} \o \cdots \o \circled{\varepsilon_m}$ and $g_2 = \circled{\eta_1} \o \circled{\eta_2} \o \cdots \o \circled{\eta_m}$ Without loss of generality, let one of the multiplication maps $m_1$ merge the first two circles. Then $0 = m_1(g_1+g_2)$ can only occur when $\varepsilon_1 = -\eta_1$, $\varepsilon_2 = -\eta_2$, and $\varepsilon_j = \eta_j$ for $j \neq 1,2$. Thus, two elements can only vanish under multiplication if the corresponding labels for the merging circles are opposite in sign and all other signs are equal. But this means that under any other multiplication map that involves the $j$th circle, we cannot have $g_1+g_2$ vanish since $\varepsilon_j = -\eta_j$ is necessary condition for vanishing under this second map, but we have $\varepsilon_j = \eta_j$.
\end{proof}
We now provide stronger counterexamples that definitively answer Audoux's question for code distances.
\begin{example} \label{ex:RII code dist cex}
    Figure \ref{fig:RIIcex} shows that the code distance does not double under an RII move. Note that since the diagrams are planar isotopic to their mirror image, the complex and its mirror have the same homological distances. 
    For the top diagram, it is relatively easy to compute that the code distance is 2. For the bottom diagram, we exhibit a nontrivial element of homology of weight 2 at the bottom of the figure. It is relatively straightforward to check by hand that this element is indeed nontrivial in homology.
\end{example}

\begin{example}\label{ex:RIII code dist cex}
Figure \ref{fig:RIIIcex} shows that code distances are not always preserved under RIII. The code distance for the diagram on the top can be found from the previous example, since it is simply the top diagram of \ref{fig:RIIcex} with an additional RI twist. Hence the code distance is still 2. Using computer aid \cite{program}, the code distance of the bottom diagram was found to be 4. However, it is still feasible to check by hand that there are no elements in the kernel of weight 2 by drawing out the cube of resolutions and using Lemma \ref{lemma:dist=2condition}.
\end{example}

\begin{example}
Though we know RIII moves do not preserve code distance in general, it may still be that distance is preserved in nicer diagrams, such as those derived from braid closures. We demonstrate that this is not the case by considering braid-closures of length 4 words in $\mathcal B_3$ (the braid group on three strands) as shown in the right hand side of Figure \ref{fig:RIIIcexbraid}. The left hand side of Figure \ref{fig:RIIIcexbraid} simply demonstrates what an RIII move would look like on braids for visualization purposes.

By computations using \cite{program}, we are able to compute the code distances before and after to be 2 and 4 respectively. Thus, RIII moves do not preserve code distance.
\end{example}

Given these examples, we try to find sufficient conditions for which Audoux's conjecture holds. In particular, we study when code distances do double under RII moves. We begin by noting the following fact.

\begin{proposition}
    \label{prop:RIIdoesntmatter}
    The two ways to pick the overstrand in a RII move as shown in Figure \ref{fig:RIIleftandrightover} correspond to two diagrams with the same homological distances everywhere.
    \begin{figure}
        \centering
        \begin{subfigure}{0.3\textwidth}
            \centering
            \includestandalone[scale=0.6]{RIIdifferenttypes}
            \caption{Two possible RII moves can be performed on the left diagram.}
            \label{fig:RIIleftandrightover}
        \end{subfigure}
        \hfill
        \begin{subfigure}{0.6\textwidth}
            \centering
            \begin{tikzpicture}[scale = 0.4];
                \begin{scope}
                \begin{scope}
                    \node[dotted,circle,draw=black, fill=white, inner sep=0pt,minimum size=1cm] (a) at (0,0){};
                    \draw (.75,1) to [curve through = {(-3/8,0)}] (.75,-1);
                    \draw[black,preaction={draw,line width=3pt, white}] (-0.75,1) to [curve through = {(3/8,0)}] (-.75,-1);
                \end{scope}
                
                \begin{scope}[shift = {(5,0)}]
                    \node[dotted,circle,draw=black, fill=white, inner sep=0pt,minimum size=1cm] (b) at (0,0){};
                    \draw (.75,1) to [curve through = {(-3/8,0)}] (.75,-1);
                    \draw (-0.75,1) to [curve through = {(3/8,0)}] (-.75,-1);
                    \draw[dotted] (0,0) circle (1.25); 
                    \filldraw[white] (0,.785) circle (0.2);
                    \filldraw[white] (0,-.785) circle (0.2);
                    \draw (-.178,.899) to [curve through = {(0,0.855)}] (.178,.899);
                    \draw (-.148,.637) to [curve through = {(0,.707)}] (.148,.637);
                    \draw (-.177,-.899) to [curve through = {(-.102,-.77)}] (-.148,-.637);
                    \draw (.177,-.899) to [curve through = {(.102,-.77)}] (.148,-.637);
                \end{scope}
            
                \begin{scope}[shift = {(11.5,1.5)}]
                    \node[dotted,circle,draw=black, fill=white, inner sep=0pt,minimum size=1cm] (c) at (0,0){};
                    \draw (.75,1) to [curve through = {(-3/8,0)}] (.75,-1);
                    \draw (-0.75,1) to [curve through = {(3/8,0)}] (-.75,-1);
                    \draw[dotted] (0,0) circle (1.25); 
                    \filldraw[white] (0,.785) circle (0.2);
                    \filldraw[white] (0,-.785) circle (0.2);
                    \draw (-.177,.899) to [curve through = {(-.102,.77)}] (-.148,.637);
                    \draw (.177,.899) to [curve through = {(.102,.77)}] (.148,.637);
                    \draw (-.177,-.899) to [curve through = {(-.102,-.77)}] (-.148,-.637);
                    \draw (.177,-.899) to [curve through = {(.102,-.77)}] (.148,-.637);
                \end{scope}
                
                \begin{scope}[shift = {(9.5,-1.5)}]
                    \node[dotted,circle,draw=black, fill=white, inner sep=0pt,minimum size=1cm] (d) at (0,0){};
                    \draw (.75,1) to [curve through = {(-3/8,0)}] (.75,-1);
                    \draw (-0.75,1) to [curve through = {(3/8,0)}] (-.75,-1);
                    \draw[dotted] (0,0) circle (1.25); 
                    \filldraw[white] (0,.785) circle (0.2);
                    \filldraw[white] (0,-.785) circle (0.2);
                    \draw (-.178,.899) to [curve through = {(0,0.855)}] (.178,.899);
                    \draw (-.148,.637) to [curve through = {(0,.707)}] (.148,.637);
                    \draw (-.178,-.899) to [curve through = {(0,-0.855)}] (.178,-.899);
                    \draw (-.148,-.637) to [curve through = {(0,-.707)}] (.148,-.637);
                \end{scope}
            
                \begin{scope}[shift = {(16,0)}]
                    \node[dotted,circle,draw=black, fill=white, inner sep=0,minimum size=1cm] (e) at (0,0){};
                    \draw (.75,1) to [curve through = {(-3/8,0)}] (.75,-1);
                    \draw (-0.75,1) to [curve through = {(3/8,0)}] (-.75,-1);
                    \draw[dotted] (0,0) circle (1.25); 
                    \filldraw[white] (0,.785) circle (0.2);
                    \filldraw[white] (0,-.785) circle (0.2);
                    \draw (-.178,-.899) to [curve through = {(0,-0.855)}] (.178,-.899);
                    \draw (-.148,-.637) to [curve through = {(0,-.707)}] (.148,-.637);
                    \draw (-.177,.899) to [curve through = {(-.102,.77)}] (-.148,.637);
                    \draw (.177,.899) to [curve through = {(.102,.77)}] (.148,.637);
                \end{scope}
                \node[fit = (a)] (A){};
                \node[fit = (b)] (B){};
                \node[fit = (c)] (C){};
                \node[fit = (d)] (D){};
                \node[fit = (e)] (E){};
            
                \draw[->,thick] (B) -- (C);
                \draw[->,thick] (B) -- (D);
                \draw[->,thick] (C) -- (E);
                \draw[->,thick] (D) -- (E);
            
                \node at (2,0) {$:$};
                \end{scope}
            
                \begin{scope}[shift={(0,-8)}]
                \begin{scope}
                    \node[dotted,circle,draw=black, fill=white, inner sep=0pt,minimum size=1cm] (f) at (0,0){};
                    \draw (-0.75,1) to [curve through = {(3/8,0)}] (-.75,-1);
                    \draw[black,preaction={draw,line width=3pt, white}] (.75,1) to [curve through = {(-3/8,0)}] (.75,-1);
                \end{scope}
                
                \begin{scope}[shift = {(5,0)}]
                    \node[dotted,circle,draw=black, fill=white, inner sep=0pt,minimum size=1cm] (g) at (0,0){};
                    \draw (.75,1) to [curve through = {(-3/8,0)}] (.75,-1);
                    \draw (-0.75,1) to [curve through = {(3/8,0)}] (-.75,-1);
                    \draw[dotted] (0,0) circle (1.25); 
                    \filldraw[white] (0,.785) circle (0.2);
                    \filldraw[white] (0,-.785) circle (0.2);
                    \draw (-.178,-.899) to [curve through = {(0,-0.855)}] (.178,-.899);
                    \draw (-.148,-.637) to [curve through = {(0,-.707)}] (.148,-.637);
                    \draw (-.177,.899) to [curve through = {(-.102,.77)}] (-.148,.637);
                    \draw (.177,.899) to [curve through = {(.102,.77)}] (.148,.637);
                \end{scope}
            
                \begin{scope}[shift = {(11.5,1.5)}]
                    \node[dotted,circle,draw=black, fill=white, inner sep=0pt,minimum size=1cm] (h) at (0,0){};
                    \draw (.75,1) to [curve through = {(-3/8,0)}] (.75,-1);
                    \draw (-0.75,1) to [curve through = {(3/8,0)}] (-.75,-1);
                    \draw[dotted] (0,0) circle (1.25); 
                    \filldraw[white] (0,.785) circle (0.2);
                    \filldraw[white] (0,-.785) circle (0.2);
                    \draw (-.177,.899) to [curve through = {(-.102,.77)}] (-.148,.637);
                    \draw (.177,.899) to [curve through = {(.102,.77)}] (.148,.637);
                    \draw (-.177,-.899) to [curve through = {(-.102,-.77)}] (-.148,-.637);
                    \draw (.177,-.899) to [curve through = {(.102,-.77)}] (.148,-.637);
                \end{scope}
                
                \begin{scope}[shift = {(9.5,-1.5)}]
                    \node[dotted,circle,draw=black, fill=white, inner sep=0pt,minimum size=1cm] (i) at (0,0){};
                    \draw (.75,1) to [curve through = {(-3/8,0)}] (.75,-1);
                    \draw (-0.75,1) to [curve through = {(3/8,0)}] (-.75,-1);
                    \draw[dotted] (0,0) circle (1.25); 
                    \filldraw[white] (0,.785) circle (0.2);
                    \filldraw[white] (0,-.785) circle (0.2);
                    \draw (-.178,.899) to [curve through = {(0,0.855)}] (.178,.899);
                    \draw (-.148,.637) to [curve through = {(0,.707)}] (.148,.637);
                    \draw (-.178,-.899) to [curve through = {(0,-0.855)}] (.178,-.899);
                    \draw (-.148,-.637) to [curve through = {(0,-.707)}] (.148,-.637);
                \end{scope}
            
                \begin{scope}[shift = {(16,0)}]
                    \node[dotted,circle,draw=black, fill=white, inner sep=0pt,minimum size=1cm] (j) at (0,0){};
                    \draw (.75,1) to [curve through = {(-3/8,0)}] (.75,-1);
                    \draw (-0.75,1) to [curve through = {(3/8,0)}] (-.75,-1);
                    \draw[dotted] (0,0) circle (1.25); 
                    \filldraw[white] (0,.785) circle (0.2);
                    \filldraw[white] (0,-.785) circle (0.2);
                    \draw (-.178,.899) to [curve through = {(0,0.855)}] (.178,.899);
                    \draw (-.148,.637) to [curve through = {(0,.707)}] (.148,.637);
                    \draw (-.177,-.899) to [curve through = {(-.102,-.77)}] (-.148,-.637);
                    \draw (.177,-.899) to [curve through = {(.102,-.77)}] (.148,-.637);
                \end{scope}
                \node[fit = (f)] (F){};
                \node[fit = (g)] (G){};
                \node[fit = (h)] (H){};
                \node[fit = (i)] (I){};
                \node[fit = (j)] (J){};
            
                \draw[->,thick] (G) -- (H);
                \draw[->,thick] (G) -- (I);
                \draw[->,thick] (H) -- (J);
                \draw[->,thick] (I) -- (J);
                \node at (2,0) {$:$};
                \end{scope}
            
                \draw[->,densely dashed] (B) -- (G);
                \draw[->,densely dashed] (E) -- (J);
                \draw[->,densely dashed] (C) -- (H);
                \draw[->,densely dashed] (D) -- (I);
            
                \draw [decorate,decoration={brace,amplitude=5pt,mirror,raise=4ex}]
              (3.55,-11) -- (6.45,-11) node[midway,yshift=-3em]{$i-1$};
                \draw [decorate,decoration={brace,amplitude=5pt,mirror,raise=4ex}]
              (14.55,-11) -- (17.45,-11) node[midway,yshift=-3em]{$i+1$};
                \draw [decorate,decoration={brace,amplitude=5pt,mirror,raise=4ex}]
              (8,-11) -- (13,-11) node[midway,yshift=-3em]{$i$};
            \end{tikzpicture}
            \caption{Dotted arrows represent isotopies that map elements of $C(\RIIleftover)$ into $C(\RIIrightover)$. Solid lines represent the differential. The labels $i+1$, $i$, $i-1$ give the homological degree. }
            \label{fig:RIIisotopies}
        \end{subfigure}
        \hfill
        \caption{ Schematics relevant to the proof of Proposition \ref{prop:RIIdoesntmatter}. }
    \end{figure}
\end{proposition}

\begin{proof}
In the cube of resolutions for \RIIleftover, there is a natural isotopy of pictures to obtain the cube of resolutions for \RIIrightover as shown in Figure \ref{fig:RIIisotopies}. This induces an isomorphism $C^i(\RIIleftover) \rightarrow C^i(\RIIrightover)$ that preserves weight, so that $\dist^i (\RIIleftover) = \dist^i(\RIIrightover)$ for all $i$.
\end{proof}

We now note an important case under which RII moves do double the distance.

\begin{proposition} The local relation 
\label{prop:distance in RII unknot reduced}
    \begin{equation*}
    \begin{aligned}
    \label{eq:m_thmstatement}
       \includestandalone{m_thmstatement}  
    \end{aligned}
    \end{equation*}
    holds for all $i\in \Z$. 
\end{proposition}
\begin{proof}
Let us write this equality as follows: 
\begin{equation*}
    \begin{aligned}
    \label{eq:m_proof1}
       \includestandalone{m_proof1}  
    \end{aligned}
\end{equation*}

Let $D_1 =

\begin{tikzpicture}[baseline=-5pt, scale=1.4];
    \draw (1.5/8,0) circle (4.5pt);
    \draw[black,preaction={draw,line width=3pt, white}] (0,0) circle (4.5pt);
    \draw[fill=black] (11/32,0) circle (1pt);
\end{tikzpicture} 

$ and $D_2 = 

\begin{tikzpicture}[baseline=-5pt, scale=1.4];
    \draw (-0.1,0) circle (4.5pt);
    \draw (3/8,-1/4) .. controls (1/8,-1/8) and (1/8,1/8) .. (3/8, 1/4);
    \draw[fill=black] (3/16,0) circle (1pt);
\end{tikzpicture}

$, and construct
\begin{equation*}
    \begin{aligned}
    \label{eq:m_csumprocess}
        \includestandalone{m_csumprocess}  
    \end{aligned}
\end{equation*}

The left inequality follows from Proposition \ref{prop:connected sum} and \cite[Proposition 2.15]{Audoux-Couvreur} (note $C(D_1)$ is balanced since $D_1$ represents the unlink, and the differential on Khovanov complexes never sends a basis vector to zero). 

The right inequality follows from \cite[Corollary 2.11]{Audoux} since the difference between the diagram $D_1 \# D_2$ and $D_2$ is a Reidemeister II move. 
\end{proof}

\begin{theorem}
\label{thm:unknot RII}
    Sliding an unknot under a link will double the code distance so that
    \begin{equation}
    \begin{aligned}
       d^i ( \RIIcomplexmerged ) = 2d^i (\RIIcomplexdisjoint{}{}).
    \end{aligned}
    \end{equation}
\end{theorem}
\begin{proof}
    Using Proposition \ref{prop:distance in RII unknot reduced} and Theorem \ref{thm:reduced equals unreduced}, we can drop the basepoint to immediately obtain $\dist^i(\RIIcomplexmerged) = 2 \dist^i(\RIIcomplexdisjoint{}{})$. Observe that the statement on code distances then follows from Proposition \ref{prop:RIIdoesntmatter}, since in the mirror image, we would be sliding the unknot over the diagram, which would double the homological distance in the dual complex.
\end{proof}

Using the fact that sliding an unknot doubles distances, we are able to prove a much more powerful result.
\begin{corollary}
\label{cor:RII for disjoint diagrams}
    Applying a Reidemeister II move between two disjoint link diagrams (Figure \ref{fig:RIIcomplexprocess}) will double the code distance. This described by the equation
    \begin{center}
        $d^i($
        \begin{tikzpicture}[baseline = -3, scale = 0.5];   
            \draw [black] (1/4,-1/2) .. controls (-1/4,-1/4) and (-1/4,1/4) .. (1/4, 1/2);
            \draw[black,preaction={draw,line width=3pt, white}] (-1/4,-1/2) .. controls (1/4,-1/4) and (1/4,1/4) .. (-1/4, 1/2);
            \draw[densely dotted] (-1/4,-5/8) rectangle ++(-3/4,5/4) node[pos=.5] {};
            \draw[densely dotted] (1/4,-5/8) rectangle ++(3/4,5/4) node[pos=.5] {};
        \end{tikzpicture}
        $)= 2d^i($
        \begin{tikzpicture}[baseline = -3, scale = 0.5];
            \draw[densely dotted] (-1/4,-5/8) rectangle ++(-3/4,5/4) node[pos=.5] {};
            \draw[densely dotted] (1/4,-5/8) rectangle ++(3/4,5/4) node[pos=.5] {};
            \draw (1/4,-1/2) .. controls (1/16,-1/4) and (1/16,1/4) .. (1/4, 1/2);
            \draw (-1/4,-1/2) .. controls (-1/16,-1/4) and (-1/16,1/4) .. (-1/4, 1/2);
        \end{tikzpicture}
        $).$
    \end{center}
    \begin{figure}
        \centering
        \begin{tikzpicture}[baseline = -2];
            \draw (-3/8,-5/8) rectangle ++(-3/4,5/4) node[pos=.5] {$D_1$};
            \draw (3/8,-5/8) rectangle ++(3/4,5/4) node[pos=.5] {$D_2$};
            \draw (3/8,-1/2) to [curve through = {(1/8,0)}] (3/8, 1/2);
            \draw (-3/8,-1/2) to [curve through = {(-1/8,0)}] (-3/8, 1/2);
        
            \draw [->] (1.5,0) -- (2.5,0) node[midway, above] {RII};
            \begin{scope}[shift = {(4,0)}]
                \draw (3/8,-1/2) to [curve through = {(-1/8,0)}] (3/8, 1/2);
                \draw[black,preaction={draw,line width=3pt, white}] (-3/8,-1/2) to [curve through = {(1/8,0)}] (-3/8, 1/2);
                \draw (-3/8,-5/8) rectangle ++(-3/4,5/4) node[pos=.5] {$D_1$};
                \draw (3/8,-5/8) rectangle ++(3/4,5/4) node[pos=.5] {$D_2$};
            \end{scope}
        \end{tikzpicture}
        \caption{An RII move applied to disjoint link diagrams.}
        \label{fig:RIIcomplexprocess}
    \end{figure}
\end{corollary}
\begin{proof}
    We view the proof as a series of pictures. 
    \begin{center}
    $\dist^i( C($
    \begin{tikzpicture}[scale = 0.5, baseline = -3];
        \begin{scope}[shift = {(.6,0)}]
            \draw (0,-1/2) .. controls (-1/2,-1/4) and (-1/2,1/4) .. (0, 1/2);
            \draw[densely dotted] (0,-5/8) rectangle ++(3/4,5/4) node[pos=.5] {};
        \end{scope}
        \draw[black,preaction={draw,line width=3pt, white}] (-1/2,-1/2) to [curve through = {(-1/8,-1/4) (0.2,-.35) (.48,0) (.2,.35) (-1/8,1/4)}] (-1/2,1/2);
        \filldraw (-1/8,.25) circle (2pt);
        \draw[densely dotted] (-1/2,-5/8) rectangle ++(-3/4,5/4) node[pos=.5] {};
    \end{tikzpicture}
    $)) = \dist^i ( C($
    \begin{tikzpicture}[scale = 0.5, baseline = -3];
        \draw [black] (-1/2,-1/2) .. controls (0,-1/4) and (0,1/4) .. (-1/2, 1/2);
        \draw[densely dotted] (-1/2,-5/8) rectangle ++(-3/4,5/4) node[pos=.5] {};
        \filldraw (-1/8,0) circle (2pt);
    \end{tikzpicture} 
    $\#$
    \begin{tikzpicture}[scale = 0.5, baseline = -3];
        \draw (0,-1/2) .. controls (-1/2,-1/4) and (-1/2,1/4) .. (0, 1/2);
        \draw[black,preaction={draw,line width=3pt, white}] (-1/2,0) circle (3/8);
        \draw[densely dotted] (0,-5/8) rectangle ++(3/4,5/4) node[pos=.5] {};
        \filldraw (-7/8,0) circle (2pt);
    \end{tikzpicture} 
    $)) = \dist^i( C($
    \begin{tikzpicture}[scale = 0.5, baseline = -3];
        \draw [black] (-1/2,-1/2) .. controls (0,-1/4) and (0,1/4) .. (-1/2, 1/2);
        \draw[densely dotted] (-1/2,-5/8) rectangle ++(-3/4,5/4) node[pos=.5] {};
        \filldraw (-1/8,0) circle (2pt);
        \begin{scope}[shift = {(1.3,0)}]
            \draw (0,-1/2) .. controls (-1/2,-1/4) and (-1/2,1/4) .. (0, 1/2);
            \draw[black,preaction={draw,line width=3pt, white}] (-1/2,0) circle (3/8);
            \draw[densely dotted] (0,-5/8) rectangle ++(3/4,5/4) node[pos=.5] {};
            \filldraw (-7/8,0) circle (2pt);
        \end{scope}
    \end{tikzpicture} 
    $)) = 2 \dist^i( C ($
    \begin{tikzpicture}[baseline = -3, scale = 0.5];
        \draw[densely dotted] (-1/4,-5/8) rectangle ++(-3/4,5/4) node[pos=.5] {};
        \draw[densely dotted] (1/4,-5/8) rectangle ++(3/4,5/4) node[pos=.5] {};
        \draw (1/4,-1/2) .. controls (1/16,-1/4) and (1/16,1/4) .. (1/4, 1/2);
        \draw (-1/4,-1/2) .. controls (-1/16,-1/4) and (-1/16,1/4) .. (-1/4, 1/2);
        \filldraw (-.11,0) circle (2pt);
    \end{tikzpicture}
    $)).$
    \end{center}
Note that the last equality follows from Theorem \ref{thm:unknot RII}, where sliding an unknot always doubles the distance. Moreover, since the dot has no effect on distance, we can view this as a statement on unpointed diagrams. Lastly, since the way we apply an RII move doesn't matter by Proposition \ref{prop:RIIdoesntmatter}, the above equalities hold for the mirror image as well, so this statement does in fact hold for the code distance, not just the homological distance.
\end{proof}

The relation between connected sums and disjoint unions inspires us to ``disconnect" and reconnect components in already-known codes. One such way is to rearrange unknots in the shape of a graph-theoretic tree. If $G$ is a tree in the plane as in Figure \ref{fig:tree diagram}, we can consider all associated unlink diagrams $D_G$ as in Figure \ref{fig:unlink tree} by replacing vertices of the trees with unknots and replacing edges with overlap by a RII move (although there are two options for each RII move, it does not matter which is chosen). We wish to show that codes from $D_G$ all share the same code. Such a tree with $l+1$ vertices necessarily has $l$ edges. Note that Audoux's unlink code from \cite[Section 4]{Audoux} is a special case.

\begin{corollary}
A diagram with $\ell+1$ unmarked unknots joined together into one component with $\ell$ RII moves as shown in Figure \ref{fig:unlink tree figure} all produce the same code with $\llbracket n_\ell, k_\ell, d_\ell\rrbracket$ with $n_\ell \sim \sqrt{\frac{6}{ \pi \ell}} 6^\ell$, as $\ell$ tends to infinity, $k_\ell = 2^{\ell+1}$, and $d_\ell = 2^{\ell}$.
\end{corollary}

\begin{figure}[h]
    \centering
    \begin{subfigure}{0.4\textwidth}
        \centering
        \includestandalone[scale=1]{tree}
        \caption{An example of a graph-theoretic tree.}
        \label{fig:tree diagram}
    \end{subfigure}
    \hfill
    \begin{subfigure}{0.4\textwidth}
        \centering
        \includestandalone[scale=0.2]{unlinktree}
        \caption{One possible corresponding link diagram. The marked point can be placed anywhere.}
        \label{fig:unlink tree}
    \end{subfigure}
    \hfill
    \caption{A graph-theoretic tree alongside one possible link diagram representation.  The nodes in the tree represent unknots and the edges represent overlap by an RII move.}
    \label{fig:unlink tree figure}
\end{figure}

\begin{proof}
    Specifically, let $S_\ell$ denote the set of all such link diagrams. For all link diagrams $D_\ell \in S_\ell$, examining $\left( C^{-1}(D_\ell) \xlongrightarrow{\partial^{-1}} C^0(D_\ell) \xlongrightarrow{\partial^0} C^1(D_\ell) \right)$ produces the same code.

    We examine $n_\ell$, $k_\ell$, and $d_\ell$ separately.

    First note that all elements of $S_\ell$ have the unlinks overlap together as a graph-theoretic tree, with the $\ell+1$ ``nodes" being the unlinks and the $\ell$ ``edges" being the joining of two unlinks with a RII move. Note that we can apply Theorem \ref{thm:connectSumToDisjointUnionParameters} to split unlinks in half. One can verify inductively that any tree consisting of $\ell+1$ unlinks can be broken apart into $l$ components, each containing exactly two unlinks. This is accomplished using $\ell-1$ connected sum splits. Then, letting $D_{ul}$ refer to the link diagram shown in \eqref{eq:Dul},
    \begin{equation}
    \begin{aligned}
    \label{eq:Dul}
        \begin{tikzpicture}
    \draw (0,0) circle [radius=.5];
    \draw[preaction={draw,line width=3pt, white}] (.75,0) circle [radius=.5];
    \node at (-1.5,0) {$D_{ul} \ = $};
        \end{tikzpicture}
        \end{aligned}
    \end{equation} 
    by Theorem \ref{thm:connectSumToDisjointUnionParameters}, we have $n^0(C(D_\ell)) = 2^{-\ell+1} n^0(C(D_{ul})^{\otimes \ell}).$ Then, since $C(D_{ul})$ contains two basis elements with homological degree $-1$, eight basis elements with homological degree $0$, and two with homological degree $1$, we can compute the length using generating functions. Let $[t^k] P(t)$ denote the coefficient of $t^k$ in the Laurent polynomial $P(t)$. Then, $n^0(C(D_\ell)) = 2^{-\ell+1} [t^0] (2t^{-1} + 8 + 2t)^{\ell} = 2 [t^0] (t^{-1} + 4 + t)^\ell$, for all $D_\ell \in S_\ell$. From \cite[Proposition 4.1, A.2]{Audoux}, the asymptotic satisfies $n_\ell \sim \sqrt{\frac{6}{\pi \ell}} 6^\ell$ as $\ell$ tends to infinity.
    
    Since $k_\ell$, the dimension of the Khovanov homology, is a link invariant, and all diagrams from the equivalence class $[D_{\text{ul}}^{\ell}]$ are isotopic under Reidemeister moves, we must have $k_\ell$ be equivalent for all link diagrams in the equivalence class $[D_\ell^{\text{ul}}]$. Let $D_{uk}$ be the diagram of the unknot. We have $k^0(C(D_\ell)) = k^0(C(D_{uk})^{\otimes (\ell+1)}) = 2^{\ell+1}.$

    Since the diagram consisting of $\ell$ disjoint unmarked circle and one disjoint pointed circle has distance 1, then performing $\ell$ joining RII moves to join the disjoint circles into one connected component will always result in $d^\ell = 2^\ell$ from Theorem \ref{thm:unknot RII}.
\end{proof}
One may also note that this method of disconnecting a diagram into disjoint constituents and recombining them allows us to show that many different families of diagrams produce identical codes.

We end this section with another example, a generalization of the unknot code in \cite[Section 3]{Audoux}. As an example,
consider the 2-branch unknot with $\ell = 3$ crossings on each branch shown in Figure \ref{fig:branched unknot}. 

\begin{figure}
  \centering
  \includegraphics[scale=0.25]{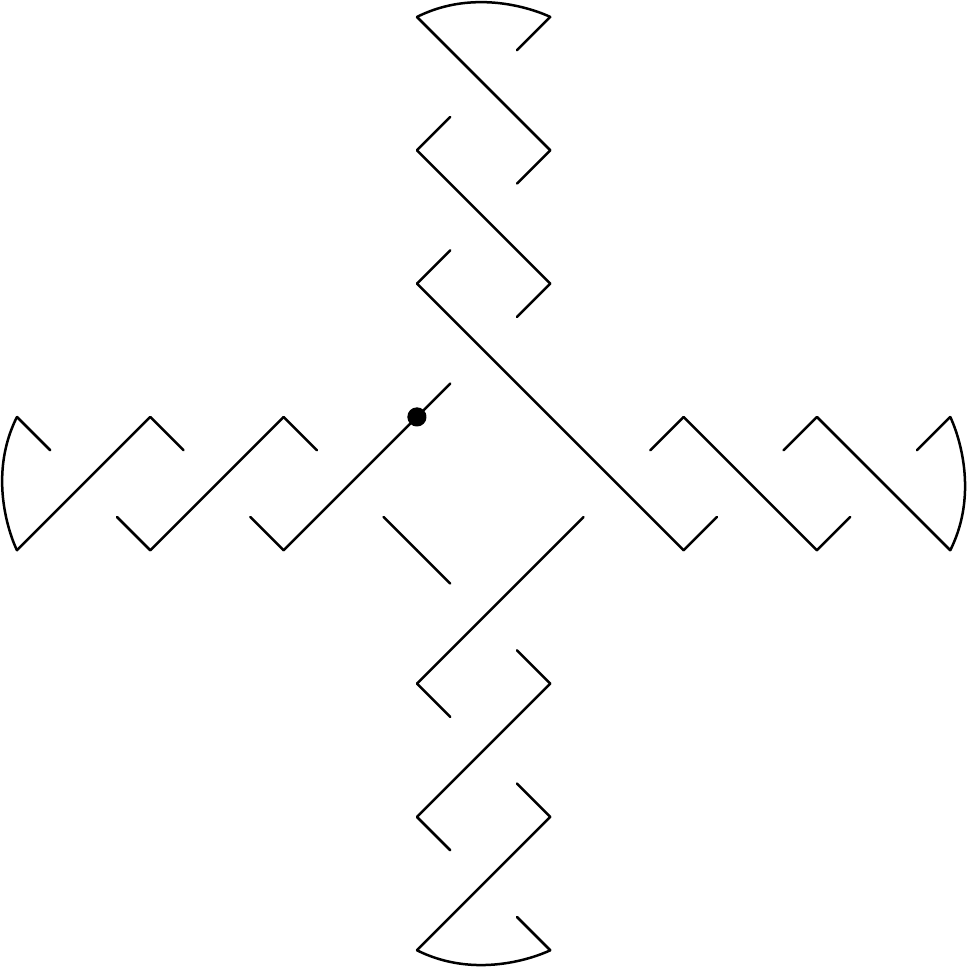}
  \caption{The $2$-branch unknot with $\ell=3$ crossings on each branch. Notice that each branch has positive crossings on one side and negative crossings on the
  other.}
  \label{fig:branched unknot}
\end{figure}

Let us denote the $b$-branched unknot code with $\ell$ crossings on each branch (it is actually possible to have varied lengths on the branches, where the $b\ell$ term is now a sum of the lengths of each branch. However, each branch must have the same number of positive and negative crossings in order for the diagram to be planar isotopic to its mirror image) by $U_b^\ell$ and
its diagram $D_b^\ell$.

By a similar argument in section 3 of \cite{Audoux}, we see that resolving every crossing leaves us
with $b\ell$ undotted circles and one dotted circle. Changing a 1 resolution to a 0 resolution
decreases the number of undotted circles by 1 and changing a 0 resolution to a 1 resolution
decreases the number of undotted circles by 1. Thus in a similar manner we may collect generators: 
\[ {n_b}_\ell = \sum_{r=0}^{b\ell} \binom{b\ell}{r}\binom{b\ell}{b\ell-r}2^{2r} = \sum_{r=0}^{b\ell}\left(\binom{b\ell}{r}2^r\right )^2. \]

Asymptotically we have (from \cite{Audoux} appendix A)
\[ {n_b}_\ell \sim \frac{3^{2b\ell +1}}{\sqrt{2\pi b\ell}}. \] 
From\cite[Theorem 2.3]{Audoux}, we know that $Kh(D_b^\ell) \cong Kh(D_b^{\ell - 1})\{b\}$ as
there is one positive RI move for every branch in $D_b^\ell$. Thus we see that ${k_b}_\ell = 1$. 
From the above argument, we also see that ${d_b}_\ell = 2^{b\ell}$, as there is once again one
positive RI move for every branch.

\begin{remark}
    In fact, from Lemma 2.11 in \cite{Audoux}, it follows that by making only Reidemeister I moves, it is not possible to get a resulting CSS code that is asymptotically more desirable than a code of the above form, up to some constant factor (depending on the initial homological distance of the link diagram). From our results in Section \ref{sec:properties of distance from Khovanov homology}, we cannot make any general conclusions about Reidemeister moves, but local modifications are not an efficient way to generate desirable CSS codes. 
\end{remark}

\section{Annular Khovanov homology}\label{sec:annular}

Let $\A = S^1 \times [0,1]$ denote the annulus. Fix an embedding $\A \hookrightarrow \mathbb{R}^2$ and identify $\A$ with $\{x\in \mathbb{R}^2 \mid 1\leq x \leq 2 \}$. We will consider link diagrams in $\A \times [0,1],$ drawing them in $\A$ (rather, in $\mathbb{R}^2 \backslash (0,0)$). In the annulus, all the usual Reidemeister moves hold \textit{away} from the puncture. 

Given an annular link diagram $D$, we can form its cube of resolutions as usual. For each resolution $D_{u}, u \in \{0,1\}^{\text{\# of crossings}},$ define another grading on $\F(D_{u}) = A^{\otimes \text{\# of circles in }D_{u}}$. There are two types of circles in $D_{u}$. \textit{Trivial} circles do not contain the puncture, and \textit{essential} circles do. 

On an essential circle, let $V$ be the graded vector space assigned to it, with basis $\{v_{+}, v_{-}\}$. For a trivial circle, we continue to use $A$ as the vector space assigned to it. 

\begin{definition}
    We define the annular degree, denoted $\mathrm{adeg}$, by: 
    \begin{itemize}
        \item All elements on a trivial circle are in $\mathrm{adeg}$ zero.
        \item On an essential circle, $\mathrm{adeg}(v_{-}) = -1$ and $\mathrm{adeg}(v_{+}) = 1$.  
    \end{itemize}
\end{definition}

We introduce the following maps in annular homology in addition to the non-annular $m$ and $\Delta$ maps. There are 4 types of saddles involving at least one essential circle. We denote them $I, II, III, $ and $IV,$ as shown in Figure~\ref{fig:annularmaps}. 

\begin{figure}
    \centering
    \begin{tabular}{|c|c|}
        \hline
        \includestandalone{m_1map} & \includestandalone{1map}\\
        \hline
        \includestandalone{m_2map} & \includestandalone{2map}\\
        \hline
        \includestandalone{m_3map} & \includestandalone{3map}\\
        \hline
        \includestandalone{m_4map} & \includestandalone{4map}\\
        \hline
    \end{tabular}
    \caption{The maps in annular Khovanov homology assigned to saddles involving at least one essential circle.}
    \label{fig:annularmaps}
\end{figure}

Let $C_{\A}(D)$ be the annular Khovanov complex of $D$ with differential $\partial_{\A}$ and $H_{\A}(D)$ be its homology. Note that $\partial_{\A}$ does not increase annular degree. Furthermore, denote by $C_{\A}(D;k)$ the subcomplex generated by elements in annular degree $k$. 

\begin{theorem}\cite{Asaeda_2004}
    The homology $H_{\A}$ of $(C_{\A}(D); j)$ is an invariant of annular links for each $j$. 
\end{theorem}

Let $D$ be the closure of a $(1,1)-$tangle diagram with a puncture placed next to the arc forming the closure. From $D$ construct $D^{\bullet}$ by removing the puncture and placing a dot on the arc closing the tangle as shown in Figure~\ref{fig:m_tangles}.   
\begin{figure}
        \centering
        \includestandalone{m_tangles} 
        \caption{The $(1,1)-$tangle $T$, followed by $D$ and $D^{\bullet}$.}
        \label{fig:m_tangles}
\end{figure}

\begin{proposition}\label{tanglesprop}
    Using the above notation, we have an isomorphism of chain complexes $C_{\A}(D;\pm 1) \cong C(D^{\bullet})$. 
\end{proposition}

\begin{proof}
    We begin by considering the cube of resolutions of $D$, observing that in any given resolution there is exactly one essential circle. For example, we could have the following, where $T'$ is an arbitrary resolution of the crossings inside the tangle itself, shown in \eqref{eq:m_tangleres}. 

    \begin{equation}\label{eq:m_tangleres}
    \begin{aligned}
        \includestandalone{m_tangleres} 
        \end{aligned}
    \end{equation}
    
    Fix annular degree $1$ or $-1$. Then, there is only one label assigned to the essential circle. Similarly, in $T^{\bullet},$ the circle containing the dot in a given resolution is labeled by $X$ by definition. 

    We will look at the maps that can involve an essential circle or a dotted circle, since those involving trivial circles or circles without dots are certainly the same; more precisely, they are the typical $m$ and $\Delta$ maps seen in the non-annular case. 

    A single essential circle could involve either the $I$ or the $II$ map. In either case, the annular degree remains unchanged. Under $I$, $v_{\pm} \mapsto v_{\pm} \otimes X = v_{\pm} \otimes (\circled{+} + \circled{-})$. Notice that the basis element on the essential circle remains unchanged. Under $II,$ $v_{\pm} \otimes \circled{\eps} \mapsto v_{\pm}$. Here, the map ``absorbs'' the trivial circle, with the basis element on the essential circle again remaining unchanged.
    In the $D^{\bullet}$ case, $m$ sends $X \otimes \circled{\pm} \mapsto X$ and $\Delta$ sends $X \mapsto X \otimes ( \circled{+} + \circled{-})$. 

    We now wish construct an isomorphism between $C_{\A}(D;\pm 1)$ and $C(D^{\bullet})$. We want this isomorphism to map the essential circle in any resolution $D_u$ of $D$ to the dotted circle in the resolution $D^\bullet_u$. Consider the map $\iota: V \to A$ such that $v_{\pm} \mapsto X$ and extend it to a map $\iota: 
    C_{\A}(D;\pm 1)\to C(D^{\bullet})$ (with the same notation) given by the identity on other tensor factors. Since $\iota$ is an isomorphism of vector spaces at each homological grading, it suffices to show that $\iota$ is a chain map. 

    We show that the maps in Figure~\ref{fig:m_isomorphism} commute. Denote the label assigned to the essential circle by $v_{\alpha}$. Starting at the top left corner of the left diagram, we trace each possible path of maps to obtain $\iota \circ I(v_{\alpha}) = \iota(v_{\alpha} \o \circled{\eps}) = X \o \circled{\eps}$, and that $\Delta \circ \iota(v_{\alpha}) = \Delta(X) = X \o \circled{\eps}$. 

    Now, from the top left corner of the right diagram, we denote the essential circle again by $v_{\alpha}$ and the trivial circle by $\circled{\eps}$. Here, we have $\iota \circ II(v_{\alpha} \o \circled{\eps}) = \iota(v_{\alpha}) = X$ and $m \circ \iota(v_{\alpha} \o \circled{\eps}) = m(X \o \circled{\eps}) = X$. 

    Thus, we have shown that these maps commute and as a result that $\iota$ is an isomorphism. 

    \begin{figure}
        \centering
        \includestandalone{m_isomorphism}
        \caption{The relevant chain maps in the proof of Proposition \ref{tanglesprop}.}
        \label{fig:m_isomorphism}
    \end{figure}
\end{proof}

\begin{remark}
    The statement of the previous result only applies to (1,1)-tangles. 
\end{remark}

\begin{corollary}
\label{cor:a closure of (2,n) knot}
    The CSS parameters for the annular link diagram in Figure~\ref{fig:m_annularclosure} are the same as those in \cite{Audoux} for the non-annular, reduced case. 

    \begin{figure}
    \centering
    \includestandalone{m_annularclosure}
    \caption{An annular closure of the $(2,n)$ torus link.}
    \label{fig:m_annularclosure}
    \end{figure}

\end{corollary}

In Proposition~\ref{tanglesprop}, we fixed a specific annular degree. In general, this is advantageous because it prevents basis vectors being sent to zero under the $III$ map. 

In the non-annular case, choosing the basis $\{\circled{-}, \circled{+}\}$ prevents each possible map (namely $m$ and $\Delta$) from sending a basis vector to zero. However, in the annular case, simply choosing a different basis is not sufficient: while no basis vectors are sent to zero under $I, II,$ or $IV$, $III$ sends $v_{\pm} \o v_{\pm} \mapsto 0$. Here, we use the basis $\{v_{+}, v_{-}\}$. There are two other possible bases we could choose, $\{v_{+}, v_{-} + v_{+}\}$ or $\{v_{-}, v_{-} + v_{+}\}$. However, in each of these cases, $III$ will send a basis vector to zero still. But if we fix the annular degree to be 0 in this map, $v_{\pm} \o v_{\pm}$ is no longer an option, so there are no basis vectors sent to zero. 

\begin{example}
\label{ex:annular link example}
Consider the annular link diagram $D_{\ell}$, where we place $\ell - 1$ concentric circles around the puncture and place an ellipse atop the puncture and the left half of the concentric circles. $D_5$ is shown in Figure~\ref{fig:m_annularunlink}. Experimental evidence from \cite{program} for small cases has shown that the minimum distance for $D_{\ell}$ appears to grow when fixing annular degree 0 for $\ell$ even and annular degree 1 for $\ell$ odd, as shown in the table below. 

\begin{center}
\begin{tabular}{|c|c|c|c|c|c|}
\hline
$\ell$ & 1 & 2 & 3 & 4 & 5\\
\hline
$d$ & 1 & 2 & 3 & 5 & $\geq 3$\\
\hline
\end{tabular}
\end{center}

\begin{figure}
    \centering
    \includestandalone[scale=1.5]{m_annularunlink}
    \caption{The annular link diagram $D_\ell$ when $\ell=5$ considered in Example \ref{ex:annular link example}.}
    \label{fig:m_annularunlink}
\end{figure}

\end{example}

We hope that annular homology could be used to find more quantum codes with desirable properties. 

\section{Tensor products of Khovanov chain complexes}\label{sec:tensor}

In this section, we discuss the CSS codes obtained from a connected sum of two link diagrams. As per Proposition \ref{prop:connected sum}, this corresponds to a tensor product of chain complexes. Although Proposition \ref{prop:connected sum} only holds for pointed link diagrams, Theorem \ref{thm:connectSumToDisjointUnionParameters} allows us to consider the connected sum of unpointed link diagrams as well. Nevertheless, we will restrict our attention to pointed link diagrams in this section.

As discussed in Remark \ref{remark: tensor}, our work in this section supports the conjecture that the upper bound of Proposition \ref{prop: TensorUpperBound} holds as an equality in general.

\subsection{Setting up notation}
\label{sec:Connect Sum Notation}
Let us set up some general notation that will serve as a process for the computation of distances in connected sum families. Our main strategy will be to leverage Equation \eqref{eq:homological equation}. The notation used there will be explained shortly, and we also include a summary in Table \ref{table:summary of notation}.
Let $D_1$ and $D_2$ be two pointed knot diagrams, and let $D$ denote $D_1 \# D_2$. Since we only work with pointed diagrams and reduced Khovanov homology in this section, we omit the $\bullet$ from the notation to avoid clutter.

\bgroup
\def\arraystretch{1.4}
\begin{table}
    \begin{tabular}{|c|c|}
    \hline
  Notation  &  Meaning \\ \hline
  $b_i$       & The dimension of $C^i(D_1)$ \\  \hline
     $a^i_j$    & Basis elements of $C^i(D_1)$ for $1\leq j \leq b_i$ \\  \hline
       $h_i$  & The dimension of $Kh^i(D_1)$ \\  \hline
        $[g^i_j]$ &  Basis elements of $Kh^i(D_1)$ for $1\leq j \leq h_i$ \\  \hline
        $\alpha^i_j$ & Elements of $C^{m-i}(D_2)$ for $1\leq j \leq b_i$ \\  \hline
        $[\varphi^i_j]$ &  Elements of $Kh^{m-i}(D_2)$ for $1\leq j \leq h_i$ \\  \hline
        $\beta^i_j$ &  Elements of $C^{m-1-i}(D_2)$ for $1\leq j \leq b_i$  \\  \hline
    \end{tabular}
    \caption{A summary of the notation in Equation \eqref{eq:homological equation}. Note, in the last three rows, Greek letters are reserved for elements related to $D_2$, and these are not necessarily basis elements. In Equation \eqref{eq: generalTensorEq}, subscripts of $g^i_j$ and $\varphi^i_j$ are omitted when $h_i=1$.}
    \label{table:summary of notation}
\end{table}
\egroup 

Immediately, from Corollary \ref{cor: ConnectUpperBound}, we have the naive upper bound 
\begin{equation*}
    \dist^m (D) \leq \min_{i\in \mathbb{Z}} \{ \dist^i(D_1) \dist^{m-i}(D_2) \}.
\end{equation*}
Our goal will be to match this with the same lower bound in certain cases, thus proving an exact relation for the connected sum distance.

Now, suppose $D_1$ has $e$ crossings, $n_-$ of which are negative, with $C^i(D_1)$ having dimension $b_i$ for $i \in \{-n_-, \dots, e-n_-\}$, with a basis of $a^i_j$ for $j \in \{1, \dots, b_i\}$. Furthermore, let $Kh^i(D_1)$ have dimension $h_i$ for $i \in \{-n_-, \dots, e-n_-\}$ with a basis of $[g^i_j]$ for $j \in \{1, \dots, h_i\}$.
Then, take $$x_0 \in C^m(D)$$ such that $$[x_0] \neq 0 \in Kh^m(D).$$ Now, by \cite[Definition 1.5]{Audoux}, since $$C(D) \cong C(D_1) \otimes C(D_2),$$ we can write $$x_0 = \sum_{i=-n_-}^{e-n_-} \sum_{j=1}^{b_i} a^i_j \otimes \alpha^i_j,$$ where $a^i_j$ are basis vectors for $C^i(D_1)$ as discussed, and $\alpha^i_j \in C^{m-i}(D_2)$ (not necessarily basis vectors). Clearly, $$|x_0| = \sum_{i=-n_-}^{e-n_-} \sum_{j=1}^{b_i} |\alpha^i_j|.$$ Since $[x_0] \in Kh^m(D)$, we can also write $$[x_0] = \sum_{i=-n_-}^{n-e_-} \sum_{j=1}^{h_i} [g^i_j] \otimes [\varphi^i_j],$$ as per the Kunneth Theorem \eqref{eq:Kunneth theorem}, where $[g^i_j]$ are basis vectors for $Kh^i(D_1)$ as discussed, and $[\varphi^i_j] \in Kh^{m-i}(D_2)$ (not necessarily basis terms). It then follows that we can write $x_0$ in the form $$x_0 = \sum_{i=-n_-}^{e-n_-} \sum_{j=1}^{h_i} g^i_j \otimes \varphi^i_j + \partial(y),$$ for some $y \in C^{m-1}(D)$. Putting everything together, 
\begin{equation}
\label{eq:homological equation}
\sum_{i=-n_-}^{e-n_-} \sum_{j=1}^{b_i} a_j^i \otimes \alpha_j^i = \sum_{i=-n_-}^{e-n_-} \sum_{j=1}^{h_i} g_j^i \otimes \varphi_j^i + \partial \left(\sum_{i=-n_-}^{e-n_-} \sum_{j=1}^{b_i} a_j^i \otimes \beta_j^i \right),
\end{equation}
where $\beta^i_j \in C^{m-1-i}(D_2)$. Note, we have used the isomorphism $C(D) \cong C(D_1) \otimes C(D_2)$ to express $y$ in the form $\sum_{i=-n_-}^{e-n_-} \sum_{j=1}^{b_i} a_j^i \otimes \beta_j^i $, hence the homological gradings of the $\beta^i_j$ terms.

We now consider the specific case where $h_i$ is $0$ or $1$ for every $i \in \{-n_-,\dots, e-n_-\}$, as will be the case in what follows. Letting $S$ denote the degrees at which $D$ has homology, our equation reduces to  
\begin{equation}
\label{eq: generalTensorEq}
\sum_{i=-n_-}^{e-n_-} \sum_{j=1}^{b_i} a^i_j \otimes \alpha^i_j = \sum_{i \in S} g^i \otimes \varphi^i + \partial \left(\sum_{i=-n_-}^{e-n_-} \sum_{j=1}^{b_i} a^i_j \otimes \beta^i_j\right),
\end{equation}
where we omit the subscript in the $g^i_j$ and $\varphi^i_j$ terms since $h_i = 1$ for each $i\in S$. Now, the key observation is that since $[x_0] \neq 0$, at least one of the $\varphi^i$'s must be nonzero in homology. If we assume that one particular $\varphi^i$ is nonzero in homology, then we consider terms on both sides of \eqref{eq: generalTensorEq} of the form $a^i_j \otimes (\text{something})$ for each $j \in \{1, \dots, b_i\}$, and use this to derive a lower bound on $|x_0|$.

\subsection{Hopf link connected sum results}
\label{sec:Hopf link connected sum}
We now go through these computations where $D_1$ is a pointed Hopf link, denoted $D^\text{hl}$. There are two forms of the oriented Hopf link to consider: either the $(2,2)$ or the $(2,-2)$ torus link, shown in \eqref{eq:two Hopf links}. 
\begin{equation}
\begin{aligned}
\label{eq:two Hopf links}
    \includestandalone{TwoHopf}
    \end{aligned}
\end{equation}
Although they have slightly different cubes of resolutions, it is sufficient to consider them together. 
Indeed, consider that both satisfy 
\begin{align*}
    \dim(C^{-n_-}(D_1)) &= 2, \dim(C^{1-n_-}(D_1)) = 2, \dim(C^{2-n_-}(D_1)) = 2, \\
    \dim(Kh^{-n_-}(D_1)) &= 1, \dim(Kh^{1-n_-}(D_1)) = 0, \dim(Kh^{2-n_-}(D_1)) = 1,
\end{align*}
 where $n_-=0$ for the $(2,2)$ torus link and $n_-=2$ for the $(2,-2)$ torus link, and the complexes are identical except for a grading shift. For instance, the cube of resolutions for the $(2,2)$ torus link is shown in Figure \ref{fig:Hopf cube}. Recall that in our basis for the reduced Khovanov complex, the marked circle  in any resolution is labeled by $X$ and any unmarked circle is labeled by $\m=1$ or $\p=1+X$.  Using the notation of the previous section, for $i\in \{-n_-, 1-n_-, 2-n_-\}$ and $j\in \{1,2\}$, we have basis vectors $a^i_j$ in homological grading $i$ (for $i=1-n_-$, one of the basis vectors is in the top resolution and the other is in the bottom resolution).
 For $j\in \{1,2\}$, the differential is given by  
\begin{equation}
\label{eq:Hopf cube differential}
\partial(a^{-n_-}_j) = a^{1-n_-}_1 + a^{1-n_-}_2,\ \ 
\partial(a^{1-n_-}_j) = a^{2-n_-}_1 + a^{2-n_-}_2, \ \ 
\partial(a^{2-n_-}_j) = 0. 
\end{equation}
The complex for the $(2,-2)$ torus link is identical, except the homological grading is shifted down by $2$.

\begin{figure}
    \centering
    \includestandalone{Hopf_cube}
    \caption{The cube of resolutions for the $(2,2)$ torus link, where the left-most term is in homological grading zero.}
    \label{fig:Hopf cube}
\end{figure}

Furthermore, as long as we enumerate their respective bases in a consistent manner, we will have that the differential maps are the same for both Hopf links (that is, they can both be represented by the same matrix). Therefore, our calculations will be analogous regardless of which Hopf link we consider.

Now, we may consider our preliminary calculations from Section \ref{sec:Connect Sum Notation} with $D_1 = D^{\text{hl}}$ and $D_2 = D$ as an arbitrary pointed knot diagram. That is, considering $x_0 \in C^m(D^\text{hl} \# D)$, in nonzero homological class, Equation \eqref{eq: generalTensorEq} becomes 
\begin{equation}
\sum_{i-n_-}^{2-n_-} \sum_{j=1}^{b_i} a^i_j \otimes \alpha^i_j = \sum_{i \in \{-n_-,2-n_-\}} g^i \otimes \varphi^i + \partial \left(\sum_{i=-n_-}^{2-n_-} \sum_{j=1}^{b_i} a^i_j \otimes \beta^i_j\right),
\end{equation}
where $b_i = 2$ for $i \in \{-n_-,1-n_-,2-n_-\}$. Furthermore, it is easy to verify from \eqref{eq:Hopf cube differential} that we can let 
\begin{align*}
    g^{-n_-} &= a^{-n_-}_1 + a^{-n_-}_2, \\
    g^{2-n_-} &= a^{2-n_-}_1,
\end{align*} as these are both non-image terms in the kernel. We now consider each case in turn.

\paragraph{\textbf{Case 1:}} $[\varphi^{-n_-}] \neq 0$
\newline
Note that the only way to obtain a term whose first tensor factor is $a^{-n_-}_j$ from within the image term is from $\sum_{j=1}^2 a^{-n_-}_j \otimes \partial(\beta^{-n_-}_j)$. Then, our two equations here become 
\begin{align*}
    \alpha^{-n_-}_1 &= \varphi^{-n_-} + \partial(\beta^{-n_-}_1), \\
    \alpha^{-n_-}_2 &= \varphi^{-n_-} + \partial(\beta^{-n_-}_2).
\end{align*} However, since $[\varphi^{-n_-}] \neq 0$ and $[\partial(\beta^{-n_-}_j)] = 0$, it follows that $[\alpha^{-n_-}_j] \neq 0$. Thus, $|\alpha^{-n_-}_1|, |\alpha^{-n_-}_2| \geq \dist^{m+n_-}(D_2)$. So, in this case, $$|x_0| = \sum_{i=-n_-}^{2-n_-} \sum_{j=1}^{b_i} |\alpha^i_j| \geq \sum_{j=1}^{2} |\alpha^{-n_-}_j| \geq 2\dist^{m+n_-}(D_2).$$
\paragraph{\textbf{Case 2:}} $[\varphi^{2-n_-}] \neq 0$
\newline
There are now two ways to obtain a term whose first tensor factor is $a^{2-n_-}_j$ from within the image term; $\sum_{j=1}^2 a^{2-n_-}_j \otimes \partial(\beta^{2-n_-}_j)$ and $\sum_{j=1}^2 \partial(a^{1-n_-}_j) \otimes \beta^{1-n_-}_j$. Note that $\partial(a^{1-n_-}_j) = a^{2-n_-}_1+a^{2-n_-}_2$ for $j = 1,2$. Thus, we now obtain 
\begin{align*}
    \alpha^{2-n_-}_1 &= \varphi^{2-n_-} + \beta^{1-n_-}_1 + \beta^{1-n_-}_2 + \partial(\beta^{2-n_-}_1), \\
    \alpha^{2-n_-}_2 &= \beta^{1-n_-}_1 + \beta^{1-n_-}_2 + \partial(\beta^{2-n_-}_2).
\end{align*} Adding our equations, we obtain 
$$\alpha^{2-n_-}_1 + \alpha^{2-n_-}_2 = \varphi^{2-n_-} + \partial(\beta^{2-n_-}_1+\beta^{2-n_-}_2).$$ Once again, since $[\varphi^{2-n_-}] \neq 0$ and $[\partial(\beta^{2-n_-}_1+\beta^{2-n_-}_2)] = 0$, we have that $[\alpha^{2-n_-}_1 + \alpha^{2-n_-}_2] \neq 0$. So, 
\begin{align*}
    |x_0| = \sum_{i=-n_-}^{2-n_-}\sum_{j=1}^{b_i} |\alpha^i_j| \geq \sum_{j=1}^2 |\alpha^{2-n_-}_j| \geq |\alpha^{2-n_-}_1 + \alpha^{2-n_-}_2| \geq \dist^{m-2+n_-}(D_2).
\end{align*}
Since at least one of these two cases must be true, we finally obtain $$|x_0| \geq \min\{2\dist^{m+n_-}(D_2), \dist^{m-2+n_-}(D_2)\}.$$ Since this is a lower bound on the weight of any nonzero term in $Kh^m(D)$, it is also a lower bound on $\dist^m(D)$. We can easily compute that $\dist^{-n_-}(D_1) = 2$ and $\dist^{2-n_-}(D_1) = 1$. Thus, this lower bound becomes $$\dist^m(D) \geq \min_{i\in \mathbb{Z}} \{ \dist^i(D_1) \dist^{m-i}(D_2)\}.$$ However, note that this lower bound exactly matches our upper bound from  Corollary \ref{cor: ConnectUpperBound}. Thus, we finally have the following result.
\begin{theorem}
    Given any pointed knot diagram $D$ and a pointed Hopf link $D^\text{hl}$, we have $$\dist^m(D \# D^\text{hl}) = \min\left\{2\dist^{m+n_-}(D), \dist^{m-2+n_-}(D)\right\}.$$
\end{theorem} Effectively, this allows us to ``slide'' an unknot onto any strand. 
\begin{corollary}
\label{prop:HopfLinkCorollary}
    We have the following:
    \begin{center}
        \includestandalone[scale=1]{HopfLinkConnect}
    \end{center}
\end{corollary}
We can use Corollary \ref{prop:HopfLinkCorollary} to obtain a sequence of quantum error-correcting codes by connected summing $\ell$ pointed $(2,2)$ torus links, in the following manner:
\begin{center}
     \includestandalone[scale=0.5]{HopfLinkSequence}
\end{center}
For convenience, let $D^\text{hl}_\ell$ denote $(D^\text{hl})^{\# \ell}$, and let $\dist^m_\ell$ denote $\dist^m(D^\text{hl}_\ell)$.
\begin{proposition}
    The homological distance at degree $m$ of the diagram obtained by taking the connected sum of $\ell$ copies of the pointed $(2,2)$ torus link is $$\dist^m_\ell = 2^{\ell-\frac{m}{2}}$$ for all even $0 \leq m \leq 2\ell$. 
\end{proposition}
\begin{proof}
    We use induction on $\ell$. For our base case of $\ell = 1$, we have that $\dist^0_1 = 2$ and $\dist^2_1 = 1$, as desired. Then, assuming our result is true for $\ell'-1$, consider $\dist^m_{\ell'}$.
    
    Within this specific value of $\ell'$, we apply induction on $m$. Our base case here is $m = 0$, in which case it is easy to see that $\dist^0_{\ell'}= 2^{\ell'}$ (the sum of all $2^{\ell'}$ basis vectors is the only non-zero element in the homology). Then, assume the desired statement is true for all $(m,\ell)$ with $\ell < \ell'$ or $\ell = \ell'$ and $m < m'$. We have 
    \begin{align*}
        \dist^m_{\ell'} &= \min\left\{2\dist^{m'}_{\ell'-1}, \dist^{m'-2}_{\ell'-1}\right\} \\
        &= \min\left\{2\cdot 2^{\ell'-1-\frac{m'}{2}}, 2^{\ell'-1-\frac{m'-2}{2}}\right\} \\
        &= 2^{\ell'-\frac{m'}{2}}.
    \end{align*} 
    This completes both inductions. Finally, we must also manually verify that $\dist^{2\ell}_{\ell} = 1$. However, this is indeed true; the final homological degree consists only of co-multiplication maps. Thus, it is impossible for any element of odd weight to be obtained in the image. Thus we finally have that $$\dist^m(D^{\text{hl}}_\ell) = 2^{\ell - \frac{m}{2}}. $$ 
\end{proof}
We now consider the $\ell^{\text{th}}$ \textit{iterated Hopf link code}, obtained from 
\[
\left(C^{2\ell-1}(D^{\text{hl}}_{2\ell}) \rightarrow C^{2\ell}(D^\text{hl}_{2\ell}) \rightarrow C^{2\ell+1}(D^\text{hl}_{2\ell})\right),
\]
with parameters $\llbracket n_\ell; k_\ell; d_\ell\rrbracket$.
\begin{proposition}
    We have $d_\ell = 2^{\ell}$.
\end{proposition}
\begin{proof}
    As per \cite[Proposition 2.5]{Audoux}, the distance is simply $$d_\ell = \min\{\dist^{2\ell} (D^{\text{hl}}_{2\ell}), \dist^{-2\ell}(\overline{D^{\text{hl}}_{2\ell}})\}.$$ However, $\overline{D^{\text{hl}}_{2\ell}}$ is simply the connected sum of $2\ell$ copies of the negatively-oriented Hopf link (recall that the bar denotes mirror image). As discussed, and as we can see from our above calculations, it does not make a difference to the distance whether we consider the positive or negative Hopf link. Thus, $$d_\ell = \dist^{2\ell}_{2\ell} = 2^\ell.$$
\end{proof}
\begin{proposition}
    We have $k_\ell = \binom{2\ell}{\ell}$.
\end{proposition}
\begin{proof}
    As per the Künneth Formula, \cite[Proposition 1.2]{Audoux}, we have that $Kh(D^\text{hl}_{2\ell}) \cong Kh(D^\text{hl}_1)^{\otimes 2\ell}.$ Since the dimensions of $Kh(D^\text{hl}_1)$ can be represented by the generating function $1+t^2$, it follows that $k_\ell$ is the coefficient of degree $2\ell$ in $(1+t^2)^{2\ell}$, which is $\binom{2\ell}{\ell}$ by the Binomial Theorem.
\end{proof}
\begin{proposition}
    $n_\ell \sim \frac{\sqrt{3}\cdot 6^{2\ell}}{2\sqrt{2\pi\ell}}$ as $\ell$ tends to infinity.
    \label{prop:iterated hopf asymptotic}
\end{proposition}
\begin{proof}
    As per \cite[Definition 1.5]{Audoux}, we have that $C(D^\text{hl}_{2\ell}) \cong C(D^\text{hl}_1)^{\otimes 2\ell}$. Since the dimensions of $C(D^\text{hl}_1)$ can be represented by the generating function $2+2t+2t^2$, it follows that $n_\ell$ is the coefficient of degree $2\ell$ in $(2+2t+2t^2)^{2\ell}$. Note that this is equivalent to finding the constant term in the expansion $2^{2\ell}(t^{-1}+1+t)^{2\ell}$. We then have 
    \begin{align*}
        2^{\ell}(t^{-1}+1+t)^{\ell} &= 2^{\ell}\left((t^{-\frac12}+t^{\frac12}) - 1\right)^{\ell} \\
        &= 2^{\ell} \sum_{r=0}^{\ell} \binom{\ell}{r} (t^{-1/2}+t^{1/2})^{2r} (-1)^{\ell-r} \\
        &= (-2)^{\ell} \sum_{r=0}^{\ell} \binom{\ell}{r} (-1)^r \sum_{i=0}^{2r} \binom{2r}{i} t^{r-i}.
    \end{align*} 
    Defining $$c_\ell \stackrel{\text{def}}{=} (-2)^{\ell}\sum_{r=0}^{\ell}\binom{\ell}{r}\binom{2r}{r} (-1)^r,$$ the coefficient of the zero-degree term is then $c_{2\ell}$. Since $c_\ell \sim \frac{\sqrt{3} \cdot 6^\ell}{2\sqrt{\pi \ell}}$ by Proposition \ref{prop:hopf link asymptotic}, and since $n_\ell = c_{2\ell}$, our desired result follows.
\end{proof}

\subsection{Torus link connected sum results}
\label{sec:torus knot connected sum}
Consider the pointed $(2,\ell)$ positively-oriented torus link, which we denote $D^\text{tl}_\ell$. As per \cite[Section 5.1]{Audoux}, homology exists at every degree besides $r=1$, with dimension $1$. Furthermore, as per \cite[Proposition 5.2]{Audoux}, the distances are $\dist^r({D^\text{tl}_\ell}) = \binom{\ell}{r}$ for $2 \leq r \leq \ell$ and $\dist^0({D^\text{tl}_\ell}) = 2$. We then claim the following.
\begin{proposition}
    Given a pointed  knot diagram $D$, we have that $$\dist^m(D \# D^\text{tl}_\ell) = \min_{i\in \mathbb{Z}} \{ \dist^i(D^\text{tl}_\ell)\dist^{m-i}(D) \}.$$
\end{proposition}
\begin{proof}
    Given Corollary \ref{cor: ConnectUpperBound} it is sufficient to prove that $$\dist^m(D \# D^\text{tl}_\ell) \geq \min_{i\in \mathbb{Z}} \{ \dist^i(D^\text{tl}_\ell)\dist^{m-i}(D) \}.$$ Now, our homological equation \eqref{eq: generalTensorEq} becomes $$\sum_{i=0}^\ell \sum_{j=1}^{b_i} a^i_j \otimes \alpha^i_j = \sum_{i\in S} g^i \otimes \varphi^i + \partial\left(\sum_{i=0}^\ell \sum_{j=1}^{b_i} a^i_j \otimes \beta^i_j\right),$$ where $S = \{0, 2, 3, \dots, \ell\}$. We consider the cube of resolutions for the $(2,4)$ torus link as a helpful reference, shown in Figure \ref{fig:(2,4) cube}.

    \begin{figure}
        \centering
        \includestandalone[scale = 0.8]{TorusCube}
        \caption{The cube of resolutions for the $(2,4)$ torus link.}
        \label{fig:(2,4) cube}
    \end{figure}

    As usual, we consider which of the $\varphi^i$ terms are in nonzero homological class.
    \paragraph{\textbf{Case 1:}} $[\varphi^0] \neq 0$. 
    \newline
    Here, we let $g^0 = a^0_1 + a^0_2$. This is clearly not in the image of $\partial^{-1}$, and it is in the kernel of $\partial^0$; since every map from degree $0$ to $1$ is a multiplication map, we will have $\partial^0(a^0_1) = \partial^0(a^0_2)$. Now, note that the only way to obtain a term of the form $a^0_j \otimes$ from within the image term is from $\sum_{j=1}^2 a^0_j \otimes \partial(\beta^0_j)$. Then, our two equations here become 
    \begin{align*}
        \alpha^0_1 &= \varphi^0 + \partial(\beta^0_1), \\
        \alpha^0_2 &= \varphi^0 + \partial(\beta^0_2).
    \end{align*} However, since $[\varphi^0] \neq 0$ and $[\partial(\beta^0_j)] = 0$, it follows that $[\alpha^0_j]$ is a nontrivial element of $Kh^m(D)$. Thus, in this case, $$|x_0| = \sum_{i=0}^\ell \sum_{j=1}^{b_i} |\alpha^i_j| \geq \sum_{j=1}^2 |\alpha^0_j| \geq 2\dist^m(D).$$
    \paragraph{\textbf{Case 2:}} $[\varphi^i] \neq 0$ for some $2 \leq i \leq \ell$. 
    \newline
    Using notation from \cite[Section 5.3]{Audoux}, define $$E_i = \left\{\phi: \{\text{crossings of } D^\text{tl}_\ell \} \rightarrow \{0,1\} \mid \phi^{-1}(1)| = i\right\}.$$ Then, for each $\phi \in E_i$, define $D^-_\phi$ as the basis vector in $C^i(D^\text{tl}_\ell)$ where all (non-reduced) circles in the resolution $\phi$ are labeled with $\m$.
    \begin{claim}
        The element $\sum_{\phi \in E_i} D^-_\phi$ of $C^i(D^\text{tl}_\ell)$ is a homology element belonging to a non-zero class.
    \end{claim}
    \begin{proof}
        As per notation in the proof of \cite[Proposition 5.2]{Audoux}, we define $D_{\underline{\varepsilon}}$ as a labeling of the diagram formed by all 1 resolutions, with the labels given by $\underline{\varepsilon} = (\varepsilon_1, \dots, \varepsilon_{\ell-1})$. Then, we define $D^i_{\underline{\varepsilon}}$ as the image of $D_{\underline{\varepsilon}}$ under the partial maps $\left(\partial^{i+1}\right)^{-1}\circ \dots \circ \left(\partial^{\ell-1}\right)^{-1}$. As proven in the proof of \cite[Proposition 5.2]{Audoux}, $\partial(D^i_{\underline{\varepsilon}}) = 0$. If we let $\underline{\varepsilon} = (\m, \dots, \m)$, then it is clear that $D^i_{\underline{\varepsilon}} = \sum_{\phi \in E_i} D^-_\phi$: every map above degree $0$ in the cube of resolutions is a comultiplication map, and thus, the inverse maps are all multiplication. However, since multiplication maps $\m \otimes \m$ to $\m$, all basis vectors in $D^i_{\underline{\varepsilon}}$ will contain only $\m$ (and it is clear that any such term can be obtained, by simply following the maps to that particular resolution).
        
        Then, all that is left to do is show that $\sum_{\phi \in E_i} D^-_\phi$ cannot lie in the image of $\partial^{i-1}$. Since all maps from $C^{i-1}$ to $C^i$ must be comultiplication maps, each basis vector in $C^{i-1}$ maps to an even number of basis elements within each $\phi \in E_i$. Thus, any element in $C^{i-1}$ can only map to an even number of basis elements within each $\phi \in E_i$; although certain terms may cancel, this will not affect the parity of the number of basis elements. Since $\sum_{\phi \in E_i} D^-_\phi$ contains exactly one basis element in each $\phi \in E_i$, it cannot be in the image.
    \end{proof}
    Thus, we will let $g^i = \sum_{\phi \in E_i} D^-_\phi$. Note that, by this construction, $|g^i| = \dist^i(D^{\text{tl}}_\ell)$. Now, for any given $\phi \in E_i$, let $A_\phi$ be the set of the indices for all basis vectors of $C^i(D^\text{tl}_\ell)$ formed by labelings of the circles in $\phi$. For each $j \in A_\phi$, we consider the terms of the form $a_j^i \otimes\text{(something)}$ on both sides of our homological equation \eqref{eq: generalTensorEq}. If we form these equations and add over all $j \in A_\phi$, note that we will obtain 
    \begin{equation}
    \label{eq: torusHom}
        \sum_{j \in A_\phi} \alpha_j^i = \varphi^i + \partial\left(\sum_{j \in A_\phi} \beta^i_j\right).
    \end{equation}
    This is because only the $a_j^i$ corresponding to $\m \otimes \dots \otimes \m$ will contain a $\varphi^i$ term. For the image contribution on the right side of each such equation, we will have a contribution from $a^i_j \otimes \partial(\beta^i_j)$ and a contribution from the relevant terms of $\sum_{j=1}^{b_{i-1}} \partial(a^{i-1}_j) \otimes \beta^{i-1}_j$. 
    
    However, since the the map $\partial^{i-1}$ for $i \geq 2$ consists of only comuliplication maps, note that every $\partial^{i-1}(a^{i-1}_j)$ for $1 \leq j \leq b^{i-1}$ will contain exactly $0$ or $2$ terms in each $\phi \in E_i$. Thus, terms of the form $\sum_{j=1}^{b_{i-1}} \partial(a^{i-1}_j) \otimes \beta^{i-1}_j$ will cancel in pairs as we add the equations. Then, considering \eqref{eq: torusHom}, we see that if $[\varphi^i] \neq 0$, then $\left[\sum_{j\in A} \alpha^i_j\right] \neq 0$. So, 
    \begin{align*}
        |x_0| &= \sum_{i=0}^\ell \sum_{j=1}^{b_i} |\alpha^i_j| \geq \sum_{j=1}^{b_i} |\alpha^i_j| \geq \sum_{\phi \in E_i} \left|\sum_{j \in A_\phi} \alpha^i_j\right| = \dist^i(D^\text{tl}_\ell)\dist^{m-i}(D).
    \end{align*}
    Since at least one of these cases must be true, we have our desired result.
\end{proof}

\section{Generalization to \texorpdfstring{$\mathfrak{sl}_3$}{sl3} link homology} \label{sec:sl3 homology}

\subsection{\texorpdfstring{$\mathfrak{sl}_3$}{sl3} link homology}

One of the many generalizations of Khovanov homology is the $\mathfrak{sl}_3$ link homology, first introduced by Khovanov in \cite{Khovanov_sl3}. In this section, we use this formulation to derive CSS codes with desirable properties. We first recall the definition of $\mathfrak{sl}_3$ link homology, and refer the readers to \cite{Khovanov_sl3} for more details. 

Let $L\subset S^3$ be a link, and let $D$ be be a diagram of $L$. In the $\mathfrak{sl}_3$ case, we resolve the crossings of $D$ as in Figure \ref{fig: smooothings}.
\begin{figure}[!hbt]
\centering

\begin{tikzcd}[column sep=tiny]
& & \begin{tikzpicture}[baseline=5pt,scale=1.2]
\draw[thick, -{Stealth[length=2mm, width=1mm]}] (0.5,0) -- (0,.5);
\draw[thick] (0,0) -- (0.2,0.2); 
\draw[thick, -{Stealth[length=2mm, width=1mm]}] (0.3,0.3) -- (0.5,0.5); 
\end{tikzpicture} \ar[drr, "1"] \ar[dll, swap, "0"]& &\\
  \begin{tikzpicture}[baseline=5pt,scale=1.2]
\draw[thick, -{Stealth[length=2mm, width=1mm]}] (0,0) .. controls (.2,.25) .. (0,.5);

\draw[thick, -{Stealth[length=2mm, width=1mm]}] (.5,0) .. controls (.3,.25) .. (.5,.5);

\end{tikzpicture}
    & & &
      & \begin{tikzpicture}[baseline=5pt,scale=0.9, inward/.style={postaction={decorate}, decoration={
        markings, mark=at position 0.8 with {\arrow{Stealth[length=2mm, width=1mm]}}}}]
\draw[thick, -{Stealth[length=2mm, width=1mm]}] (0,0.5) -- (-0.5,0.75);
\draw[thick, -{Stealth[length=2mm, width=1mm]}] (0,0.5) -- (0.5,0.75);
\draw[thick, inward] (0,0.5) -- (0,0);
\draw[thick, -{Stealth[length=2mm, width=1mm]}] (-0.5,-0.25) -- (0,0);
\draw[thick, -{Stealth[length=2mm, width=1mm]}] (0.5,-0.25) -- (0,0);
\end{tikzpicture} \\
& & \begin{tikzpicture}[baseline=5pt,scale=1.2]
\draw[thick, -{Stealth[length=2mm, width=1mm]}] (0,0) -- (.5,.5);
\draw[thick] (0.5,0) -- (0.30,0.20); 
\draw[thick, -{Stealth[length=2mm, width=1mm]}] (0.2,0.3) -- (0,0.5); 
\end{tikzpicture} \ar[urr, swap, "0"] \ar[ull,"1"] & &
\end{tikzcd}
\caption{$0$ and $1$ resolutions of the crossing}
\label{fig: smooothings}
\end{figure}

Whereas each resolution of $D$ consists of disjoint circles in Khovanov homology (which is associated with the Lie algebra $\mathfrak{sl}_2$), it consists of \textit{webs} in the $\mathfrak{sl}_3$ case. Formally, a \textit{web} $\Gamma$ is an oriented trivalent graph embedded in a plane, possibly with verticeless loops, such that at each vertex edges are oriented in one of the ways shown in \eqref{eq:orientations at vertices}. 
   \begin{equation}
   \label{eq:orientations at vertices}
    \begin{aligned}
        \includestandalone[scale=0.6]{web}
    \end{aligned}
\end{equation}
For instance, the following are webs \begin{equation*}
    \begin{aligned}
        \includestandalone[scale=0.6]{web_example}
    \end{aligned}
\end{equation*}
A \textit{closed foam} $U$ is a closed oriented 2-dimensional CW complex embedded in $\mathbb{R}^3$ satisfying the following properties:
\begin{itemize}
    \item Every point in $U$ has a neighborhood homeomorphic to the letter Y times an interval, as shown in Figure \ref{fig: singular circle}, or to $\mathbb{R}^2$. The first type of points are called \textit{singular points}, and a connected component in the collection of all singular points form a \textit{singular circle}. 
    \item A \textit{facet} of $U$ is a connected component in the complement of singular circles in $U$. Each facet is oriented such that every two of the three facets meeting at a singular circle $C$ are oriented incompatibly, which induces an orientation on $C$, as shown in the left side of Figure \ref{fig: singular circle}. Furthermore, the orientation on $C$ induces a cyclic ordering of the three facets meeting at $C$, as shown in the right side of Figure \ref{fig: singular circle}. Note that there are two possible cyclic orderings at each $C$; we choose the cyclic ordering according to the left-hand rule.
    \item Each facet of a foam can be decorated with dots that can move freely on the facet to which they belong, but are not allowed to cross over any part of singular circles. 
\end{itemize}

A \textit{foam with boundary} is the intersection of a closed foam $U$ and a thickened plane $T\times [0,1]\cong \mathbb{R}^2\times [0,1]$ (embedded into $\mathbb{R}^3$ in the standard way), such that $T\times\{0, 1\}\cap U$ are webs. We refer to foams with boundary simply as \textit{foams}. Interpreted as morphisms between webs, foams are read from bottom to top by convention, and we denote $U:\Gamma_1\to \Gamma_2$ for a foam $U$ that has $\Gamma_1$ as its bottom boundary (i.e. $\Gamma_1 = (T\times \{0\})\cap U$) and $\Gamma_2$ as its top boundary (i.e. $\Gamma_2 = (T\times \{1\})\cap U$).
As such, a closed foam $U$ is a foam $U:\varnothing \to \varnothing$. Figure \ref{fig: theta foam} shows an example of a closed foam.

We use \textbf{Foam} to denote the category where the objects are webs and the morphisms are $\mathbb{F}_3$-linear combinations of isotopy classes of foams. We can compose two foams $U_1:\Gamma_0\to \Gamma_1$ and $U_2: \Gamma_1\to \Gamma_2$ the usual way by gluing along their common boundary $\Gamma_1$, and we denote the composition by $U_2U_1$. We will elaborate more on the choice of the ground field $\mathbb{F}_3$ later. 

\begin{figure}[!hbt]
    \centering
    \begin{minipage}{0.52\textwidth}
        \centering
        \includestandalone[scale = 1]{singular_circle}
        \caption{The orientations of the three facets meeting at a singular circle $C$ induce an orientation on $C$ (left). The orientation on $C$ in turn induces a cyclic order of the three facets (right).}
        \label{fig: singular circle}
    \end{minipage}
    \begin{minipage}{0.47\textwidth}
        \centering
        \begin{equation*}
            \theta(a,b,c) = \begin{aligned}
                \includestandalone[scale = 0.6]{standard_theta}
            \end{aligned}
        \end{equation*}
        \caption{The theta foam $\theta(a,b,c)$, where $a,b,c$ represent the number of dots decorating each facet. The cyclic order of facets induced by the singular circle at the equator is $1\to 2\to 3\to 1$.}
        \label{fig: theta foam}
    \end{minipage}
\end{figure}

We now introduce a functor $\mathcal{F}$ from \textbf{Foams} to $\textbf{Vect}_{\mathbb{F}_3}$, the category of $\mathbb{F}_3$-vector spaces. Let $A = \mathbb{F}_3[X]/(X^3)$. This is a vector space over $\mathbb{F}_3$ equipped with a trace map 
\[\varepsilon: A\to \mathbb{F}_3, \ \ \varepsilon(X^2) = -1, \ \ \varepsilon(X) = \varepsilon(1) = 0\]
and a unit map 
\[\iota: \mathbb{F}_3\to A,\ \  1\to 1\]
along with a multiplication map $m:A\otimes A\to A$ and a comultiplication map $\Delta: A\to A\otimes A$, 
\begin{align*}
    \Delta(1) &= -X^2\otimes 1 - X\otimes X - 1\otimes X^2\\
    \Delta(X) &= -X\otimes X^2 - X^2\otimes X\\
    \Delta(X^2) &= - X^2\otimes X^2. 
\end{align*}
First, we define the functor $\mathcal{F}$ on the category of dotted 2-dimensional oriented cobordisms, which is a subcategory of $\textbf{Foams}$. For an object $M$, which is a 1-manifold consisting of $J$ disjoint copies of $S^1$ for some $J\in \mathbb{Z}_{\ge 0}$, the image $\mathcal{F}(M)$ is defined as $A^{\otimes J}$. For a 2-cobordism $U$, the image $\mathcal{F}(U)$ is defined similarly to the TQFT constructed in Section \ref{sec:Khovanov homology}. For example, if $U$ is the ``pants cobordism" from two circles to one circle, the associated linear map $\mathcal{F}(U)$ is the multiplication map $m: A\otimes A\to A$, and so on. Note that a dot on a 2-cobordism denotes multiplication by $X$. For instance, the cylinder $U = S^1\times [0,1]$ induces the identity map $\id: A\to A$, whereas adding a dot to $U$ induces the multiplication by $X$ endomorphism on $A$.

Next, we define the functor $\mathcal{F}$ on a closed foam $U$. To each $U$ we assign an element $\mathcal{F}(U)$ in the ground field $\mathbb{F}_3$, called the \textit{evaluation} of $U$, according to the following rules:
\begin{figure}
        \centering
\includestandalone[scale = 1]{neck_cutting_sl3}
        \caption{The neck-cutting relations. The foams in the picture are shown without an orientation to indicate that the relation holds for both orientations.}
        \label{fig: neck cutting}
    \end{figure}
\begin{itemize}
    \item If $U$ is a closed surface, i.e., $U$ doesn't have any singular circles, then $\mathcal{F}(U)$ is zero unless $U$ is a 2-sphere decorated by 2 dots, in which case $\mathcal{F}(U) = -1$.
    \item $\mathcal{F}$ is multiplicative with respect to disjoint union of closed foams, i.e., 
    \[\mathcal{F}(U_1\sqcup U_2) = \mathcal{F}(U_1)\mathcal{F}(U_2)\]
    \item For an annulus inside a facet $U$, as in Figure \ref{fig: neck cutting}, do a surgery to obtain foams $U_1, U_2, U_3$, as depicted on the right side of Figure \ref{fig: neck cutting}. Then we have the \textit{neck-cutting relations}
    \[\mathcal{F}(U) = -\mathcal{F}(U_1) - \mathcal{F}(U_2)- \mathcal{F}(U_3).\]
    \item The theta foam $\theta(a,b,c)$ as shown in Figure \ref{fig: theta foam} has the following evaluation 
    \[
        \mathcal{F}(\theta(a,b,c)) = \begin{cases}
			1, & \text{if $(a,b,c) = (0,1,2)$, up to cyclic permuation}\\
            -1, & \text{if $(a,b,c) = (0,2,1)$, up to cyclic permutation}\\
            0, & \text{otherwise}
		 \end{cases} 
    \]
\end{itemize}
These rules are consistent with the definition of $\mathcal{F}$ on oriented 2-cobordisms and they uniquely determine the evaluation of any closed foam $U$. For a proof of consistency and uniqueness, see \cite[Proposition 3]{Khovanov_sl3}.

\begin{figure}
    \centering
    \includestandalone{bubble_bursting}
    \caption{Bubble-bursting relations}
    \label{fig: bubble bursting}
\end{figure}
As a consequence of these rules, we obtain a particularly useful family of relations, called the \textit{bubble-bursting relations}, which were first introduced in \cite[Figure 18]{Khovanov_sl3}. Let $U$ be a foam, and let $U_{n,m}$ denote the foam obtained by adding a bubble to a facet of $U$, such that the two new facets are decorated with $m$ dots and $n$ dots respectively, and the facet with $n$ dots directly precedes the facet with $m$ dots in the cyclic ordering. Let $U_n$ denote the foam obtained from $U$ by adding $n$ dots to the same facet. A schematic of $U_n$ and $U_{n,m}$ are shown in Figure \ref{fig: bubble bursting}. Then we have 
\begin{align*}
    & \mathcal{F}(U_{n,n}) = 0, \ \ \ \ \mathcal{F}(U_n) = 0\textrm{ for }n\ge 3,\ \ \ \mathcal{F}(U_{1,0}) = \mathcal{F}(U) = -\mathcal{F}(U_{0,1}),\\
    & \mathcal{F}(U_{0,2}) = \mathcal{F}(U_1) = -\mathcal{F}(U_{2,0}),\ \ \ \ \ \mathcal{F}(U_{2,1}) = \mathcal{F}(U_2) = -\mathcal{F}(U_{1,2}).
\end{align*}

Next, we define the functor $\mathcal{F}$ on a web $\Gamma$: it is a $\mathbb{F}_3$-vector space freely generated by all foams $U:\varnothing\to \Gamma$ modulo the relation $\sim$: for $a_i\in \mathbb{F}_3$ and $U_i:\varnothing \to \Gamma$, we have $\sum_{i}a_iU_i \sim 0$  if and only if 
    \[\sum_i\ a_i\mathcal{F}(VU_i) = 0\]
for any foam $V:\Gamma\to \varnothing$, where $\mathcal{F}(VU_i)$ is the evaluation of a closed foam. Note that $\F(\Gamma)$ is finitely generated, and its dimension can be computed recursively via the relations in \cite[Figure 8]{Khovanov_sl3}.

Finally, for each foam $U:\Gamma_1\to \Gamma_2$, there exists a natural homomorphism 
\[\mathcal{F}(U) : \mathcal{F}(\Gamma_1)\to \mathcal{F}(\Gamma_2)\]
which takes a foam $V:\varnothing \to \Gamma_1$ to $UV:\varnothing \to \Gamma_2$. Functoriality follows directly from this definition. Furthermore, by an argument in \cite[Corollary 3]{Khovanov_sl3}, one can also show that the natural map $\mathcal{F}(\Gamma_1)\otimes \mathcal{F}(\Gamma_2)\to \mathcal{F}(\Gamma_1\sqcup \Gamma_2)$ is an isomorphism.

To a link diagram $D$, we use $\mathcal{F}$ to associate a chain complex $C(D)$ over $\mathbb{F}_3$ as follows. First, we form the cube of resolutions as in \ref{sec:Khovanov homology}, except that we now resolve each crossing of $D$ according to Figure \ref{fig: smooothings}. Label the crossings of $D$ from $1$ to $n$. To a vertex $u = (u_1,\cdots, u_n)\in \{0,1\}^n$, we associate an abelian group $\mathcal{F}(D_u)$, where $D_u$ is the web obtained from performing the $u_i$-th resolution at the $i$th crossing. To an edge from $u$ to $v$, where $u=(u_1,\cdots, u_n)$ and $v = (v_1,\cdots, v_n)$ are vertices that differ only on the $i$th entry such that $v$ is the immediate successor of $u$, we associate the group homomorphism $\mathcal{F}(D_u)\to \mathcal{F}(D_v)$ induced by the ``singular saddle" cobordism in Figure \ref{fig: singular saddle} if the $i$-th crossing is positive, or its reflection across the horizontal plane if the $i$-th crossing is negative. In this way, we obtain an $n$-dimensional cube of abelian groups. Add minus signs to some of the maps as in \cite[Section 3.3]{Khovanov} such that the cube anticommutes, and then form the total complex $C(D)$ of this cube, where 
\[C^i(D)= \bigoplus_{|u| = i + n_+}\mathcal{F}(D_u)\]
with $|u|=\sum_i u_i$ and $n_+$ denoting the number of positive crossings in $D$. In this way, we obtain the $\mathfrak{sl}_3$ chain complex $C(D)$ associated to a link diagram. It is shown in \cite[Section 5]{Khovanov_sl3} that the homology of $C(D)$ is a link invariant.
\begin{figure}
    \centering
\includestandalone[scale = 0.7]{singular_saddle}
    \caption{The ``singular saddle" cobordism}
    \label{fig: singular saddle}
\end{figure}

For example, the chain complex associated to the following diagram of the unknot is
\begin{equation*}
\begin{tikzpicture}[scale = 0.4, baseline=0pt]
    \begin{knot}
    \strand (-3,1) .. controls  +(-1,0) and +(+1,0).. (-4.5, -1) .. controls  +(-1,0) and +(0,0) .. (-5.5,0);
    \strand (-5.5,0) .. controls  +(0,0) and +(-1,0).. (-4.5, 1) .. controls  +(1,0) and +(-1,0) .. (-3,-1);
    \strand[
  only when rendering/.style={
    postaction=decorate,
  },
  decoration={
    markings,
    mark=at position .45 with {\arrow{<}}
  }] (-3,1) .. controls  +(1.5,0) and +(+1.5,0).. (-3, -1) ;
    \end{knot}
    \end{tikzpicture}
    \rightsquigarrow \left(\mathcal{F}
   \left(
   \begin{tikzpicture}[scale = 0.3, baseline=0pt, inward/.style={postaction={decorate}, decoration={
        markings, mark=at position 0.5 with \arrow{<}}}]
    \draw (2,-1)--(2,1);  
    \draw[inward] (2,1) arc (90:-90:2 and 1);
    \draw (2,-1) arc (-90:90:-2 and 1);
\end{tikzpicture}\right)\rightarrow\mathcal{F}\left(
\begin{tikzpicture}[scale = 0.3, baseline=0pt]
\tikzset{arrow circle/.style={
        postaction={decorate, decoration={
            markings, mark=at position #1 with {\arrow{<}}
        }}
    }}
 \tikzset{reverse arrow circle/.style={
    postaction={decorate, decoration={
        markings, mark=at position #1 with {\arrow{>}}
    }}
}}
\draw[arrow circle=0.04] (0,0) circle (1);
\draw[reverse arrow circle=0.5] (3,0) circle (1);
\end{tikzpicture}
\right)\right)
\raisebox{-4ex}{\makebox[0pt][l]{\hspace{-26ex}$-1$}}\raisebox{-4ex}{\makebox[0pt][l]{\hspace{-8ex}$0$}}
\end{equation*}
where $-1,0$ beneath the diagrams indicate their homological gradings. One can check that the homology of this complex is the same as that obtained from a circle oriented counterclockwise.

We conclude with a remark about gradings. As in \cite{Audoux} and Section \ref{sec:Khovanov homology}, we omit the second (quantum) grading in this discussion because in what follows, we will adopt a non-homogeneous basis with respect to this grading. We refer interested readers to \cite{Khovanov_sl3} for a discussion of the quantum grading.

In the following subsection, we will calculate the CSS codes associated with the $\mathfrak{sl}_3$ chain complex of the unknot diagrams.

\subsection{Unknot codes}
For every $k,l\in \mathbb{N}$, we consider the following diagram $D_{k,l}$ of the oriented unknot with $k$ positive crossings and $l$ negative crossings, as shown in Figure \ref{fig: unknot}.
\begin{figure}[!hbt]
  \centering
  \includestandalone{unknot}
  \caption{The unknot code $D_{k,l}$, with $k$ positive crossings on the left, and $l$ negative crossings on the right. We fix the orientation so that the rightmost arc is oriented counterclockwise.}
  \label{fig: unknot}
\end{figure}

We call the \textit{$l$th unknot code with basis $\mathcal{B}$} the code obtained from 
\[\Big(C^{-1}(D_{l,l}); \mathcal{B}\Big)\xrightarrow{\partial}\Big(C^0(D_{l,l}); \mathcal{B}\Big)\xrightarrow{\partial}\Big(C^{1}(D_{l,l}); \mathcal{B}\Big)\]
where $\mathcal{B}$ is a basis for the chain complex, and we denote its parameters by $[[n_l;k_l;d_l^\mathcal{B}]]$. Note that only the distance depends on the basis. 

For $s\in \mathbb{N}$, we denote by $\Theta_s$ a \textit{generalized theta web} with $s$ rungs, that is,
\begin{equation*}
  \Theta_s = \begin{aligned}
  \includestandalone[scale=0.5]{ThetaBasis0dot}
  \end{aligned}
  \end{equation*}
In particular, the web $\Theta_0$ is just a circle. The orientation of $\Theta_s$ is not specified in the diagram, as it can represent either orientation. Note that every vertex in the cube of resolutions of $D_{k,l}$ is a disjoint union of generalized theta webs. In order to compute the distance of $D_{k,l}$, we need to specify a basis for $\F(\Theta_s)$, where $s\ge 0$. First, we endow the following basis for $\F(\Theta_0) = A = \mathbb{F}_3[X]/(X^3)$:
\[\framebox{0} = 1, \ \ \ \ \framebox{1} = 1-X, \ \ \ \ \framebox{2} = 1+X+X^2\]
which we refer to as the \textit{box basis}.
\begin{lemma}
\label{lem:box basis properties}
    The box basis $\{\framebox{0}, \framebox{1},\framebox{2}\}$ satisfies the following properties
\begin{itemize} \label{properties of box}
        \item (Closure) For any $i,j\in \{0,1,2\}$, we have \[\framebox{$i$}\cdot \framebox{$j$} = \framebox{i+j}\]
        where the integers in the boxes are taken modulo 3. 
        \item (Negative duality) For any $i\in \{0,1,2\}$, the dual basis (as defined in Section \ref{sec:hom alg}) is $\framebox{$i$}^{*} = -\framebox{2-i}$. A schematic is shown in Figure \ref{fig: self dual}.
\end{itemize}
\end{lemma}
\begin{proof}
    Closure is straightforward to check. As for negative duality, one needs to check that for any $i,j\in \{0,1,2\}$, \[\framebox{$i$}^*\Big(\framebox{$j$} \Big)= -\delta_{ij}.\]
    Indeed, the left hand side of the equation is the evaluation of a sphere decorated with $\boxed{i}\cdot \boxed{j} = \boxed{i+j}$, which is $-1$ if $i+j=2$, and $0$ otherwise. 
\end{proof}
\begin{figure}[!hbt]
  \centering
  \includestandalone[scale=1]{Box_basis}
  \caption{The dual of $\framebox{$i$}$ is $-\framebox{$2$-$i$}$.}
  \label{fig: self dual}
\end{figure}
\begin{remark} \label{choice of box basis}
As we will see, efficiently computing the distance depend on three properties on the chosen basis, up to a sign: closure under multiplication, closure under taking duals, and the inclusion of the identity element 1. Notably, the box basis is the only basis of $A$ that satisfies these three properties. Furthermore, if the ground field is set to $\mathbb{F}_2$ so that $A = \mathbb{F}_2[X]/(X^3)$, no such basis exists at all.
\end{remark}

For $s\ge 1$, we introduce two bases $\mathcal{B}_1$ and $\mathcal{B}_2$ for $\F(\Theta_s)$, each consisting of foams from $\varnothing$ to $\Theta_s$. In basis $\mathcal{B}_1$, each foam is decorated by $\boxed{j}$ on its leftmost facet, where $j\in \{0,1,2\}$, and $i_1,\cdots,i_s$ dots in each facet between two adjacent rungs, where each $i_k\in \{0,1\}$. We denote this foam by $F(\framebox{$j$}, i_1,\cdots,i_s; \mathcal{B}_1)$. In $\mathcal{B}_2$, each foam is similarly decorated by $\boxed{j}$ on its leftmost facet, but the $i_1,\cdots,i_s$ dots are placed in each facet that contains a rung. We denote this foam by $F(\framebox{$j$}, i_1,\cdots,i_s; \mathcal{B}_2)$. A schematic of $\mathcal{B}_1$ and $\mathcal{B}_2$ is shown in Figure \ref{fig: basis}. Note that the orientation is omitted in Figure \ref{fig: basis} because the bases $\mathcal{B}_1$ and $\mathcal{B}_2$ are chosen using the description above, regardless of the orientation of $\Theta_s$. Finally, one can verify that $\mathcal{B}_1$ and $\mathcal{B}_2$ are indeed bases of $F(\Theta_s)$ using an argument similar to \cite[Proposition 8]{Khovanov_sl3}, in which we reduce the bigons in $\Theta_s$ from right to left. It should be apparent from this defintion that the dimension of $\F(\Theta_s)$ is $3\cdot 2^s$.

\begin{figure}[!hbt]
\centering
\begin{subfigure}{.49\linewidth} \label{basis 1}
  \centering
  \includestandalone[scale=0.8]{ThetaBasisB1}
  \caption{The basis $\mathcal{B}_1$.}
  \label{fig: b1}
\end{subfigure}
\begin{subfigure}{.49\linewidth}
\label{basis 2}
\centering
  \includestandalone[scale =0.8]{ThetaBasisB2}
  \caption{The basis $\mathcal{B}_2$.}
  \label{fig: b2}
\end{subfigure}
\caption{Two bases for $\F(\Theta_s)$}
\label{fig: basis}
\end{figure}
For ease of notation, we also use $\mathcal{B}_i$ $(i=1,2)$ to denote the basis for a resolution of $D_{k,l}$ where each disjoint generalized theta web is assigned the basis $\mathcal{B}_i$. Furthermore, $\mathcal{B}_i$ is used to denote the basis for the cube of resolutions of $D_{k,l}$ where each resolution is endowed with $\mathcal{B}_i$.

\begin{remark}
There are many possible choices of bases that one can make for a generalized theta web. For example, in the basis $\mathcal{B}_2$ for some $\F(\Theta_s)$, if we move the decoration $i_1$ so that it lies on the facet between two rungs, we obtain a completely different basis, as illustrated in the following diagram
\begin{equation*}
    \begin{aligned}
        \includestandalone[scale = 0.8]{ThetaBasismodify}
    \end{aligned}
\end{equation*}
Furthermore, each resolution consists of disjoint unions of generalized theta webs $\Theta_s$, and we may choose different bases for each $\Theta_s$. We can also select different bases for each resolution in the cube of resolution of $D_{k,l}$. Each choice may lead to a potentially different distance for $D_{k,l}$. We use the basis $\mathcal{B}_1$, $\mathcal{B}_2$ specified above because they are particularly amenable to computation.
\end{remark}

Recall from Sections \ref{sec:hom alg} and \ref{sec:Khovanov homology} that $\ol{D}_{k,l}$ denotes the mirror image of $D_{k,l}$, and $\ol{C}(D_{k,l})$ denotes the dual complex of $C(D_{k,l})$ with grading negated. We define a map $\varphi: C(\ol{D}_{k,l})\to \ol{C}(D_{k,l})$ such that for each foam $F:\varnothing \to \Gamma$ in $C(\ol{D}_{k,l})$, we have $\varphi(F) = \ol{F}: \Gamma \to \varnothing$, where $\ol{F}$ is the reflection of $F$ across $\Gamma$. A benefit of the bases $\mathcal{B}_1, \mathcal{B}_2$ is that they allow us to relate the dual distance of $D_{k,l}$ to the distance of $\ol{D}_{k,l}$. The dual distance is usually difficult to compute, whereas the distance of $\ol{D}_{k,l}$ is more straightforward. The following proposition is the first step towards establishing this result.

\begin{proposition} \label{chain iso} The map $\varphi: C(\ol{D}_{k,l})\to \ol{C}(D_{k,l})$ as defined above is an isomorphism of chain complexes.
\end{proposition}
\begin{proof}
First, we show that $\varphi$ is a degree-preserving map. Specifically, we need to show that a resolution $D_u$ of $D$ lies in the same homological grading in both $C(\ol{D}_{k,l})$ and $\ol{C}(D_{k,l})$. Let $n_+, n_-, n$ be number of positive crossings, negative crossings and total crossings in $D_{k,l}$, respectively. In $C(\ol{D}_{k,l})$, the resolution $D_u$ has homological degree $n- |u|-n_- = n_+-|u|$. In $C(D_{k,l})$ and $C^*(D_{k,l})$, the same resolution $D_u$ has homological degree $|u|-n_+$, so with grading negated, it lies in degree $n_+-|u|$ in $\ol{C}(D_{k,l})$. Since $D_u$ lies in degree $n_+-|u|$ in both $C(\ol{D}_{k,l})$ and $\ol{C}(D_{k,l})$, the map $\varphi$ is degree-preserving.

Next, we show that $\varphi$ commutes with the differential map. In the cube of resolution of $D_{k,l}$, the differential either merges two generalized theta webs via the map
\[m_{s,r}:\F(\Theta_s\sqcup \Theta_r) \to \F(\Theta_{s+r+1}),\]
or splits a generalized theta web into two via
\[\Delta_{s,r}: \F(\Theta_{s+r+1})\to \F(\Theta_s\sqcup \Theta_r).\]
It suffices to check that the diagrams in Figure \ref{fig:comdiags} commute. This follows by observing that the reflection of the split cobordism $\Delta_{s,r}$ across $\Theta_s\sqcup \Theta_r$ is exactly $m_{s,r}^{*}$, and the reflection of the merge cobordisms $m_{s,r}$ across $\Theta_{s+r}$ is exactly $\Delta_{s,r}^{*}$.
\begin{figure}[!hbt]
\centering
\begin{subfigure}{.4\linewidth} 
\centering
    \begin{tikzcd}
 \F(\Theta_{s+r+1})^{*} \arrow{r}{m_{s,r}^*} 
    & \F(\Theta_s\sqcup \Theta_r)^{*}   \\
  \F(\Theta_{s+r+1}) \arrow{r}{\Delta_{s,r}} \arrow{u}{\varphi}
    & \F(\Theta_s\sqcup \Theta_r) \arrow{u}{\varphi} \end{tikzcd}
    \end{subfigure}
    \begin{subfigure}{.4\linewidth}
    \centering
 \begin{tikzcd}
  \F(\Theta_s\sqcup \Theta_r)^{*} \arrow{r}{\Delta_{s,r}^*} 
    & \F(\Theta_{s+r+1})^{*}  \\
  \F(\Theta_s\sqcup \Theta_r)  \arrow{r}{m_{s,r}} \arrow{u}{\varphi}
    & \F(\Theta_{s+r+1}) \arrow{u}{\varphi} \end{tikzcd}
    \end{subfigure}
    \caption{The commutativity of the two diagrams ensures that $\varphi$ is a chain map.}
    \label{fig:comdiags}
\end{figure}

Finally, we show that $\varphi$ is an isomorphism of vector spaces at each degree. Indeed, there exists an inverse chain map $\phi:\ol{C}(D_{k,l})\to C(\ol{D}_{k,l})$ which sends a foam $G: \Gamma\to \varnothing$ to its reflection $\ol{G}$ across $\Gamma$. One checks that $\phi\circ \varphi = \id_{C(\ol{D}_{k,l})}$ and  $\varphi\circ \phi = \id_{\ol{C}(D_{k,l})}$.
\end{proof}
To clarify the dependency of distance on the choice of basis, we denote the homological distance of a chain complex $C$ at degree $i$ with respect to a basis $\mathcal{B}$ as $\dist^i(C; \mathcal{B})$. 

\begin{corollary} \label{dual distance}
Let $\mathcal{B}_1, \mathcal{B}_2$ be basis defined above. Then for each $i\in \mathbb{Z}$, we have
\begin{align*}
& \dist^i(C(\ol{D}_{k,l});\mathcal{B}_1) = \dist^i(\ol{C}(D_{k,l});\mathcal{B}_2)\ = \dist^{-i}(C^{*}(D_{k,l});\mathcal{B}_2), \\
& \dist^i(C(\ol{D}_{k,l});\mathcal{B}_2) = \dist^i(\ol{C}(D_{k,l});\mathcal{B}_1)\ = \dist^{-i}(C^{*}(D_{k,l});\mathcal{B}_1).
\end{align*}
\end{corollary}
\begin{proof}
    We will prove the equation in the first line; the second line follows analogously. Consider the chain isomorphism $\varphi: C(\ol{D}_{k,l})\to \ol{C}(D_{k,l})$ in Proposition \ref{chain iso}. For a basis element $F(\framebox{$j$},i_1,\cdots,i_s; \mathcal{B}_1)$ of $\F(\Theta_s)$, we claim that up to a sign, its image under $\varphi$ is the dual vector basis $F(\framebox{2-$j$}, ,1-i_1,\cdots, 1-i_s; \mathcal{B}_2)^*$. This is because if we evaluate the closed foam
    \[\ol{F}(\framebox{$j$},i_1,\cdots, i_s; \mathcal{B}_1)\circ F(\framebox{$k$},l_1,\cdots, l_s; \mathcal{B}_2)\]
    using the bubble-bursting relations in Figure \ref{fig: bubble bursting}, we find that it is equal to $\pm 1$ if $k= 2-j$ and $l_m =1-i_m$ for all $1\le m\le s$, and it is equal to $0$ otherwise. In other words, if we pick the basis $\mathcal{B}_1$ for $C(\ol{D}_{k,l})$ and $\mathcal{B}_2$ for $\ol{C}(D_{k,l})$, the map $\varphi$ is a chain isomorphism which, up to a sign, sends each basis element to a dual basis element. By \cite[Proposition 2.10]{Audoux}, we have $\dist^i(C(\ol{D}_{k,l});\mathcal{B}_1) = \dist^i(\ol{C}(D_{k,l});\mathcal{B}_2)$ for all $i\in \mathbb{Z}$. 
\end{proof}
Note that the mirror image of $D_{l,l}$ is exactly $D_{l,l}$ itself. By Corollary $\ref{dual distance}$, we can compute the dual distance of $C(D_{l,l})$ with basis $\mathcal{B}_1$ (resp. $\mathcal{B}_2$) by computing the distance with basis $\mathcal{B}_2$ (resp. $\mathcal{B}_1$). In the next two propositions, we present a strategy for calculating the distance of $C(D_{l,l})$ with either basis $\mathcal{B}_1$ or $\mathcal{B}_2$.

\begin{proposition} \label{Dkl to D0l}
For any $k,l\in \mathbb{N}$ and $\mathcal{B}$ is either $\mathcal{B}_1$ or $\mathcal{B}_2$, we have
\[\dist^0(C(D_{k,l});\mathcal{B}) = \dist^0(C(D_{0,l});\mathcal{B}).\]
\end{proposition}
\begin{proof}
It suffices to show that for any $k\ge 1$, we have
\[\dist^0(C(D_{k,l});\mathcal{B}) = \dist^{0}(C(D_{k-1,l});\mathcal{B}),\]
from which the statement follows by recursively applying this relation. Note that $D_{k,l}$ and $D_{k-1,l}$ differ by a positive Reidemeister I move. As shown in \cite[Section 5.1]{Khovanov_sl3}, there exists chain maps $g:C(D_{k,l})\to C(D_{k-1,l})$ and $-\iota: C(D_{k-1,l})\to C(D_{k,l})$, which induce isomorphisms on homology, defined as follows. Let $C(D_0)$ be the subspace of $C(D_{k,l})$ corresponding to the 0-resolution of the leftmost crossing, and let $C(D_1)$ be the subcomplex corresponding to the 1-resolution of the same crossing. Note that the 1-resolution is the disjoint union of a circle and $D_{k-1,l}$, so we have $C(D_1)\cong A\otimes C(D_{k-1,l})$. The chain map $g$ sends the subcomplex $C(D_0)$ to zero, and it merges the circle and $D_{k-1,l}$ on $C(D_1)$. The map  $\iota: C(D_{k-1,l})\to A\otimes C(D_{k-1,l})\cong C(D_1)$ is defined to be the unit map, i.e., 
$\iota(b) = 1\otimes b$. 

In both $\mathcal{B}_1$ and $\mathcal{B}_2$, the leftmost facet of any generalized theta foam is decorated by the box basis, so the map $g$ multiplies two box basis on the leftmost facet, as described in Lemma \ref{properties of box}. In basis $\mathcal{B}$, the maps $g$ and $-\iota$ are given by
\begin{equation*}
\begin{aligned}
      C\left(
\begin{tikzpicture}[baseline = -3, xscale=.6, yscale=.4]
\begin{knot}[
  clip width=5
]
\strand (-3,1) .. controls  +(-1,0) and +(+1,0).. (-4, -1) .. controls  +(-0.5,0) and +(0,0) .. (-4.5,0);
\strand (-4.5,0) .. controls  +(0,0) and +(-0.5,0).. (-4, 1) .. controls  +(1,0) and +(-1,0) .. (-3,-1);
\end{knot}
\end{tikzpicture}
  \right)&\xrightarrow{g}  C\left( 
  \begin{tikzpicture}[baseline = -3, xscale=.6, yscale=.4]
  \path (-3,-1) (-2.7,0);

\draw (-3,1) .. controls  +(-1,0) and +(-1,0).. (-3, -1) ;
\end{tikzpicture}
  \right) &  C\left( 
  \begin{tikzpicture}[baseline = -3, xscale=.6, yscale=.4]
  \path (-3,-1) (-2.7,0);

\draw (-3,1) .. controls  +(-1,0) and +(-1,0).. (-3, -1) ;
\end{tikzpicture}
  \right)&\xrightarrow{-\iota}  C\left(
\begin{tikzpicture}[baseline = -3, xscale=.6, yscale=.4]
\begin{knot}[
  clip width=5
]
\strand (-3,1) .. controls  +(-1,0) and +(+1,0).. (-4, -1) .. controls  +(-0.5,0) and +(0,0) .. (-4.5,0);
\strand (-4.5,0) .. controls  +(0,0) and +(-0.5,0).. (-4, 1) .. controls  +(1,0) and +(-1,0) .. (-3,-1);
\end{knot}
\end{tikzpicture}
  \right)\\
   \begin{tikzpicture}[baseline = -3, xscale=.4, yscale=.4]

\draw (-4.9,0) circle (23pt);
\path (-3,-1) (-2.7,0);
\draw (-3,1) .. controls  +(-1,0) and +(-1,0).. (-3, -1) (-2.9,0)node{\tiny $\framebox{$j$}$};
\draw(-4.9,0)node{\tiny $\framebox{$i$}$};

\end{tikzpicture}
&\mapsto 
\begin{tikzpicture}[baseline = -3, xscale=.4, yscale=.4]
\path (-3,-1) (-2.7,0);
\draw (-2.9,1) .. controls  +(-1,0) and +(-1,0).. (-3, -1) (-2.4,0)node{\tiny $\framebox{$i$+$j$}$};
\end{tikzpicture} &  
\begin{tikzpicture}[baseline = -3, xscale=.4, yscale=.4]
\draw (-2.9,1) .. controls  +(-1,0) and +(-1,0).. (-3, -1) (-2.8,0)node{\tiny $\framebox{$j$}$};
\end{tikzpicture}
&\mapsto - 
\begin{tikzpicture}[baseline = -3, xscale=.4, yscale=.4]
\draw (-4.9,0) circle (23pt);
\draw (-3,1) .. controls  +(-1,0) and +(-1,0).. (-3, -1) (-2.9,0)node{\tiny $\framebox{$j$}$};
\draw(-4.9,0)node{\tiny $\framebox{0}$};
\end{tikzpicture}
  \end{aligned}
  \end{equation*}
  where the crossing in the local picture represents the leftmost crossing of $D_{k,l}$. By \cite[Proposition 2.10]{Audoux}, we obtain \[\dist^0(C(D_{k,l});\mathcal{B}) = \dist^0(C(D_{0,l});\mathcal{B}),\]
as desired.
\end{proof}

\begin{proposition} \label{D0l}
    For any $l\in \mathbb{N}$, the distance of the unknot code $D_{0,l}$ is given by
    \[\dist^0(C(D_{0,l}; \mathcal{B}) = 3^l,\]
    where $\mathcal{B}$ is either $\mathcal{B}_1$ or $\mathcal{B}_2$.
\end{proposition}
\begin{proof}
The distance $\dist^0(C(D_{0,l}; \mathcal{B}))$ is obtained from the minimally weighted element in
\[\ker \bigg(\Big(C^{0}(D_{0,l}); \mathcal{B}\Big)\xrightarrow{\partial}\Big(C^1(D_{0,l}); \mathcal{B}\Big)\bigg) = H^0(C(D_{0,l}))\cong \mathbb{F}_3^{\otimes 3}.\]
Our goal is to find a basis for $H^0(C(D_{0,l}))$, from which we can find the minimally weighted element by calculating the weight of all linear combinations of the basis. First, note that $D_{0,l}$ may be obtained from $D_{0,0}$ by successively applying $l$ negative Reidemeister I moves.
For a negative Reidemeister I move turning $D_{0,i}$ into $D_{0,i+1}$, we define a chain map $f: C(D_{0,i})\to C(D_{0,i+1})$, following the construction in \cite[Section 5.1]{Khovanov_sl3}, which induce an isomorphism on homology. Let $C(D_0)$ be the subcomplex of $C(D_{0,i+1})$ corresponding to the 0-resolution of the rightmost crossing, which consists of the disjoint union of a circle and $D_{0,i}$. The map $f$ sends $C(D_{0,i})$ into $C(D_0)$ and is induced by the foam which splits off a circle from $D_{0,i}$.

Now, a basis for $H^0(D_{0,0})$ is a cup with $j$ dots, where $j\in \{0,1,2\}$. Applying the chain isomorphism $f$ successively, we obtain an induced basis for $H^0(D_{0,l})$. They are the foams $F_j$ shown below, where $j\in \{0,1,2\}$, whose underlying surface is a sphere with $l+1$ punctures decorated by $j$ dots:
\begin{equation*}
   F_j =  \begin{aligned}
        \includestandalone[scale=.5]{General_Cups}
    \end{aligned}
\end{equation*}
Using the neck-cutting relation in Figure \ref{fig: neck cutting} on each of the $l+1$ top circles, we obtain the following expressions, up to a sign:
\begin{align*}
    F_0 & = \sum_{\substack{a_1+\cdots +a_{l+1} = 2l \\  a_i \in \{0,1,2\}}}X^{a_1}\otimes \cdots \otimes X^{a_{l+1}}\\
    F_1 &= \sum_{\substack{a_1+\cdots +a_{l+1} = 2l+1 \\  a_i \in \{0,1,2\}}}X^{a_1}\otimes \cdots \otimes X^{a_{l+1}}\\
    F_2 & = (X^2)^{\otimes {l+1}}
\end{align*}
Applying the change of basis $1 = \framebox{0}$, $X = \framebox{0}-\framebox{1}$, $X^2 = \framebox{0}+\framebox{1}+\framebox{2}$, we can write $F_0, F_1, F_2$ as:
\begin{align*}
    F_0 & =  \sum_{\substack{k_1+k_2= l, \\k_1,k_2\ge 0}}(\framebox{0}+\framebox{1}+\framebox{2})^{\otimes k_1}\otimes \framebox{0}\otimes (\framebox{0}+\framebox{1}+\framebox{2})^{\otimes k_2}\\
    & + \sum_{\substack{k_1+k_2+k_3= l-1\\k_1,k_2,k_3\ge 0}}(\framebox{0}+\framebox{1}+\framebox{2})^{\otimes k_1}\otimes (\framebox{0}-\framebox{1})\otimes (\framebox{0}+\framebox{1}+\framebox{2})^{\otimes k_2}\otimes(\framebox{0}-\framebox{1})\otimes(\framebox{0}+\framebox{1}+\framebox{2})^{\otimes k_3}\\
    F_1 & =  \sum_{\substack{k_1+k_2= l, \\k_1,k_2\ge 0}}(\framebox{0}+\framebox{1}+\framebox{2})^{\otimes k_1}\otimes(\framebox{0}-\framebox{1})\otimes(\framebox{0}+\framebox{1}+\framebox{2})^{\otimes k_2}\\
    F_2 & = (\framebox{0}+\framebox{1}+\framebox{2})^{\otimes (l+1)}.
\end{align*}
It remains to compute the minimum weight of any linear combinations (over $\mathbb{F}_3$) of the basis elements $F_0,F_1,F_2$. By Lemma \ref{lemma: convergence} below, this minimum weight is given by $3^l$, as desired. 
\end{proof}
\begin{lemma} \label{lemma: convergence}
The minimum weight of any linear combination of the basis $\{F_0,F_1,F_2\}$ over $\mathbb{F}_3$ is $3^\ell$, where $F_0,F_1,F_2$ are defined above.
\end{lemma}
\begin{proof}
    We first show that individually, the weight of $F_0$ is $3^\ell$, the weight of $F_1$ is $2 \cdot 3^\ell$, and the weight of $F_2$ is $3^{\ell+1}$. Notice that the tensor products in each summand of $F_0, F_1, F_2$ can be represented by length $\ell+1$ sequences containing only $\{0, 1, 2\}$. Then, we can express each basis as the monomial \[F_i = \sum_{S \in \{0, 1, 2\}^{l+1}} c_{S, i} S,\] where each $c_{S, i} \in \mathbb F_3$. The weight of $F_i$ is given by the number of non-zero $c_{S, i}$.

    To calculate the weight of $F_0$, we compute $c_{S, 0}$ for an arbitrary length $\ell+1$ sequence. Let $n_0, n_1, n_2$ be the number of zeros, ones, and twos in $S$, respectively. Note that the expression for $F_0$ includes two summations. The first summation counts the sequence $S$ exactly $n_0$ times, once whenever $k_1$ equals one less than the index of each zero in the sequence. By an $(a, b)$ subsequence we mean a pair of indices $i < j$ such that $S_i = a$ and $S_j = b$. For each $(0, 0)$ or $(1, 1)$ subsequence in $S$, the second summation produces $S$ once. Similarly, for each $(0, 1)$ or $(1, 0)$ subsequence in $S$, the second summation produces $-S$ once. Note that the number of $(1, 0)$ subsequences in $S$ plus the number of $(0, 1)$ subsequences equals exactly $n_0 n_1$. Then, the coefficient of $S$ in $F_0$ is exactly
    \[c_{S, 0} = n_0 + \binom{n_0}{2} + \binom{n_1}{2} - n_0 n_1.\] One can verify by checking all cases for $n_0$ and $n_1$ mod 3 that $c_{S,0} \equiv 1 \pmod 3$ iff $n_0 - n_1 \equiv 1 \pmod 3$, and equals zero otherwise.
    
    We claim that the number of sequences of length $\ell + 1$ satisfying $n_0 - n_1 \equiv 1 \pmod 3$ is exactly $3^\ell$. Consider all sequences $S'$ of length $\ell$, and let $S'$ have counts $n_0', n_1', n_2'$. If $n_0' - n_1' \equiv 0$, we must append a zero to obtain a non-trivial sequence counted by $F_0$. If $n_0' - n_1' \equiv 1$, we must append a two. If $n_0' - n_1' \equiv 2$, we must append a one. Then, $F_0$ has weight $3^\ell$.

    A similar argument for $F_1$ yields $c_{S, 1} = n_0 - n_1$, and that $F_1$ has weight exactly equal to $2 \cdot 3^\ell.$ We also note that $F_2$ has weight $3^{l+1}$, because when $F_2$ is expanded, we observe $c_{S,2} = 1$ for all sequences $S$. The coefficients and weights of the basis are summarized in Table \ref{table:basisWeights}.
\begin{table}[h]
\centering
\begin{tabular}{l|c|c}
      & $c_{S, i}$                                                                         & Weight           \\ \hline
$F_0$ & \begin{tabular}[c]{@{}c@{}}1 if $n_0-n_1 \equiv 1$\\ 0 otherwise\end{tabular} & $3^\ell$         \\ \hline
$F_1$ & $n_0 - n_1$                                                                   & $2 \cdot 3^\ell$ \\ \hline
$F_2$ & 1                                                                             & $3^{\ell+1}$    
\end{tabular}
\caption{Coefficients and weights of basis}
\label{table:basisWeights}
\end{table}
    
    Note that among all linear combinations over $\mathbb F_3$ of $F_0, F_1, F_2$, terms of the form $F_i \pm F_j$ with $i\neq j$, $i, j \in \{0, 1, 2\}$ must have weight at least $3^\ell$. Then, we only need to examine elements of the form $\varepsilon_0 F_0 + \varepsilon_1 F_1 + \varepsilon_2 F_2$, where $\varepsilon_i \in \{-1, 1\}$. Letting $\varepsilon_0 = 1$ without loss of generality yields the four cases $F_0 + \varepsilon_1 F_1 + \varepsilon_2 F_2$. Consider first $F_0 + \varepsilon_1 F_1$. A sequence $S$ in $F_0 + F_1$ has coefficient $c_{S, 0} + c_{S, 1} \equiv 2$ if $n_0 \not \equiv n_1$, and zero otherwise, so $F_0 + F_1$ has weight $2 \cdot 3^\ell$ by considering length $\ell$ sequences. Similarly, a sequence $S$ in $F_0 - F_1$ has coefficient $c_{S, 0} - c_{S, 1} \equiv 1$ if $n_1 - n_0 \equiv 2$, and zero otherwise. Then, $F_0 - F_1$ has weight $3^\ell$ also by considering length $\ell$ sequences. In either case $\varepsilon_1 = \pm1$, adding or subtracting $F_2$ results in a term of weight no less than $3^\ell$. In conclusion, all non-trivial linear combinations of $F_0, F_1, F_2$ have weight at least $3^\ell$.
\end{proof}

We are now ready to state the main theorem of this section.
 \begin{theorem} \label{main theorem sl3}
    For any $l\in \mathbb{N}$ and $\mathcal{B}$ is either $\mathcal{B}_1$ or $\mathcal{B}_2$, the $l$th unknot code with basis $\mathcal{B}$ has parameters $[[ n_l; k_l; d_l^\mathcal{B}]]$, where 
    \[
    n_l\sim \frac{15 \cdot 25^l}{2 \sqrt{6 \pi l}}, \ \ k_l = 3, \ \ d_l^{\mathcal{B}} = 3^l,
    \]
    where $\sim$ denotes asymptotic convergence as $l$ tends to infinity.
\end{theorem}
\begin{proof}
    Since $D_{l,l}$ is a diagram of the unknot, the homology at degree $l$ has dimension 3. Thus $k_l = 3$ for all $l$. We now show that the CSS distance is $d_l^\mathcal{B} = 3^l$.
Without loss of generality, we may assume $\mathcal{B} = \mathcal{B}_1$. By Proposition \ref{Dkl to D0l} and Lemma \ref{D0l}, the distance of $D_{l,l}$ is
\[\dist^0(C(D_{l,l});\mathcal{B}_1) = \dist^0(C(D_{0,l});\mathcal{B}_1) = 3^l.\]
Moreover, by Corollary \ref{dual distance}, Proposition \ref{Dkl to D0l} and Lemma \ref{D0l}, the dual distance 
\[\dist^{0}(C^{*}(D_{l,l});\mathcal{B}_1) = \dist^{0}(C(\ol{D}_{l,l});\mathcal{B}_2) = \dist^{0}(C(D_{l,l});\mathcal{B}_2) = \dist^0(C(D_{0,l});\mathcal{B}_2) = 3^l,\]
so the CSS distance is $d_l^\mathcal{B} = 3^l$. 

Finally, we compute the parameter $n_l$. When $0$-resolving all the positive crossings and $1$-resolving all the negative crossings, we obtain $\Theta_{2l}$. Every resolution in $C^0(D_{l,l})$ can be obtained by swapping the resolution of $k$ positive crossings and $k$ negative crossings, where $0\le k\le l$. For a fixed $k$, such a resolution has $2k+1$ generalized theta foams, which we denote as $\Theta_{a_1}, \Theta_{a_2}, \cdots, \Theta_{a_{2k+1}}$, where $a_1,\cdots, a_{2k+1}$ are nonnegative integers that sum up to $2l-2k$. The dimension of this resolution is
    \[(3\cdot 2^{a_1})\cdots (3\cdot 2^{a_{2k+1}}) = 3^{2k+1}\cdot 2^{2l-2k}.\]
    Now, there are $\binom{l}{k}$ ways to swap the positive crossings in $\Theta_{2l}$ and $\binom{l}{k}$ ways to swap the negative crossings. Hence, the total dimension $n_l$ is given by \[n_l = \sum_{k=0}^{l}\binom{l}{k}^2 3^{2k+1}\cdot 2^{2l-2k}.\]
That the asymptotic $n_l \sim \frac{15 \cdot 25^l}{2 \sqrt{6 \pi l}}$ as $l$ tends to infinity is proven in Proposition \ref{prop:sl3 asymptotic}.
\end{proof}

\bibliographystyle{alpha}
\bibliography{main}

\appendix

\section{Alternate proof of Theorem \ref{thm:unknot RII}}
Here, we present a more algebraic proof of Theorem \ref{thm:unknot RII} that is inspired by \cite[Corollary 2.11]{Audoux}.
\begin{proof}
    We first show that sliding an unknot doubles homological distance. We already have the inequality $\dist^i ( \RIIcomplexmerged ) \leq 2\dist^i (\RIIcomplexdisjoint{}{})$ from \cite[Corollary 2.11]{Audoux}, so it suffices to prove the reverse inequality. In the spirit of Audoux, let $x \in C^i (\RIIcomplexmerged)$ represent a nontrivial element of homology. Then, grouping like pictures together, $x$ decomposes as $x = a_- + a_+ + b_- + b_+ + c + d$, where each term in the decomposition is a sum of generators of the form described in the below table.

    \begin{table}[h]
    \begin{tabular}{c|c|c|c|c|c|c}
        & $a_-$ & $a_+$ & $b_-$ & $b_+$ & $c$ & $d$ \\ 
        \hline 
        form of generators & {\RIIcomplexa{$\varepsilon$}{$-$}} & {\RIIcomplexa{$\eta$}{$+$}} & {\RIIcomplexb{$-$}{$\mu$}} & {\RIIcomplexb{$+$}{$\nu$}} & {\RIIcomplexc{}} & {\RIIcomplexd{}}
    \end{tabular}
    \end{table}

    Since $x$ represents an element of homology, we have $\partial (x) = 0$. Focusing only on the pictures which end up looking like \RIIcomplexd{} after applying the differential on $x$ (which means we may ignore the differential on $c$), we get a sum of pictures that must still be 0 after the restriction. This yields
    \begin{align}\label{eq:diffiszero}
        0 = A_- + A_+ + B_- + B_+ + \partial (d) \in C^{i+1} (\RIIcomplexmerged),
    \end{align}
    where $A_-$ is obtained from $a_-$ by applying a multiplication map to each picture \RIIcomplexa{$\varepsilon$}{$-$} in $a_-$ to get a picture like \RIIcomplexd{$\varepsilon$}. We obtain $A_+$, $B_-$, and $B_+$ in a similar fashion, noting that we negate the sign on $\eta$ and $\nu$ to get $A_+$ and $B_+$. Also observe that weights are preserved in the sense that $|a_-| = |A_-|$, $|a_+| = |A_+|$, $|b_-| = |B_-|$, and $|b_+| = |B_+|$.

    At this point, we can view Equation ~\eqref{eq:diffiszero}  inside $C^i (\RIIcomplexdisjoint{}{})$ by first applying applying a small isotopy, then disjoint unioning with a labeled circle. There are two choices of label for the circle, corresponding to two maps $\iota_-$ and $\iota_+$ from $C^{i+1}(\RIIcomplexmerged) \to C^i(\RIIcomplexdisjoint{}{})$ defined by \RIIcomplexd{$\varepsilon$} $\xmapsto{\iota_-}$ \RIIcomplexdisjoint{$\varepsilon$}{$-$} and \RIIcomplexd{$\varepsilon$} $\xmapsto{\iota_+}$ \RIIcomplexdisjoint{$-\varepsilon$}{$+$} respectively. Notice that we negated the left strand in the map $\iota_+$ for reasons that will become clear later on. Also observe that these maps are injective on homology and preserve the weight of each term. If we denote $\iota_\pm A_-$ by $A_-^\pm$ and so on, we get two equations
    \begin{align}
        0 &= A_-^- + A_+^- + B_-^- + B_+^- + \partial (d^-) \label{eq:iota-zero}\\
        0 &= A_-^+ + A_+^+ + B_-^+ + B_+^+ + \partial (d^+) \label{eq:iota+zero}.
    \end{align}
    We now make some observations about what these terms in the equations look like. For each picture \RIIcomplexa{$\varepsilon$}{$-$} in $a_-$, there is a picture \RIIcomplexdisjoint{$\varepsilon$}{$-$} in $A_-^-$. For each \RIIcomplexa{$\eta$}{$+$} in $a_+$, there is a \RIIcomplexdisjoint{$\eta$}{$+$} in $A_+^+$. For each \RIIcomplexb{$+$}{$\nu$} in $b_+$, there is a \RIIcomplexdisjoint{$-\nu$}{$-$} in $B_+^-$ and a \RIIcomplexdisjoint{$\nu$}{$+$} in $B_+^+$.

    Now recall the chain homotopy $F: C^i (\RIIcomplexmerged) \to C^i(\RIIcomplexdisjoint{}{})$ described in \cite[Figure 4]{Audoux}, which is defined by\RIIcomplexa{$\varepsilon$}{$\eta$} $\xmapsto{F}$ \RIIcomplexdisjoint{$\varepsilon$}{$\eta$} and \RIIcomplexb{$+$}{$\nu$} $\xmapsto{F}$ \RIIcomplexdisjoint{$-\nu$}{$-$} $+$ \RIIcomplexdisjoint{$\nu$}{$+$}, and taking everything else to 0. Applying this to $x$, we see that that the $a_-$ and $a_+$ undergo a small isotopy, the $b_+$ term splits under co-multiplication, and all the other terms are killed. Let $Fb_+$ decompose as $Fb_+ = Fb_+^- + Fb_+^+$, where $Fb_+^-$ is a sum of pictures of the form \RIIcomplexdisjoint{$-\nu$}{$-$} and $Fb_+^+$ is a sum of pictures of the form \RIIcomplexdisjoint{$\nu$}{$+$}. Then, we have 
    \begin{align*}
        Fx = Fa_- + Fa_+ + Fb_+^- + Fb_+^+,
    \end{align*}
    which represents a nontrivial homology class in $C^i(\RIIcomplexdisjoint{}{})$. This implies that at least one of $Fa_- + Fb_+^-$ or $Fa_+ + Fb_+^+$ must also represent a nontrivial homology class in $C^i(\RIIcomplexdisjoint{}{})$, or else their sum would also be trivial in homology (also note that both these terms are indeed in the kernel of the differential since the differential does nothing to the circle on the right). We now have two cases to consider.

    If $Fa_- + Fb_+^-$ is nontrivial in homology, we make the key insight that $A_-^- + B_+^- = Fa_- + Fb_+^-$; it is helpful to keep track of the pictures along the way to see this equality. Then, using Equation \eqref{eq:iota-zero}, we have $[A_+^- + B_-^-] = [A_-^- + B_+^-] = [Fa_- + Fb_+^-] \neq [0]$, which means that $|A_+^- + B_-^-| , |A_-^- + B_+^-| \geq \dist^i(\RIIcomplexdisjoint{}{})$. Using the fact that we have $|a_-| = |A_-| = |A_-^-|$ and so on, we get
    \begin{align*}
        |x| &\geq |a_-| + |a_+| + |b_-| + |b_+| \\
        &= |A_-^-| + |A_+^-| + |B_-^-| + |B_+^-|  \\
        &\geq |A_-^- + B_+^-| + |A_+^- + B_-^-| \\
        &\geq 2\dist^i ( \RIIcomplexdisjoint{}{})
    \end{align*}
    On the other hand, if $Fa_+ + Fb_+^+$ is nontrivial in homology, we observe that $A_+^+ + B_+^+ = Fa_+ + Fb_+^+$. Then Equation \eqref{eq:iota+zero} gives $[A_-^+ + B_-^+] = [A_+^+ + B_+^+] \neq [0]$ and $|A_-^+ + B_-^+|, |A_+^+ + B_+^+| \geq \dist^i(\RIIcomplexdisjoint{}{})$. Continuing as before then yields $|x| \geq 2 \dist^i(\RIIcomplexdisjoint{}{})$ as desired.

    Lastly, to show that the code distance also doubles, we note that Proposition \ref{prop:RIIdoesntmatter} implies that sliding the unknot over the link diagram also doubles the distance. This means that in the dual complex (represented by taking the mirror image of the given link diagram), sliding an unknot will also double the homological distance, so the code distance doubles. 
\end{proof}

\section{Asymptotics}
\begin{proposition}
    \label{prop:hopf link asymptotic}
    The expression $c_\ell = (-2)^\ell \sum_{r=0}^\ell \binom{\ell}{r} \binom{2r}{r} (-1)^r$ converges asymptotically as $c_\ell \sim \frac{\sqrt{3} \cdot 6^\ell}{2 \sqrt{\pi \ell}}.$
\end{proposition}
\begin{proof}
    To further simplify this expression, we utilize a generating function as follows: 
    \begin{align*}
        f(x) &\stackrel{\text{def}}{=} \sum_{\ell \geq 0} c_\ell x^{\ell} \\
        &= \sum_{\ell \geq 0} \sum_{r=0}^{\ell} \binom{\ell}{\ell-r}\binom{2r}{r} \cdot (-2)^{\ell} \cdot (-1)^r x^\ell \\
        &= \sum_{r \geq 0} (-1)^r \binom{2r}{r} \sum_{\ell \geq r} \binom{\ell}{r} (-2x)^{\ell}.
    \end{align*}
    The inner sum is known to be the generating function of $\frac{(-2x)^r}{(1+2x)^{r+1}}$, so we now have 
    \begin{align*}
        f(x) &= \sum_{r \geq 0} \frac{(2x)^r}{(1+2x)^{r+1}} \binom{2r}{r} \\
        &= \frac{1}{1+2x} \sum_{r \geq 0} \left(\frac{2x}{1+2x}\right)^r \binom{2r}{r}.
    \end{align*}
    However, it is known that $\sum_{n \geq 0} \binom{2n}{n} x^n = \frac{1}{\sqrt{1-4x}}$ for $|x| < \frac14$ so we have $$f(x) = \frac{1}{1+2x}\cdot \frac{1}{\sqrt{1-\frac{8x}{1+2x}}} = \frac{1}{\sqrt{1-4x-12x^2}},$$ where the original sum converges to $f$ for $|x| < \frac16$.
    We extend to complex numbers to complete our analysis. Let $h: \C \to \C$, $h(z) = f(\frac{z}{6}).$ Note that $[z^n] f(z) = 6^n \cdot [z^n] h(z).$ Then, we have
    \[h(z) = \frac{1}{\sqrt{1 + \frac{1}{3} z}} (1-z)^{-\frac{1}{2}} = \left( \frac{\sqrt{3}}{2} \cdot \frac{1}{\sqrt{1-\frac{1}{4}(1-z)}}\right) (1-z)^{-\frac{1}{2}}. \] Letting $v(z) = \frac{\sqrt{3}}{2} \cdot \frac{1}{\sqrt{1-\frac{1}{4}(1-z)}}$, we note that $v(z)$ is analytic in the disk $|z| < 3$. From the generalized binomial theorem, we have \[v(z) = \sum_{j=0}^\infty \frac{\sqrt{3}}{2} \cdot \left(- \frac{1}{4} \right)^j \binom{-\frac{1}{2}}{j} (1-z)^j = \sum_{j=0}^\infty v_j (1-z)^j.\] Then, by Darboux's Lemma or \cite[Theorem 5.3.1]{Wilf}, we have
    \[[z^n] h(z) = \sum_{j=0}^m v_j \binom{n-j-\frac{1}{2}}{n} + O(n^{-m-\frac{3}{2}}).\] To compute an asymptotic, we need only $m=0$, yielding
    \[
        [z^n] h(z) = \frac{\sqrt{3}}{2} \binom{-\frac{1}{2}}{0} \binom{n-\frac{1}{2}}{n} + O(n^{-\frac{3}{2}}) = \frac{\sqrt{3}}{2} \binom{n-\frac{1}{2}}{n} + O(n^{-\frac{3}{2}}).
    \]
    We have by \cite[Lemma 5.3.2]{Wilf} that 
    \[\binom{n-\frac12}{n} = \binom{-\frac12}{n} (-1)^n = [z^n] (1-z)^{-\frac12} \sim \frac{1}{\Gamma(\frac12) \sqrt{n}} = \frac{1}{\sqrt{\pi n}}.\] Then, using this asymptotic we have
    \[
        [z^n] h(z) = \frac{\sqrt{3}}{2} \binom{n-\frac{1}{2}}{n} + O(n^{-\frac{3}{2}}) \sim \frac{\sqrt{3}}{2 \sqrt{\pi n}}.
    \]

    Finally, since we defined $h(z)$ as $f(\frac{z}{6})$, we have $$c_\ell = [z^\ell] f(z) \sim \frac{\sqrt{3}\cdot 6^\ell}{2\sqrt{\pi \ell}}.$$
\end{proof}
\begin{proposition}
    \label{prop:sl3 asymptotic}
    The expression $n_\ell = \sum_{k=0}^\ell \binom{\ell}{k}^2 3^{2k+1} \cdot 2^{2\ell-2k}$ converges asymptotically as $n_\ell \sim \frac{15 \cdot 25^\ell}{2 \sqrt{6 \pi \ell}}.$
\end{proposition}
\begin{proof}
    Consider the generating function \[f(x) = \sum_{\ell=0}^\infty n_\ell x^\ell = 3 \cdot \sum_{\ell =0}^\infty \sum_{k=0}^\ell \binom{\ell}{k}^2 p^k (4x)^\ell,\] where $p = \frac94$. We wish to find an asymptotic for the coefficients of the generating function $f(x)$. We first examine the Legendre polynomial \[P_\ell(x) = \frac{1}{2^\ell} \sum_{k=0}^\ell \binom{\ell}{k}^2 (x-1)^{\ell-k} (x+1)^k,\] from \cite[Equation 0.2]{Koepf}. Then, we have
    \[P_\ell\left( \frac{1+p}{1-p} \right) = \frac{1}{2^\ell} \sum_{k=0}^\ell \binom{\ell}{k}^2 \left( \frac{2}{1-p}\right)^{l-k} \left( \frac{2p}{1-p}\right)^k = \frac{1}{(1-p)^\ell} \sum_{k=0}^\ell \binom{\ell}{k}^2 p^k.\] Substituting into the expression for $f(x)$, we must have
    \[f(x) = 3 \cdot \sum_{\ell=0}^\infty ((1-p) \cdot 4x)^\ell P_\ell\left(\frac{1+p}{1-p} \right).\] Then, from the generating function for the Legendre Polynomials from \cite[Equation 0.5]{Koepf}, \[\sum_{n=0}^\infty P_n(x) z^n = \frac{1}{\sqrt{1-2xz+z^2}},\] we have
    \[f(x) = 3 \cdot \frac{1}{\sqrt{1 - 2 \left( \frac{1+p}{1-p}\right) ((1-p) \cdot 4x) + ((1-p) \cdot 4x)^2}} = \frac{3}{\sqrt{1 - 26x + 25 x^2}},\] since $p = \frac{9}{4}$. Then, we have
    \[f(x) = \frac{3}{\sqrt{\left(1- x\right) \left(1-25x \right)}},\] where the original summation converges to $f$ for $|x| < \frac{1}{25}.$ Let $g(x) = f\left( \frac{x}{25} \right).$ Note that $[x^\ell] f(x) = 25^\ell [x^\ell] g(x).$ We have
    \[g(x) = \frac{3}{\sqrt{1-\frac{x}{25}}} (1-x)^{-\frac12} = \frac{15}{\sqrt{24 + (1-x)}} (1-x)^{-\frac12}.\] Letting $v(x) = \frac{15}{\sqrt{24+(1-x)}},$ note that $v$ is analytic in the disk $|x| < 25$ and so the first few coefficients of the expansion of $v(x)$ around $x=1$ by the extended binomial theorem are $v_0 = \frac{15}{2\sqrt{6}}, v_1 = \frac{5}{32 \sqrt{6}}, \cdots$. Then, by Darboux's Lemma of \cite[Theorem 5.3.1]{Wilf}, we have
    \[[x^\ell] g(x) = \sum_{j=0}^m v_j \binom{\ell-j-\frac{1}{2}}{\ell} + O(\ell^{-m-\frac32}),\] where $m$ is an arbitrary integer. To derive an asymptotic, it suffices to consider $m=0$, yielding
    \[[x^\ell] g(x) = \frac{15}{2 \sqrt{6}} \cdot \binom{\ell - \frac12}{\ell} + O(\ell^{-\frac32}) \sim \frac{15}{2 \sqrt{6 \pi \ell}}\] as $\ell \to \infty$ by the same argument in Proposition \ref{prop:hopf link asymptotic}. Then, since $n_\ell = [x^\ell] f(x) = 25^\ell [x^\ell] g(x),$ we have
    \[n_\ell \sim \frac{15 \cdot 25^\ell}{2 \sqrt{6 \pi \ell}}.\]
\end{proof}
\end{document}